\documentclass{llncs}

\usepackage{graphics}
\usepackage{graphicx}
\usepackage{subfigure}
\usepackage{fullpage}
\usepackage{color}

\newcommand{\cC}{\mathcal C}

\title{Efficient Algorithms for Distributed Detection of \\ Holes and Boundaries in Wireless Networks
\thanks{A short version of this report is presented at the $10^{th}$ Symposium on Experimental Algorithms (SEA 2011).}
}

\author{Dennis Schieferdecker \and Markus V\"olker \and Dorothea Wagner}
\institute{Karlsruhe Institute of Technology (KIT), Germany \\ \email{\{schieferdecker,m.voelker,dorothea.wagner\}@kit.edu}}

\begin{document}
\maketitle
\begin{abstract}
We propose two novel algorithms for distributed and location-free boundary recognition in wireless sensor networks.
Both approaches enable a node to decide autonomously whether it is a boundary node, based solely on connectivity information of a small neighborhood.
This makes our algorithms highly applicable for dynamic networks where nodes can move or become inoperative.

We compare our algorithms qualitatively and quantitatively with several previous approaches.
In extensive simulations, we consider various models and scenarios.
Although our algorithms use less information than most other approaches, they produce significantly better results.
They are very robust against variations in node degree and do not rely on simplified assumptions of the communication model.
Moreover, they are much easier to implement on real sensor nodes than most existing approaches.
\end{abstract}


\section{Introduction}
Wireless sensor networks have become a prominent research topic in recent years.
Their unique structure and limitations provide new and fascinating challenges.
A sensor network consists of a union of small nodes that are equipped with sensing, communication, and processing capabilities.
The nodes usually only have a limited view of the network. 
Therefore, distributed algorithms that work on local information are best suited for the emerging tasks in these environments.

Many applications in sensor networks require a certain knowledge of the underlying network topology, especially of holes and boundaries.
Examples are intrusion detection, data gathering~\cite{wgm06}, mundane services like efficient routing within the network~\cite{fgg04,rao03}, or event detection~\cite{dll09}.
In many situations, holes can also be considered as indicators for insufficient coverage or connectivity.
Especially in dynamic settings, where nodes can run out of power, fail, or move, an automatic detection of holes and boundaries is inevitable.

For this reason, many boundary recognition algorithms have been developed previously.
However, most of them have certain disadvantages.
Some rely on oversimplified assumptions concerning the communication model or on knowledge about absolute or relative node positions, which is usually not available in large-scale sensor networks.
Other algorithms are not distributed or require information exchange over long distances, so they do not scale well with network size.
And those algorithms that solely work locally usually produce many misclassifications.
Furthermore, many of the existing algorithms are too complex for an actual implementation on real sensor nodes.
So there is still demand for simple and efficient algorithms for boundary recognition.

\subsubsection{Related Work}
Since there is a wide range of applications that require boundary detection, there is an equally large number of approaches to detect and classify holes.
Based on the underlying ideas, the approaches can be classified roughly into three categories.

\emph{Geometrical approaches} use information about node positions, distances between nodes, or angular relationships to detect network boundaries and holes.
Accordingly, these approaches are limited to situations where GPS devices or similar equipment are available.
Unfortunately, in many realistic scenarios this is not the case.
Examples for geometrical approaches are Fang \emph{et al.}~\cite{fgg04}, Martincic~\emph{et al.}~\cite{martincic04}, and Deogun~\emph{et al.}~\cite{deogun05}.

\emph{Statistical approaches} try to recognize boundary nodes by low node degree or similar statistical properties.
As long as nodes are smoothly distributed, this works quite well as boundary nodes usually have less neighbors than interior nodes.
However, as soon as node degrees fluctuate noticeably, most statistical approaches produce many misclassifications.
Besides, these algorithms often require unrealistic high average node degrees. 
Prominent statistical approaches are Fekete \emph{et al.} \cite{fkkl-brgsn-05,fkpfb04}, and Bi~\emph{et al.}~\cite{bi06}.

\emph{Topological approaches} concentrate on information given by the connectivity graph and try to infer boundaries from its topological structure.
The algorithm of Kr\"oller \emph{et al.}~\cite{kfpf-dbrte-06} works by identifying complex combinatorial structures called flowers.
Such flowers exist with high probability under some assumptions on the communication model if the average node degree is above 20.
The algorithm requires that every node knows its $8$-hop neighborhood.
Funke~\cite{f05} and Funke \emph{et al.}~\cite{fk06} describe algorithms that construct iso-contours and check whether those contours are broken.
If a node recognizes that a contour is broken, it classifies the corresponding contour end-points as border nodes.
The first algorithm requires that the whole network is flooded starting from some seed nodes.
The second algorithm works distributed, based on $6$-hop neighborhoods.
An algorithm that works well even in networks with low average node degree is given by Wang \emph{et al.}~\cite{wgm06}.
The algorithm involves several steps, some of which require that the whole network is flooded.
The methods proposed by Ghrist \emph{et al.}~\cite{ghrist05} and De Silva \emph{et al.}~\cite{ghrist06} detect holes by utilizing algebraic homology theory.
They are centralized and rely on restrictive assumptions on the communication model.
In~\cite{saukh08,saukh10}, Saukh \emph{et al.} propose an algorithm that tries to identify certain patterns in the neighborhood of a node.
Under certain conditions, they can guarantee that all nodes which are classified as inner nodes lie inside of the network.
The algorithm is distributed and every node only needs information of its $h$-hop neighborhood.
The radius $h$ depends on the node density. For low density, $h=6$ is used. For higher densities, it is possible to use
smaller neighborhoods.
A recent distributed algorithm by Dong \emph{et al.}~\cite{dll09} is especially aimed at locating small holes.

In previous work, there exist several classification schemes for boundary detection.
Until recently, most boundary (or hole) definitions were based on an embedding of the connectivity graph.
In \cite{fgg04}, Fang \emph{et al.} determine the Delaunay triangulation of the embedded connectivity graph and remove edges of length greater than one.
They classify faces of this reduced Delaunay graph with at least four edges as holes of the network.
Boundary nodes are those nodes that induce these faces.
In \cite{kfpf-dbrte-06}, Kr\"oller \emph{et al.} define boundaries according to a decomposition of the plane into faces based on the embedded connectivity graph.
A face is called a hole if the circumference of its convex hull exceeds a minimum value.
Since vertices of a face usually do not correspond to network nodes, the authors define boundary nodes to be the nodes on a cycle in the network graph surrounding this face.
The approaches by Fekete \emph{et al.}~\cite{fkkl-brgsn-05,fkpfb04} only apply a basic boundary definition for the continuous case.
Given a set of holes, a closed cycle is called a boundary if it separates the area of a hole from the area occupied by the network.
A mapping of the continuous boundary to network nodes is not defined.
Saukh \emph{et al.}~\cite{saukh10} classify a node as boundary node if there exists any feasible embedding of the connectivity graph in which this node is located on the boundary of the embedded graph.
Recently, Dong \emph{et al.} \cite{dll09} proposed a topological boundary definition.
They define a cycle in the connectivity graph to be a topological boundary if, given an arbitrary embedding of the graph, the embedded cycle can be continuously transformed into a boundary of the embedded graph.

\subsubsection{Our Contributions}
We propose two novel boundary recognition algorithms.
Both enable a node to decide solely based on the \emph{connectivity information} of its \emph{local neighborhood} whether it is a boundary node.
Our first approach uses multidimensional scaling on the $2$-hop neighborhood of a node to reconstruct relative angles between neighboring nodes.
This information is then used to decide whether the considered node is surrounded by other nodes.
The second algorithm infers from connectivity information if the node is enclosed by a closed circle of nodes within its $2$-hop neighborhood.
Both algorithms have several benefits over existing approaches.
Unlike many other algorithms, they are strictly local and suited for distributed application.
They are completely based on connectivity information, so no information about absolute or relative node positions is necessary.
They do not require the underlying network to be a unit disk graph or rely on other simplistic assumptions.
Furthermore, they are much easier to understand and implement than most existing approaches.
Both algorithms are very robust to non-uniform node deployment and variations in node degree.
Finally, they are equally well suited to detect extensive boundaries or small network holes that can occur when single nodes fail or move.
All this makes them perfectly suited for application in large-scale networks.
    
In extensive simulations, we compare our algorithms with several other well-known algorithms.
Unlike most existing papers, we compare the algorithms not only qualitatively but also quantitatively.
In order to allow an objective comparison, we give an intuitive definition of \emph{network holes} and classify network nodes into three separate groups--\emph{mandatory boundary nodes}, \emph{optional boundary nodes}, and \emph{interior nodes}--based on their position relative to network holes and borders.
The simulations clearly show that our algorithms, despite their simplicity, outperform the other algorithms in most scenarios by detecting a higher percentage of boundary nodes correctly and, at the same time, misclassifying less interior nodes.

\section{Model Description}\label{sec:model}

\subsection{Network Model}\label{sec:network_model}
A sensor network consists of nodes located in the two-dimensional plane according to some distribution.
Communication links between nodes induce a \emph{connectivity graph} $\cC(V, E)$ with graph nodes $v \in V$ corresponding to sensor nodes and graph edges $(u, v) \in E;\ u, v \in V$ to communication links between sensor nodes.
An \emph{embedding} $p: V \rightarrow \bbbr^2$ of the connectivity graph $\cC$ assigns two-dimensional coordinates $p(v)$ to each node $v \in V$.
For easier reading, distances are normalized to the maximum communication distance of the sensor nodes.

\subsubsection{Communication model}
Two communication models are considered.
Both assume bidirectional communication links.
In the \emph{unit disk graph} (UDG) model, two sensor nodes $u, v \in \cC$ can communicate with each other, i.e., there exists a communication link between them, if their distance $|p(u)p(v)|$ is at most $1$.
In the \emph{quasi unit disk graph} (d-QUDG) model, sensor nodes $u, v \in \cC$ can communicate reliably if $|p(u)p(v)| \leq d$ for a given $d \in [0,1]$.
For $|p(u)p(v)| > 1$ communication is impossible.
In between, communication may or may not be possible with equal probability.

Signal strength is mostly ignored throughout this work.
However, we can exploit this additional information to improve the quality of our boundary recognition.
Section \ref{ssec:mdsbr_variants} gives more details on this subject.

\subsubsection{Node distribution}
Two node distribution strategies are considered.
Using \emph{perturbed grid placement}, nodes are placed on a grid with grid spacing $0.5$ and translated by a uniform random offset taken from $[0, 0.5]$ in both dimensions.
Using \emph{random placement}, nodes are placed uniformly at random on the plane.

Perturbed grid placement guarantees a more uniform node distribution with less variance in node degree.
It is frequently used in papers on boundary recognition.
It models deployments where sensor nodes are placed in a regular pattern without the need to closely watch the exact placement.
Random placement, on the other hand, models situations where sensor nodes are arbitrarily scattered in the environment.

\subsection{Hole and Boundary Model}
For evaluation, well-defined hole and boundary definitions are required.
The definitions introduced by previous contributions are often too complicated or not extensive enough.
Several approaches define holes but do not specify which nodes are considered boundary nodes, while others classify boundary nodes without taking into account their positions.
In our work, we take a very practical look at what to label as holes.
In short, we call large areas with no communication links crossing them holes and nodes on the borders of these areas boundary nodes.

\subsubsection{Hole Definition}
Some previous definitions are based on abstract topological definitions.
In contrast, we think that the hole definition should be based on the embedding of the actual sensor network.
Thus, for evaluation only, we take advantage of the true node positions.

All faces induced by the edges of the \emph{embedded connectivity graph} $p(\cC)$ are hole candidates.
Similarly to \cite{kfpf-dbrte-06}, we define holes to be faces of $p(\cC)$ with a circumference of at least $h_{min}$.
Figure~\ref{fig:hole_definition}(left) depicts a hole according to our definition.
Take note that the exterior of the network is an infinite face.
Thus, it is regarded as a hole for the purpose of computation and evaluation.

\subsubsection{Boundary Node Definition}
\begin{figure}[t]
\begin{minipage}{0.48\columnwidth}
\centering
\includegraphics[page=1,height=.45\columnwidth]{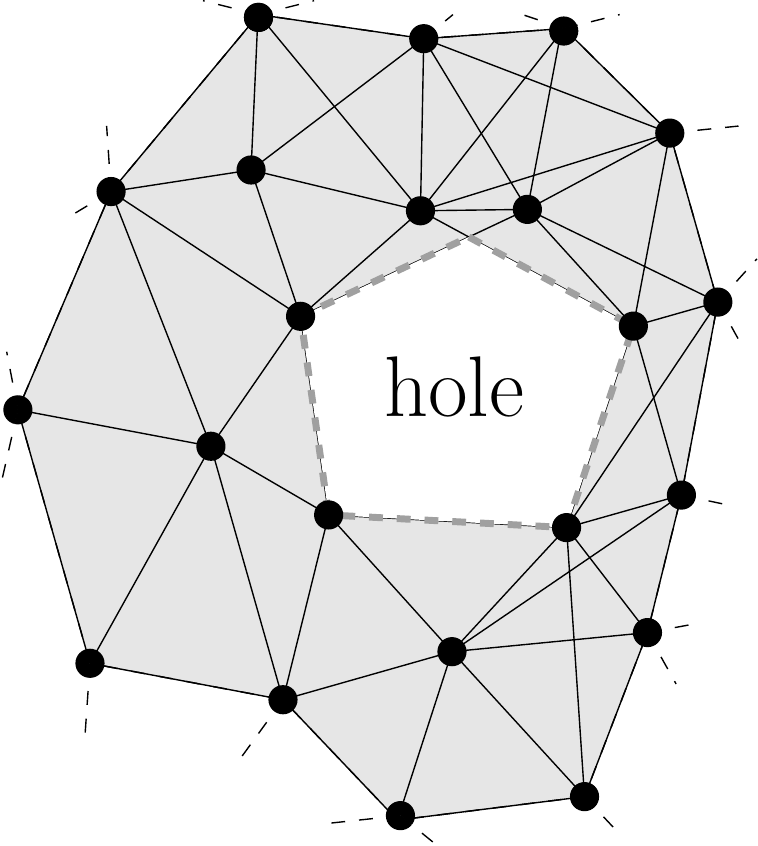}
\hspace{2em}
\includegraphics[page=2,height=.45\columnwidth]{figs_ds/mdsbr_hole_definition.pdf}
\caption{(left) Hole Definition: Border as dashed line. (right) Boundary Node Classification: Mandatory nodes (white boxes), optional nodes (gray boxes), interior nodes (black circles).}\label{fig:hole_definition}
\end{minipage}
\hfill
\begin{minipage}{0.48\columnwidth}
\centering
\includegraphics[height=.45\columnwidth]{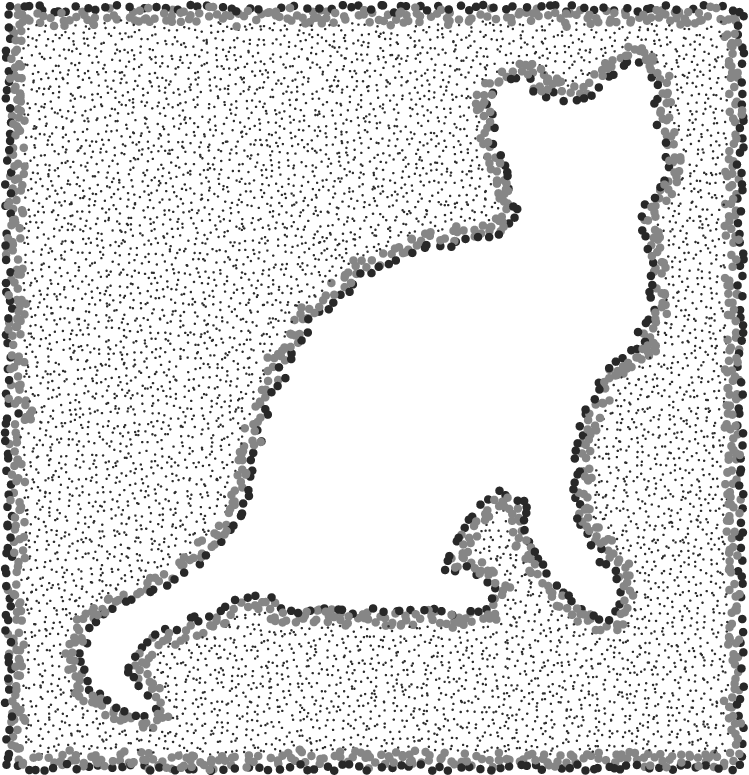}
\hfill
\includegraphics[height=.45\columnwidth]{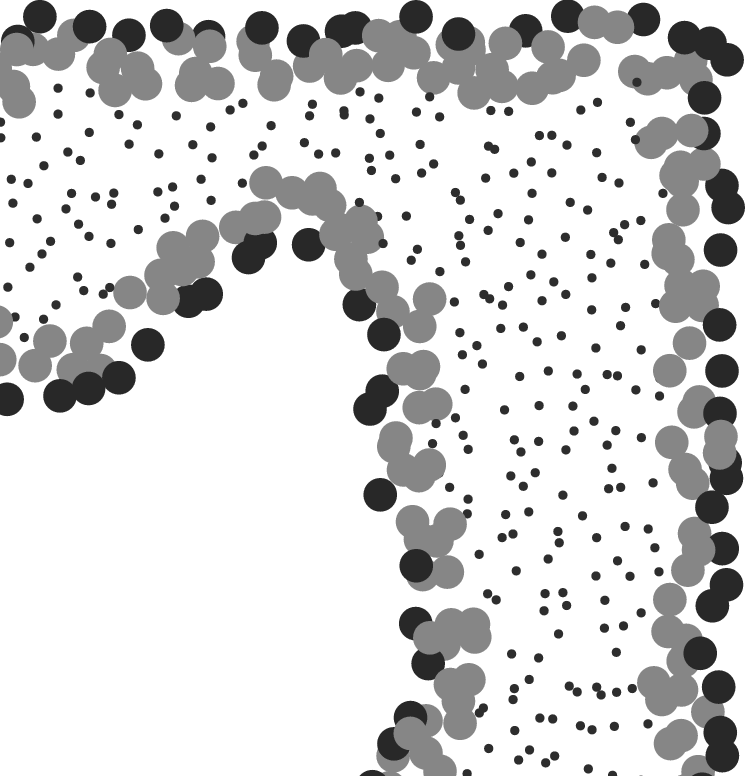}
\caption{Node Classification. Border outline of mandatory nodes (large, black), halo of optional nodes (large, gray), interior nodes (small, black). Full network and magnified upper right corner are shown.}\label{fig:hole_definition2}
\end{minipage}
\end{figure}

As seen in Figure~\ref{fig:hole_definition}(left), hole borders and node locations do not have to align.
Thus, there exists the problem which nodes to classify as boundary nodes.
For example, it can be argued whether nodes $A$ and $B$ should be boundary nodes or not.
To alleviate this problem, we classify nodes into three categories:
\begin{itemize}
	\item \emph{Mandatory Boundary Nodes.} Nodes that lie exactly on the hole border are boundary nodes.
	\item \emph{Optional Boundary Nodes.} Nodes within maximum communication distance of a mandatory node can be called boundary nodes but do not have to be.
	\item \emph{Interior Nodes.} All other nodes must not be classified as boundary nodes.
\end{itemize}
The resulting node classification is shown in Figure~\ref{fig:hole_definition}(right).
Mandatory boundary nodes form thin bands around holes, interrupted by structures like for nodes $A$ and $B$ before.
Together with the optional boundary nodes, they form a halo around each hole.
Any point within the halo is at most one maximum communication distance away from the border of the enclosed hole.
A sample classification is depicted in Figure~\ref{fig:hole_definition2}.

\section{Multidimensional Scaling Boundary Recognition (MDS-BR)}\label{sec:mdsbr}
Both of our algorithms work in a distributed fashion and only require local connectivity information.
Each node independently decides whether it is a boundary node or an interior node, solely using information gathered from a small neighborhood.

Our first algorithm is a geometrical approach at its core.
But instead of using real node coordinates, which are usually not known in sensor networks, it uses multidimensional scaling to compute virtual coordinates.
Two angular conditions are then tested to classify a node, followed by a refinement step after all nodes have classified themselves.
Subsequently, a short review of multidimensional scaling is given, followed by the description of the base algorithm and the refinement.
Additional variants of the algorithm that incorporate further information to improve classification results conclude the section.

\subsubsection{Multidimensional Scaling}
Multidimensional scaling (MDS) \cite{t-mdstm-52} is a well-known technique for finding a good embedding, given a matrix of distances between nodes.
More formally, given distances $d_{uv}$ between each pair of nodes $u, v \in V$, MDS computes an embedding $p$ that minimizes the sum $\sum_{u,v \in V, u \neq v} ( |p(u)p(v)| - d_{uv} )^2$ of squared differences between given distances $d_{uv}$ and induced distances $|p(u)p(v)|$.
The running time of MDS is in $\Theta(n^3)$, with $n$ the number of nodes.
For a detailed description and variants of MDS refer to \cite{cc-mds-01}.

\subsubsection{Base Algorithm}
\begin{figure}[t]
\centering
\begin{tabular}{llllll}
 a) \hspace{0.2cm} &
\parbox[c]{.2\columnwidth}{\includegraphics[page=1,width=.2\columnwidth]{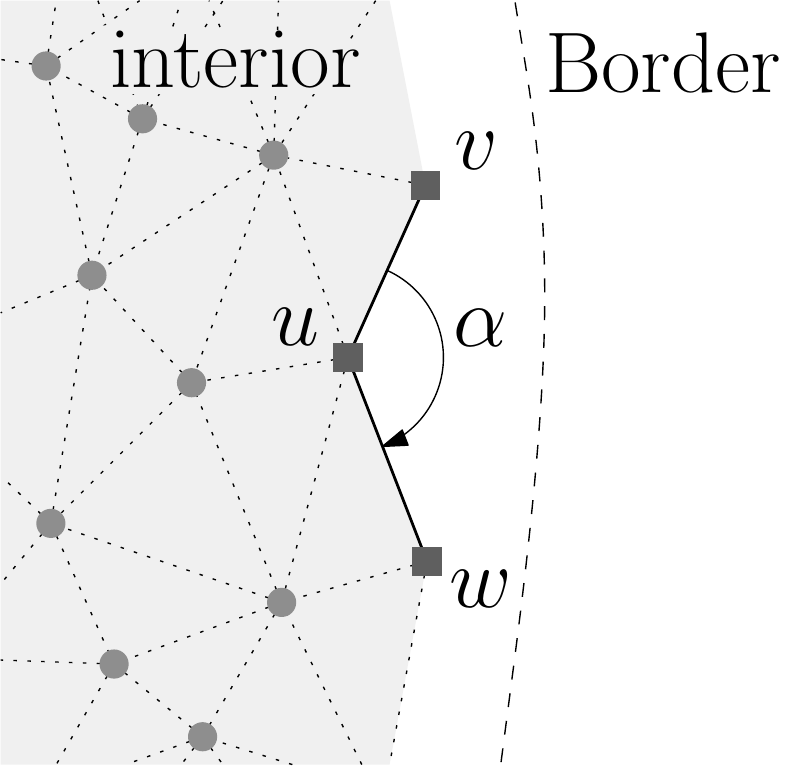}} &
\hspace{0.7cm}  b) \hspace{0.2cm} &
\parbox[c]{.2\columnwidth}{\includegraphics[page=2,width=.2\columnwidth]{figs_ds/mdsbr_boundary_condition.pdf}} &
\hspace{0.7cm}  c) \hspace{0.2cm} &
\parbox[c]{.2\columnwidth}{\includegraphics[page=3,width=.2\columnwidth]{figs_ds/mdsbr_boundary_condition.pdf}}
\end{tabular}
\caption{MDS-BR Classification Conditions. (a) Minimum opening angle. (b) Prohibited cone. (c) Micro-holes.}
\label{fig:bn_definition}
\end{figure}

Each node performs the following base algorithm to decide independently whether to classify itself as a boundary node.
At first, each node $u$ gathers its $2$-hop neighborhood $N_u^2$ and computes a two-dimensional embedding of $N_u^2 \cup \{u\}$ using hop distances to approximate true distances between nodes.
Then, using these virtual locations, node $u$ declares itself to be a boundary node if two conditions are fulfilled.
First, the maximum opening angle $\alpha$ between two subsequent neighbors $v$, $w$ of $u$ in circular order and $u$ must be larger than a threshold $\alpha_{min}$ as depicted in Figure~\ref{fig:bn_definition}(a).
This primary condition models the observation that boundary nodes exhibit a large gap in their neighborhood compared to interior nodes that are usually completely surrounded by other nodes.
Secondly, neighbors $v$, $w$ of $u$ must not have common neighbors other than $u$ in the cone opened by $(uv)$ and $(uw)$.
This is exemplified in Figure~\ref{fig:bn_definition}(b).
This condition filters micro-holes framed by $4$ nodes with a circumference of at most $4$ maximum communication distances as seen in Figure~\ref{fig:bn_definition}(c).
If such holes are to be detected, this condition can be omitted.

Both conditions only require angular information.
Thus, any embedding algorithm yielding realistic angles between nodes is sufficient -- we are not limited to MDS.
Furthermore, we do not need complex embedding techniques to compensate for problems occurring in large graphs such as drifting or foldings since we only embed very small graphs.
In particular, we only compute embeddings of $2$-hop neighborhoods around each node, i.e., graphs of diameter $4$ or less.

The running time of the base algorithm is dominated by the computation of the embedding.
The asymptotic time complexity at each node $u$ is $O(n^3)$, with $n=|N_u^2|$.
Given the maximum node degree $d_{max}$ of the graph, we can assess $|N_u^2|=O(d_{max})$ \cite{pg-fdwcd-04}.
Thus, although the asymptotic time complexity is poor, the actual computation time is short since only few nodes are considered.
Communication is limited to gathering a $2$-hop neighborhood.
If we assume synchronous communication in rounds and data aggregation, each node has to send at most $2$ messages of maximum size $O(d)$, given node degree $d$.

\subsubsection{Refinement}
\begin{figure}[b]
\centering
\begin{tabular}{llllll}
 a) \hspace{0.2cm} &
\parbox[c]{.2\columnwidth}{\includegraphics[width=.2\columnwidth]{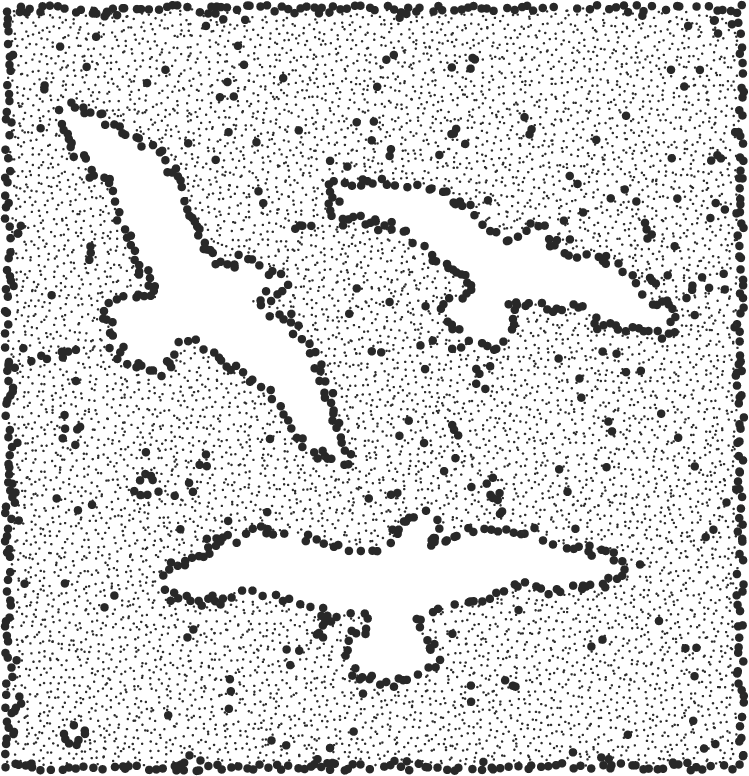}} &
\hspace{1cm}  b) \hspace{0.2cm} &
\parbox[c]{.2\columnwidth}{\includegraphics[width=.2\columnwidth]{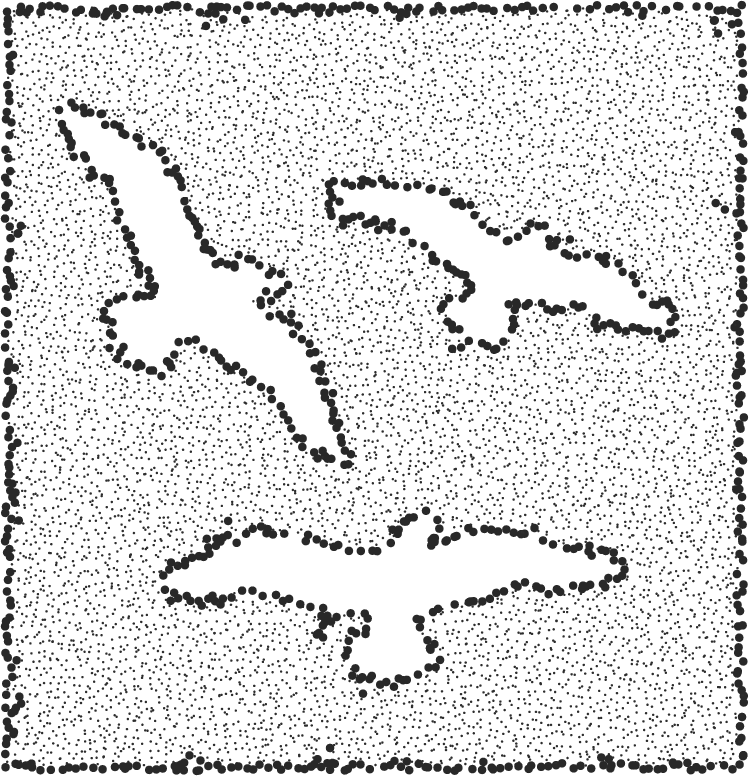}}
\hspace{0.3cm}
\end{tabular}
\caption{Classification results of MDS-BR on a sample network. Boundary nodes are marked as larger black nodes. (a) Results of the base algorithm. ``Noise'' is clearly visible. (b) Results after refinement.}
\label{fig:tlbr_refinement}
\end{figure}

The base algorithm already yields good results as shown in Figure~\ref{fig:tlbr_refinement}(a).
But it retains some ``noise'' due to detecting boundary nodes around small holes one might not be interested in and due to some misclassifications.
If desired, a refinement step can be used to remove most of these artifacts as seen in Figure~\ref{fig:tlbr_refinement}(b).

\begin{figure}[t]
\centering
\begin{tabular}{llllll}
a) \hspace{0.2cm} &
\parbox[c]{.25\columnwidth}{\includegraphics[page=4,width=.25\columnwidth]{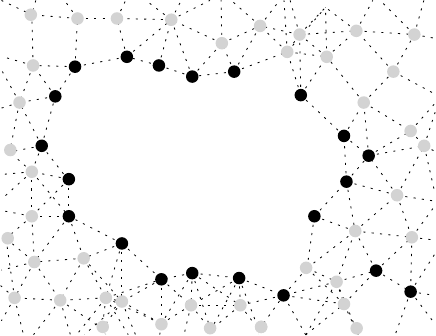}} &
\hspace{0.5cm}  b) \hspace{0.2cm} &
\parbox[c]{.25\columnwidth}{\includegraphics[page=5,width=.25\columnwidth]{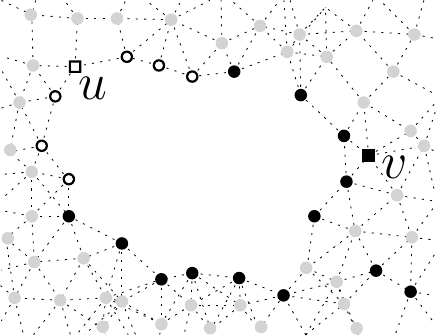}} &
\hspace{0.5cm}  c) \hspace{0.2cm} &
\parbox[c]{.25\columnwidth}{\includegraphics[page=6,width=.25\columnwidth]{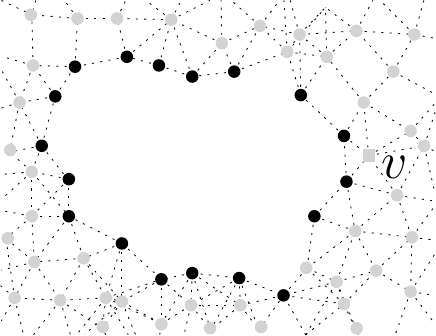}}
\end{tabular}
\caption{(a) Tentative boundary nodes after the base algorithm are shown in black. (b) An exemplary marked $3$-hop neighborhood, $\tilde N_u^3$, that is used during refinement is highlighted in white. (c) Boundary nodes remaining after a refinement step with $r_{min} = 3$ are shown in black. Note that nodes previously marked as boundary nodes in the bottom right have been removed as well as the satellite node $v$.}\label{fig:tlbr_refinement2}
\end{figure}

The refinement is performed distributed on the current set of boundary nodes.
First, each boundary node $u$ gathers its $r_{min}$-hop neighborhood $\tilde N_u^{r_{min}}$ of nodes marked as boundary nodes by the base algorithm, where $r_{min}$ is a free parameter.
Then, $u$ verifies if there exists a shortest path of at least $r_{min}$ hops in $\tilde N_u^{r_{min}} \cup \{u\}$ that contains $u$.
If no such path exists, $u$ classifies itself as interior node.
This approach removes boundary nodes that are not part of a larger boundary structure, with $r_{min}$ specifying the desired size of the structure as illustrated in Figure~\ref{fig:tlbr_refinement2}.
Note that only connectivity information is required for the refinement.

Without optimization, the time complexity of the refinement step at each node $u$ is bounded by $O(n^3)$, with $n=|\tilde N_u^{r_{min}}|$.
This complexity is uncritical as the marked $r_{min}$-hop neighborhood $\tilde N_u^{r_{min}}$ is much smaller than the full $r_{min}$-hop neighborhood.
As seen in Figure~\ref{fig:tlbr_refinement}(a), the base algorithm yields thin bands of marked nodes around holes and small groups of individual nodes elsewhere.
Larger marked neighborhood structures only occur within these bands and, evidently, they are much smaller than a full $r_{min}$-hop neighborhood.
Section~\ref{ssec:parameter_tuning} quantifies this observation.
In particular, our results suggest an average growth of $|\tilde N_u^{r_{min}}|$ linear in $r_{min}$.
Communication is only required for gathering a small neighborhood.
If we assume synchronous communication in rounds and data aggregation, each node has to send at most $r_{min}$ messages.
The maximum message size is determined by the size of the marked neighborhood and is in $O(n)$ with $n = |\tilde N_u^{r_{min}-1}|$.

\subsubsection{Variants}\label{ssec:mdsbr_variants}
We can exploit additional information to improve the graph embedding and, in turn, improve the quality of the boundary recognition.
We introduce three possible variants and present an evaluation of their performance in Section~\ref{ssec:variants}.

First, we can assume that we know the actual node positions.
With this variant, called the optimum variant, we can gauge the performance of the classification conditions independently of the quality of an embedding.
Secondly, we have the option to gather larger neighborhoods for MDS.
Using $3$-hop neighborhoods yields better embeddings at still reasonable computation costs.
Last and most interesting, signal strengths can be incorporated to estimate approximate node distances.
Since signal strength is a very fluctuating quantity, depending on many exterior factors such as node orientation or signal interference, we opted to only differentiate between two states, strong signals and weak signals.
If node $u$ receives a weak signal from another node $v$, we assume that they are far apart and set their distance for MDS to $d_{uv} = 1.0$.
Otherwise, we assume that the nodes are closer together and set $d_{uv} = 0.5$.
Distances between non-neighboring nodes are derived from these distances.

Further variants comprise replacing the embedding technique or the classification conditions.
Changing the first can, at best, yield results on par with the optimum variant.
Thus, the main reason for applying a different embedding technique would be shorter running times.
Relying on more sophisticated classification conditions should improve the quality of the results, but this would conflict with our design goal to develop a simple yet efficient algorithm.

\section{Enclosing Circle Boundary Recognition (EC-BR)}\label{sec:ecbr}
\begin{figure}[t]
\centering
\hfill
\subfigure[]{\includegraphics[page=1,width=.17\columnwidth]{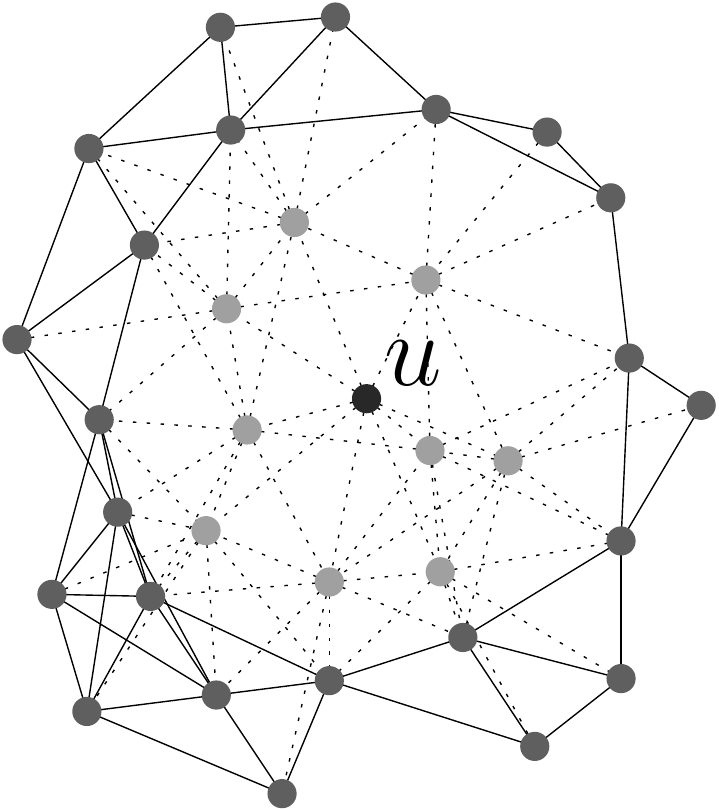}}
\hfill
\subfigure[]{\includegraphics[page=2,width=.17\columnwidth]{figs_mv/enclosing_circle.pdf}}
\hfill
\subfigure[]{\includegraphics[page=4,width=.17\columnwidth]{figs_mv/enclosing_circle.pdf}}
\hfill
\subfigure[]{\includegraphics[width=.20\columnwidth]{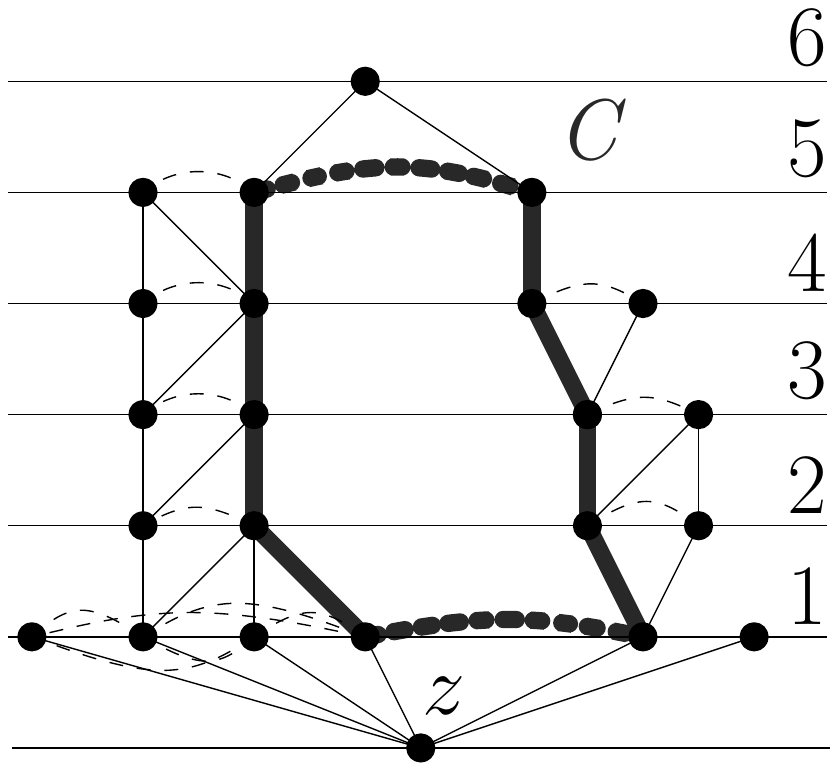}}
\hfill{}
\caption{Basic Idea of EC-BR. (a) $2$-hop neighborhood of node $u$. (b) Enclosing circle $\mathit{C}$. (c) Boundary node without enclosing circle. A non-enclosing circle is highlighted. 
(d) Modified breadth-first search.}
\label{fig::ecbr_idea}
\end{figure}

Our second algorithm, EC-BR, allows to detect reliably if a node is surrounded by other nodes without having to reconstruct node positions.
The basic idea is as simple as efficient.
The node ignores its direct neighbors and considers only nodes that are exactly two hops away.
For a node $u$, we denote the corresponding node set as $N_u^{2\setminus1}$ and the induced subgraph as $G_u^{2\setminus1} = (N_u^{2\setminus1}, E_u^{2\setminus1})$.
Based on the connectivity information in  $G_u^{2\setminus1}$, the node tries to decide if it is surrounded by a closed path $\mathit{C}$.
If such a closed path exists, one can be sure that the node is not a boundary node.
Otherwise this is seen as an indication that the node lies somewhere near a hole or border (compare Figures~\ref{fig::ecbr_idea}(a)-(c)).

\subsubsection{Enclosing Circle Detection}
Knowing the actual node positions, it would be easy to decide if an enclosing circle exists.
However, we do not have this information and we also do not want to reconstruct node positions in order to save computation time.
So, how can we distinguish between enclosing circles as the one in Figure~\ref{fig::ecbr_idea}(b) and non-enclosing circles such as the one in Figure~\ref{fig::ecbr_idea}(c)?
The length of the circle is obviously no sufficient criterion as both circles have the same length and only the first one is enclosing.
Fortunately, there is a structural difference between both types of circles: the circle in Figure~\ref{fig::ecbr_idea}(b) encloses the hole like a tight rubber band and there is no way to split it into smaller circles by adding edges from $E_u^{2\setminus1}$, whereas it is quite easy to find edges in Figure~\ref{fig::ecbr_idea}(c) that could be used to split the circle into shorter circles.
More formally, the first circle has the property that for each pair ${v,w}$ of nodes on the circle, the shortest path between them using only circle edges is also a shortest path between them in $G_u^{2\setminus1}$.
Now we try to find a preferably long circle with this property in $G_u^{2\setminus1}$.
If we assume a unit disk graph and node $u$ is enclosed by nodes of  $G_u^{2\setminus1}$, there has to exist such a circle of length at least $7$.
On the other hand, if $u$ lies somewhere near a hole then $u$ is not fully enclosed by other nodes and it is highly unlikely that a large circle with the aforementioned property exists.

To find a maximum circle with the given property, we can use a modified breadth-first search.
The corresponding search tree for $G_u^{2\setminus1}$ of Figure~\ref{fig::ecbr_idea}(b) is depicted in Figure~\ref{fig::ecbr_idea}(d).
We start the search from a random node $z$ in $G_u^{2\setminus1}$ with maximum degree.
In every step of the search, we maintain shortest path lengths for all pairs of visited nodes.
When a new edge is traversed, there are two possibilities: either a new node is visited, or a previously encountered node is revisited.
In the first case, we just set the shortest path distances between the old nodes and the new node.
This can be done efficiently, as all distances can be directly inferred from the distances to the parent node.
In the second case, we found a new circle in $G_u^{2\setminus1}$.
The length of the circle is the current shortest path between the endpoints $v$ and $w$ of the traversed edge $e=(v,w)$ plus one.
Subsequently, we update the shortest path information of all nodes.
During the search, we keep track of the maximum length of a circle encountered so far.
Depending on the maximum length that was found during the search, the considered node is classified as either a boundary node or an inner node.
For every edge in $G_u^{2\setminus1}$, the update of pairwise distances has to be performed at most once.
Thus, the asymptotic time complexity of this approach is in $O(m n^2)$, with $m=|E_u^{2\setminus1}|$ and $n=|N_u^{2\setminus1}|$.
Later in this section we will show how this complexity can be reduced to $O(m)$.

\begin{figure}[t]
\centering
\begin{tabular}{llll}
 a) \hspace{0.2cm} &
\parbox[c]{.23\columnwidth}{\includegraphics[width=.23\columnwidth]{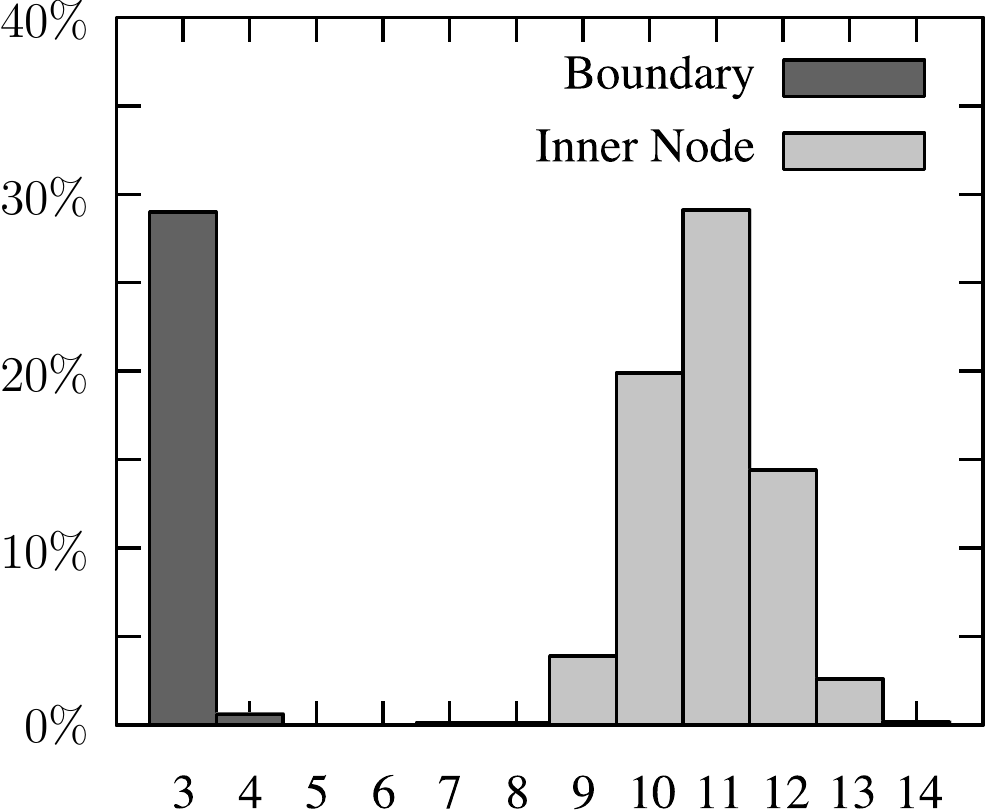}} &
\hspace{0.8cm}  b) \hspace{0.2cm} &
\parbox[c]{.23\columnwidth}{\includegraphics[width=.23\columnwidth]{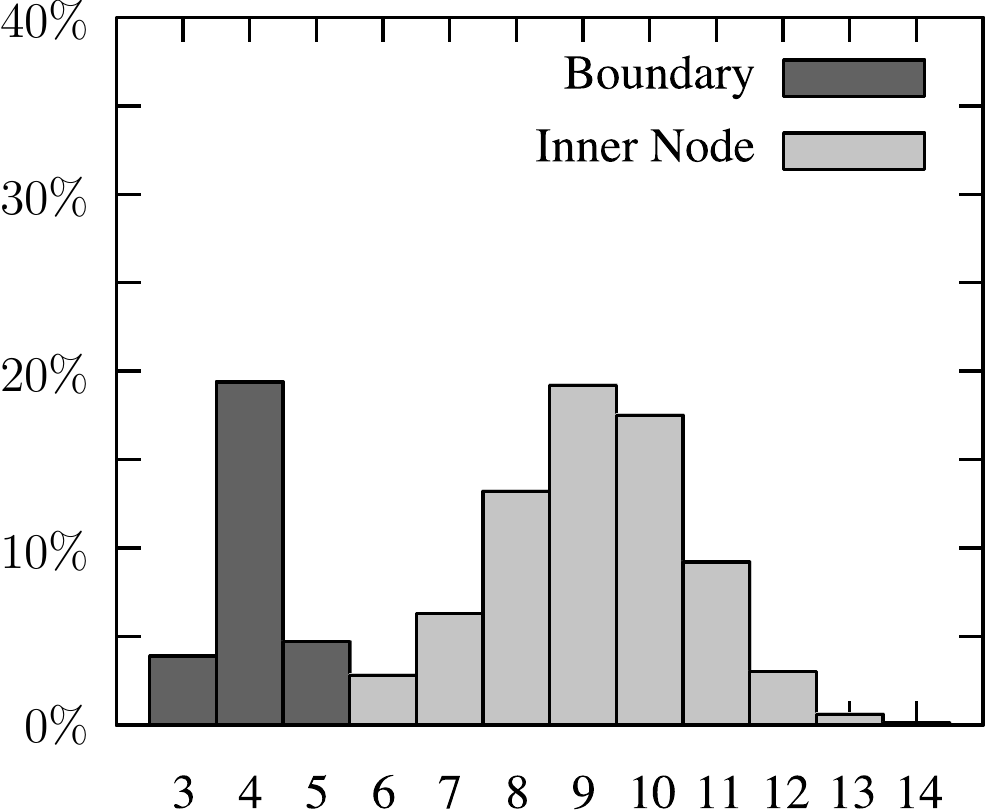}}
\end{tabular}
\caption{Distribution of Maximum Circle Lengths. (a) UDG. (b) QUDG.}
\label{fig::circle_length}
\end{figure}
Figure~\ref{fig::circle_length} depicts histograms of maximum circle lengths in our simulations with networks based on unit disk graphs and quasi unit disk graphs. There are apparently two very well defined peaks, corresponding to nodes with and without enclosing circles.
Based on this distribution, we classify all nodes $u$ that have a maximum circle in $G_u^{2\setminus1}$ with length of at least $6$ as inner nodes and all other nodes as boundary nodes.
Our simulations indicate that this statistical classification into nodes with and without enclosing circle works extremely well for both UDGs and QUDGs.
Later on, we will see how good this correlates with being in the interior or being on the boundary of the network.

It is also noteworthy that this kind of classification is extremely robust to variations in node degree: it does not matter whether $N_u^{2\setminus1}$ consists of a small number of nodes or hundreds of nodes, the same threshold 6 on the maximum circle length can be used to distinguish interior nodes from boundary nodes. 
The classification stays the same, as long as we assume that the node density is sufficiently high so that inner nodes are actually surrounded by other nodes.
This robustness distinguishes EC-BR from existing statistical approaches.

\subsubsection{Enclosing Circle Detection in Linear Time}
\label{ecbr_lintime}
The enclosing circle detection of EC-BR runs distributed on every single node and each node only has to consider its $2$-hop neighborhood, so its runtime is virtually uncritical.
Nevertheless, for the sake of completeness, we mention how the search can be improved to run in time $O(m)$, with $m=|E_u^{2\setminus1}|$, if the underlying network has properties of a quasi unit disk graph.
The key insight is that it does not make any difference for the classification if a node $u$ is enclosed by thousands of nodes or by just enough nodes so that the circle is closed.
Thus, in a first step, node $u$ can filter $N_u^{2\setminus1}$ to a small set of representatives.
By considering each edge in $E_u^{2\setminus1}$ once, a maximal independent set $\mathcal{I}$ of $G_u^{2\setminus1}$ can be computed in time $O(m)$.
Based on packing arguments, the number of nodes in set $\mathcal{I}$ is bounded by a small constant for QUDGs.
By iterating again over all edges, we assign each node $v$ to the nodes of $\mathcal{I}$ that $v$ is connected to.
Next, two nodes in $\mathcal{I}$ are connected if there exists an edge $(v,w) \in E_u^{2\setminus1}$ with $v$ and $w$ assigned to these two nodes.
As the size of $\mathcal{I}$ is asymptotically independent of the network size, this can also be achieved in time $O(m)$.
Now, node $u$ is enclosed by nodes in $\mathcal{I}$ if and only if it was enclosed in $‚G_u^{2\setminus1}$.
Thanks to the constant size of $\mathcal{I}$, the time for the enclosing circle detection on $\mathcal{I}$ is asymptotically independent of the size of $G_u^{2\setminus1}$.
However, the classification thresholds have to be adjusted as the edges in the representative graph no longer correspond to $1$-hop distances.
Altogether, the enclosing circle detection can be done in time linear in the size of $E_u^{2\setminus1}$.

\subsubsection{Classification Result}
Figure~\ref{fig::ecbr_example} shows an example of a classification with EC-BR.
By comparing this classification with Figure~\ref{fig:tlbr_refinement}, it becomes obvious that EC-BR and MDS-BR classify quite differently.
One striking difference is that EC-BR recognizes broader boundaries, meaning that even nodes which are only in proximity to a hole or the outer boundary are classified as boundary nodes.
The reason for this is that EC-BR checks whether the nodes in $2$-hop distance form a closed circle.
And even for nodes that are almost one hop away from the boundary such a closed circle does not exist.

The second apparent difference are the many small circles which do not belong to the large-scale boundaries.
By looking at the magnification in Figure~\ref{fig::ecbr_example}(b), one can see that the marked circles enclose small holes in the network.
We will see in the next paragraph how one can easily remove both kinds of artifacts, the wide borders and the circles around tiny holes in the network.
However, in many situations exactly this kind of information might be of interest.
The detection of small holes, for instance, can be used to detect node failures or areas of insufficient coverage.
And having a broader border might increase the fault tolerance and makes it easier to distribute messages along the border as it is guaranteed to be connected.
Additionally, neighboring boundary nodes can divide their workload and thus extend the lifetime of their batteries.

\subsubsection{Refinement}
Sometimes one might be only interested in large scale boundaries and not in the small holes that occur in areas with low node density.
For this situations, EC-BR can be extended with a very simple refinement, which removes most of the small holes and also makes the boundaries thinner.
The basic insight behind the refinement is that a node which lies near a hole is surrounded by other nodes that are marked as boundary nodes and by the hole itself.
So the node simply has to check whether a certain percentage $\gamma$ of its neighbors are currently classified as boundary nodes.
If this is true, the node stays a boundary node.
Otherwise, the node changes its classification to being an interior node.
Under the idealized assumption that the connectivity graph is a unit disk graph, $\gamma=100\%$ results in very precise boundaries.
For more realistic communication models, a threshold $\gamma \approx 70\%$ is more reliable.

\begin{figure}[t]
\centering
\begin{tabular}{llllll}
 a) \hspace{0.1cm} &
\parbox[c]{.2\columnwidth}{\includegraphics[width=.2\columnwidth]{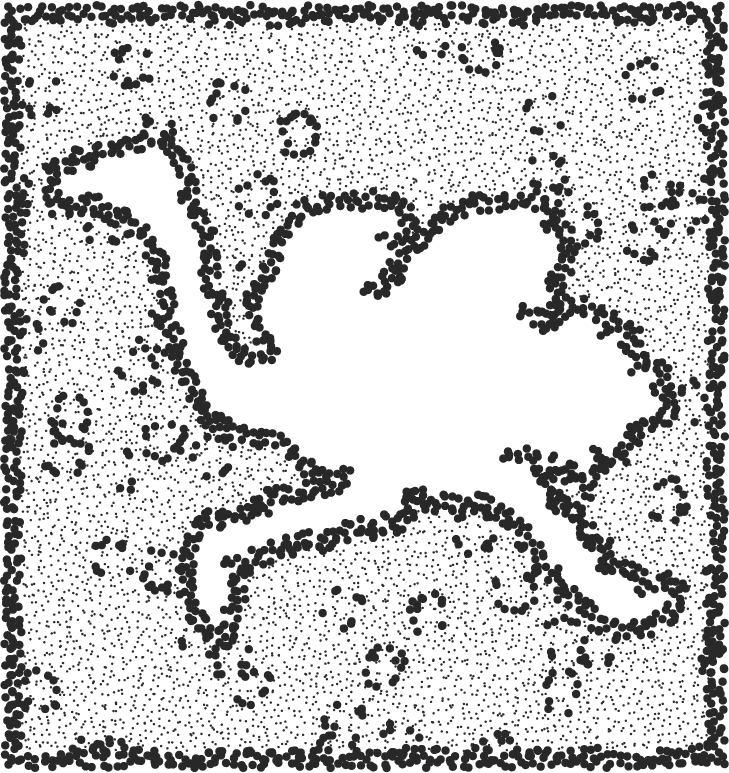}} &
\hspace{0.5cm}  b) \hspace{0.1cm} &
\parbox[c]{.2\columnwidth}{\includegraphics[width=.2\columnwidth]{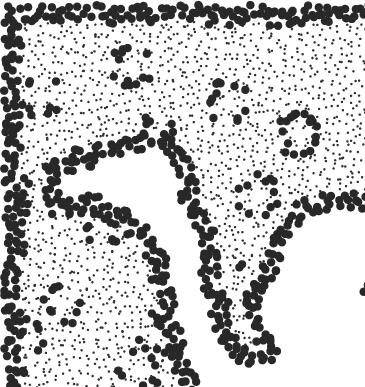}}
\hspace{0.5cm}  c) \hspace{0.1cm} &
\parbox[c]{.2\columnwidth}{\includegraphics[width=.2\columnwidth]{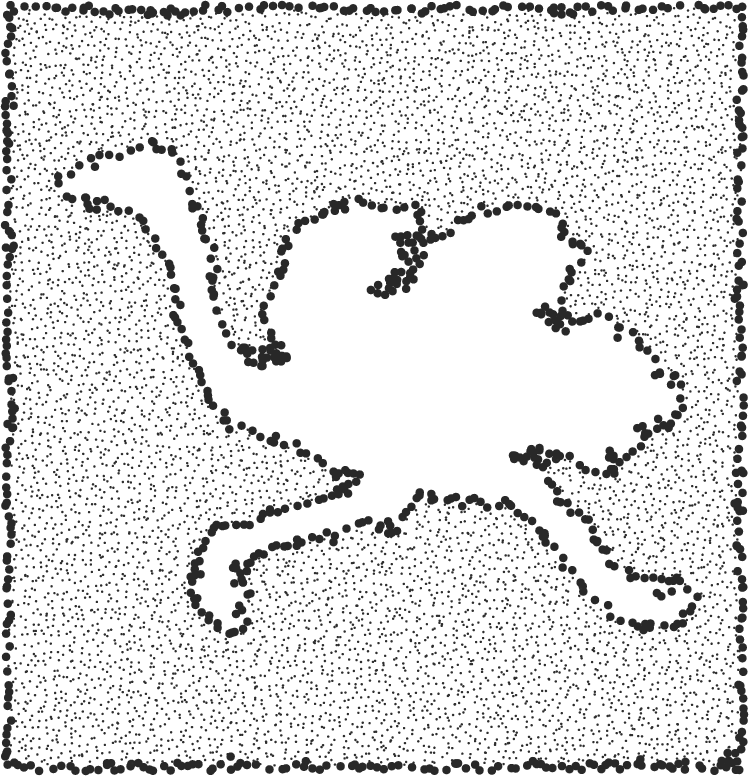}}
\end{tabular}
\caption{Classification of EC-BR. (a) Before refinement. (b) Magnification of upper left part. (c) After refinement.}
\label{fig::ecbr_example}
\end{figure}

The effect of this simple refinement strategy is depicted in Figure~\ref{fig::ecbr_example}(c).
Apparently, all nodes but the ones near large-scale holes are now classified as inner nodes and the border is very precise.


\section{Further Aspects considering both Algorithms}

\subsection{Theoretical Considerations}
Some approaches to the boundary recognition problem allow to give theoretical guarantees concerning the classification.
To achieve this, the input graph has to respect certain properties, the hole definition has to be adjusted adequately, and every node has to consider a rather big neighborhood. 
In contrast, this work focuses on efficient algorithms that are suited for real-world application in large-scale sensor networks.
This makes it necessary that the communication is restricted to a small neighborhood.
In MDS-BR, every node only needs connectivity information from at most its 3-hop neighborhood to achieve good results, and for EC-BR even the information of the immediate 2-hop neighborhood is sufficient.
This makes it rather difficult to give theoretical guarantees.
Additionally, we tried to make our hole definition as natural as possible instead of optimizing it for our algorithms.
In consequence, one can construct degenerated situations where both MDS-BR and EC-BR misclassify nodes with respect to our hole definition.
However, as our simulations show, the classifications that can be achieved with the proposed algorithms are surprisingly accurate considering how little communication they require.

\subsubsection{MDS-BR}
\begin{figure}[t]
\centering
\hfill
\subfigure[]{\includegraphics[page=1,height=7\baselineskip]{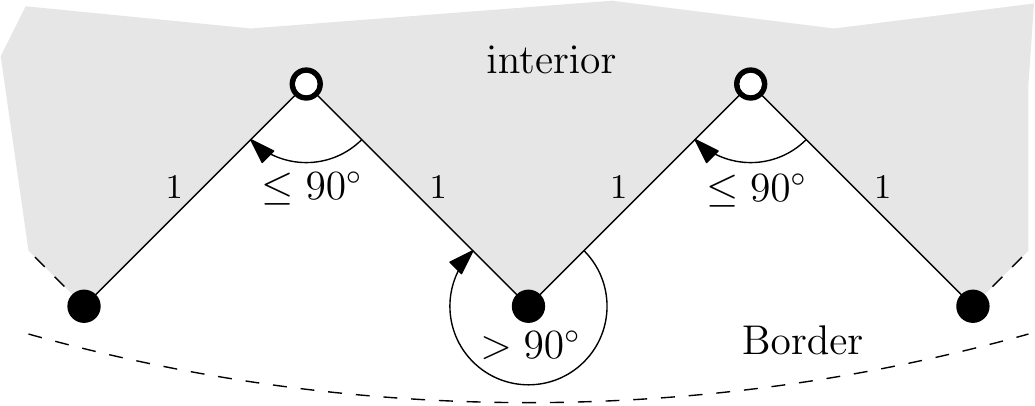}}
\hfill
\subfigure[]{\includegraphics[page=2,height=7\baselineskip]{figs_ds/mdsbr_guarantee.pdf}}
\hfill{}
\caption{(a) Sawtooth boundary. Every other node is falsely classified as interior (white) node by MDS-BR without refinement. (b) Micro-hole. Holes framed by more than four nodes but with a circumference of less than $4$ lead to false classifications under our definition.}
\label{fig:mdsbr_guarantee_1}
\end{figure}

Giving theoretical guarantees on the classification quality of MDS-BR is difficult as we need to quantify the deviation of the computed embedding from the real node positions.
But if we assume known node positions, we can asses scenarios in which MDS-BR without refinement fails.
For example, Figure~\ref{fig:mdsbr_guarantee_1}(a) shows a sawtooth boundary.
MDS-BR falsely classifies every other node as interior node according to our definition.
This case arises if the borderline is very ragged with nodes spread far apart.
With growing node density this situation will occur less likely as nodes move closer together and connections that span acute angles and put them in the interior  become possible.
Below, the worst case scenario is analyzed in more detail.
An example of the reverse situation, false classification as boundary node, is given in Figure~\ref{fig:mdsbr_guarantee_1}(b).
This case occurs if there are micro-holes that are surrounded by more than four nodes.
This also gets less likely with growing node density.
Furthermore, the refinement step will remove most of these false boundary nodes as they rarely form an extensive connected structure.

\begin{figure}[t]
\centering
\hfill
\subfigure[]{\includegraphics[page=3,width=.2\columnwidth]{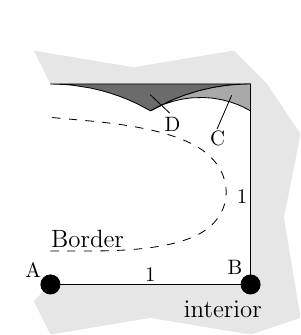}}
\hfill
\subfigure[]{\includegraphics[page=4,width=.2\columnwidth]{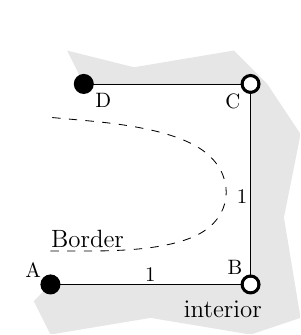}}
\hfill
\subfigure[]{\includegraphics[page=5,width=.2\columnwidth]{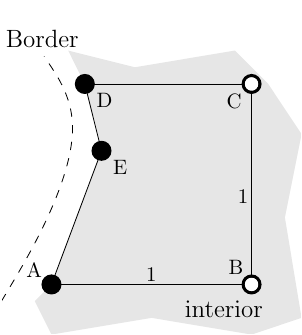}}
\hfill
\subfigure[]{\includegraphics[page=6,width=.2\columnwidth]{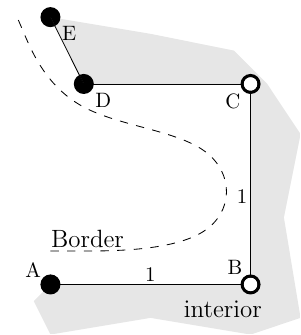}}
\hfill{}
\caption{Using MDS-BR with exact node coordinates (and w/o refinement), a boundary structure has at most two subsequent boundary nodes that are classified as interior nodes.
(a) Given nodes $A$ and $B$, the position of a node $C$ is restricted to the gray area if both nodes $B$ and $C$ are classified as interior nodes.
The darker gray area signifies all possible locations of node $D$ in this case (depending on the placement of node $C$).
Note that the colored areas only become smaller for scenarios with nodes $A$ and $B$ closer together.
(b) Possible locations of node $C$ and $D$ with nodes $B$ and $C$ falsely classified as interior nodes (white) by MDS-BR before refinement.
(c) If the opening angle at node $D$ created by node $C$ and a subsequent node $E$ is small enough, so that node $D$ would be classified as interior node, the structure coils up and a new boundary emerges.
(d) Otherwise, the angle between $(CD)$ and $(DE)$ is large enough that $D$ remains a boundary node (black).}
\label{fig:mdsbr_guarantee_2}
\end{figure}

Let's take a look at the worst case that can occur in a boundary structure under the UDG model.
If we try to construct a topology for which MDS-BR yields the most misclassifications of boundary nodes, we find that there will be at most two consecutive boundary nodes classified as interior nodes.
This is demonstrated in Figure~\ref{fig:mdsbr_guarantee_2}(a)-(d).
We can easily construct a boundary structure with two subsequent boundary nodes classified as interior nodes.
Such a structure comprises $4$ nodes and requires at least three of them to be far apart as depicted in Figure~\ref{fig:mdsbr_guarantee_2}(b).
We see that already this situation requires precise positioning and will occur less likely with growing network density as the average distance between nodes decreases.
Now, if we extend the structure by another node and try to obtain a third interior node, one of two possibilities happens.
Either the structure coils up on itself, yielding connections to previous nodes, thus creating a new boundary and turning the nodes in between into true interior nodes as shown in Figure~\ref{fig:mdsbr_guarantee_2}(c).
Or, avoiding this, the opening angle must be larger than $\alpha_{min}=90^\circ$, which results in a boundary node following the two interior nodes (see Figure~\ref{fig:mdsbr_guarantee_2}(d)).
To summarize, in a worst-case scenario at least one third of the mandatory boundary nodes will be classified correctly by MDS-BR before refinement.

\subsubsection{EC-BR}

\begin{figure}[t]
\centering
\hfill
\subfigure[]{\includegraphics[page=1,width=.2\columnwidth]{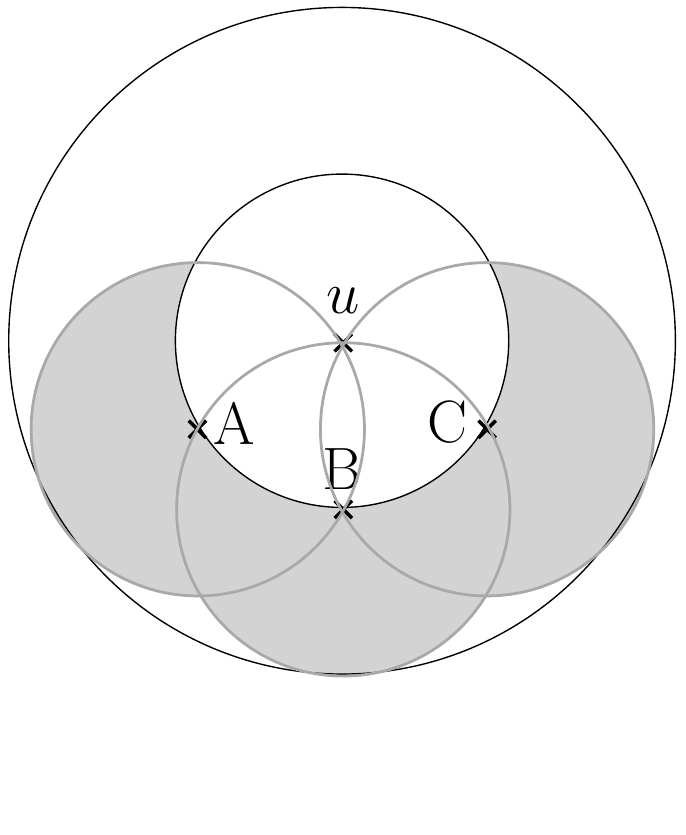}}
\hfill
\subfigure[]{\includegraphics[page=2,width=.2\columnwidth]{figs_mv/ecbr_falsepos.pdf}}
\hfill
\subfigure[]{\includegraphics[page=3,width=.2\columnwidth]{figs_mv/ecbr_falsepos.pdf}}
\hfill\hspace{0em}
\caption{
Example for a situation in which EC-BR wrongly classifies a circle to be enclosing. 
(a) Three nodes $A$, $B$, and $C$ are placed such that they are slightly more than one communication distance away from node $u$, and both $A$ and $C$ have almost distance one to node $B$.
(b) Two nodes $D$ and $E$ are added into the dark gray areas such that they are less than distance one away from each other, less than distance two away from $u$, and more than distance one away from node $B$.
(c) Finally, two nodes $F$ and $G$ are placed into the gray areas such that $F$ is less than distance one away from both $A$ and $D$, $G$ is less than distance one away from both $C$ and $E$, and $F$ and $G$ are both more than distance one away from node $B$. The enclosing circle detection would wrongly classify the depicted circle of length 7 to be enclosing. (The nodes within 1-hop distance of $u$ are omitted in this description. They have to be arranged such that the depicted nodes A-G are within 2-hop distance of $u$). 
}
\label{fig:ecbr_guarantee}
\end{figure}

EC-BR classifies nodes based on the length of a longest circle found by the enclosing circle detection.
We already mentioned that it is possible to construct degenerated settings for which a non-enclosing circle is considered to be enclosing by EC-BR.
This uncertainty is the price for the good performance of EC-BR, regarding both communication and computation efficiency.
Figure~\ref{fig:ecbr_guarantee} shows the construction of a circle of length 7 which, assuming a UDG communication model, is wrongly assumed to be enclosing. 
The nodes have to be placed very carefully in order to prevent the emergence of communication links that split the circle into smaller circles, which makes this kind of misclassification very unlikely.
Moreover, an additional node located within the depicted circle would most likely split the circle and thus prevent the misclassification.
Accordingly, for higher node densities this kind of misclassification becomes less likely.

The second type of misclassification, classifying interior nodes to be on the boundary, occurs more frequently. This happens if the node density within the 2-hop neighborhood of a node is so small that no enclosing circle exists. Thus, misclassification of an interior node can be seen as an indication of insufficient node density. 
But as long as the overall node density is sufficiently high, this kind of misclassification can be easily repaired with the refinement step.

It is important to keep in mind that misclassifications strongly depend on the considered application. There are situations where areas of low node density are of interest and other situations where only large-scale holes play a role. The hole definition used in this work includes rather small holes. This is only meaningful if the average node density is sufficiently high so that there is no excess of small holes. For low-density networks, another hole definition that aims at larger holes might be necessary. 

\subsection{Detection of Large-Scale Holes}
In some situations, one might only be interested in holes starting from a given size $r_{min}$.
In all those cases, a very good strategy would be to use one of our algorithms to identify boundary candidates and subsequently to use the refinement of MDS-BR with an appropriate value for $r_{min}$.
This is highly efficient as only information about a very small subset of all nodes, the candidate nodes, has to be communicated.
Thus, the presented algorithms can be used to recognize hole structures of arbitrary size with very low communication expense.

\subsection{Connected Boundary Cycles}
The results of MDS-BR and EC-BR can be refined to obtain connected boundary cycles if the underlying application requires such a structure.
We present the approach using EC-BR before refinement exemplarily.
Due to its nature, a boundary cycle cannot be determined locally. 
But fortunately, we only have to consider nodes previously marked by EC-BR as boundary candidates.
The strategy is quite simple.
We start from an arbitrary node in the halo of boundary nodes marked by EC-BR and perform a shortest path query that tries to find an enclosing circle.
As the halo consists of the one hop neighborhood of the boundary, we are guaranteed a connected component around the hole.
After a closed cycle is found, all nodes within the 1-hop neighborhood are classified as interior nodes.
This is repeated until all nodes either belong to a connected boundary cycle or they are classified as interior nodes.
Special care has to be taken in the case that two large holes are less than one hop distance away from each other.
In this case, the two boundary cycles have to share some nodes.


\section{Simulations}\label{sec:simulations}

\subsection{Simulation Setup}

\subsubsection{Network Layout}
\begin{figure}[b]
	\small
	\centering
	\includegraphics[width=0.19\textwidth]{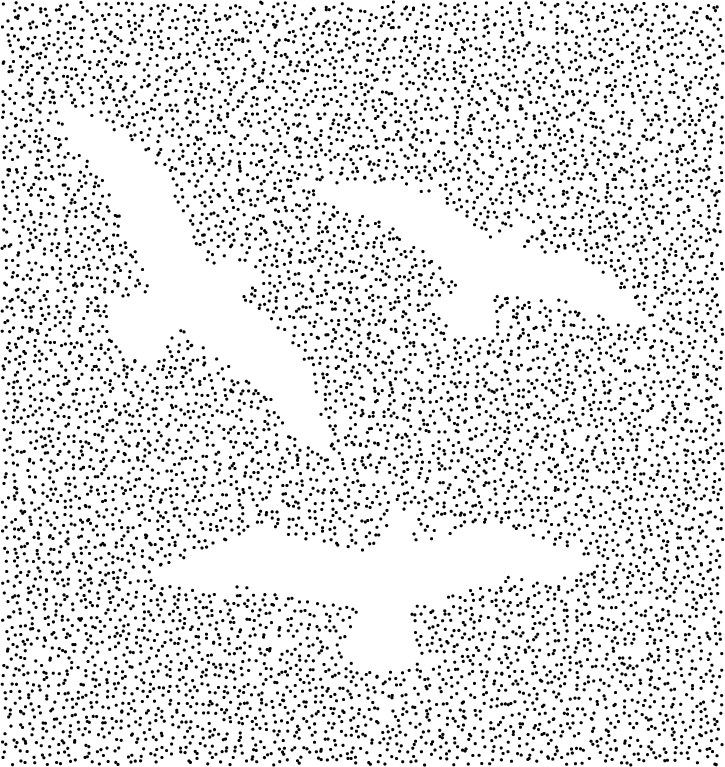}
	\includegraphics[width=0.19\textwidth]{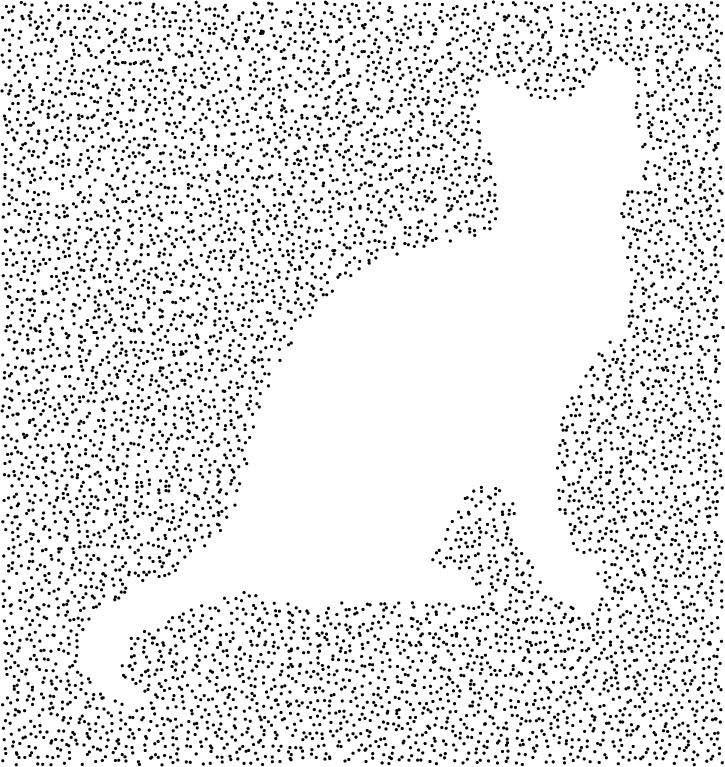}
	\includegraphics[width=0.19\textwidth]{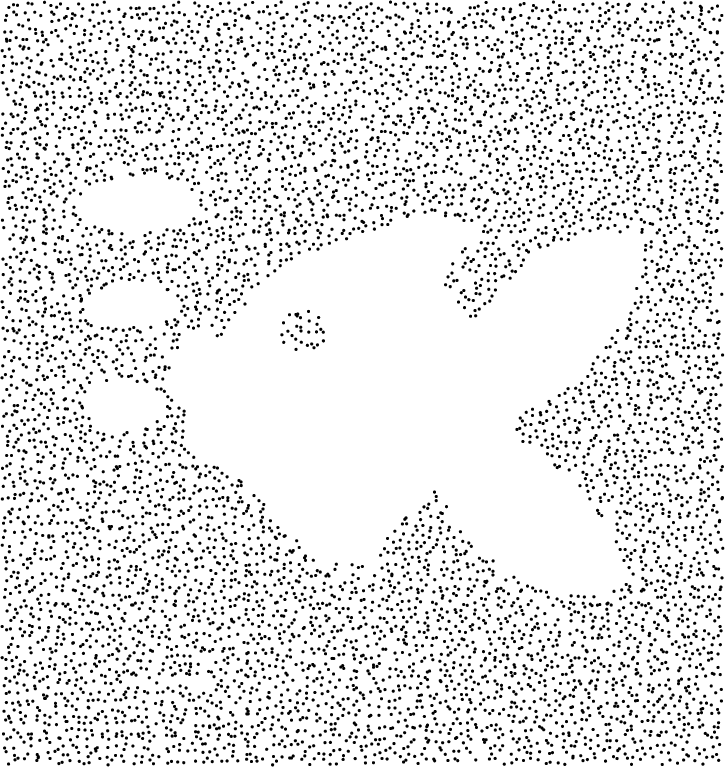}
	\includegraphics[width=0.19\textwidth]{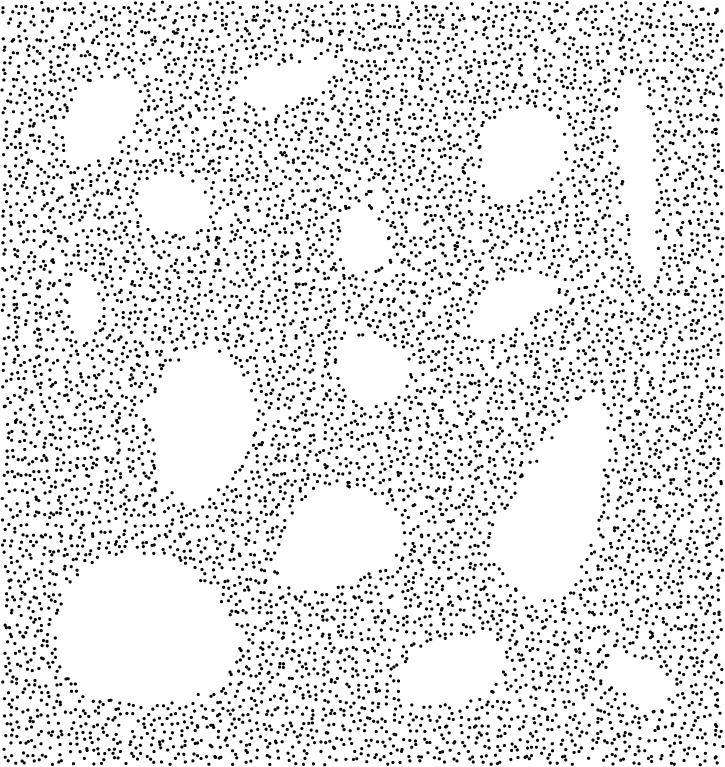}
	\includegraphics[width=0.19\textwidth]{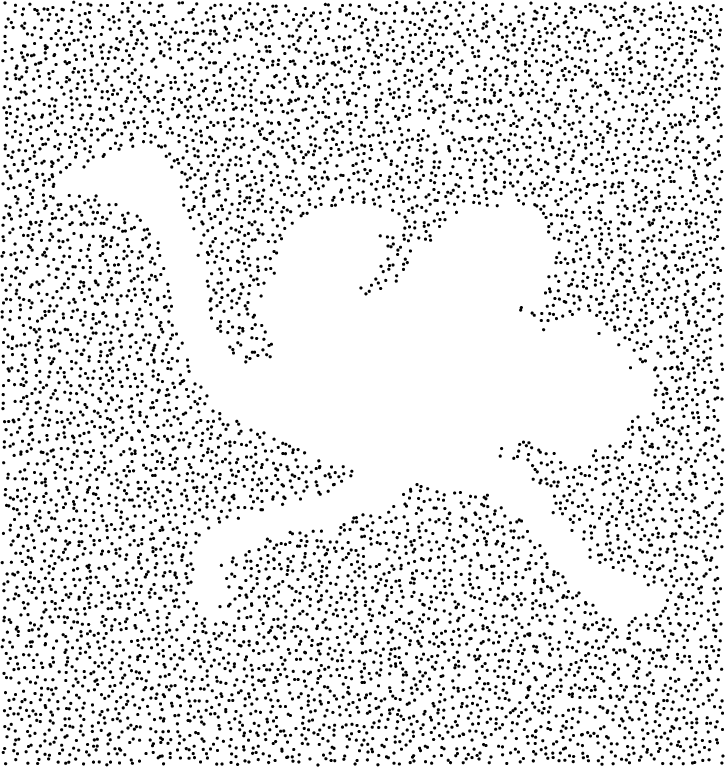}
	\caption{Hole Patterns. Node distributions with perturbed grid placement and $d_{avg}=12$.\label{fig:nets}}
\end{figure}

We generate network layouts by iteratively placing nodes on an area of $50 \times 50$ maximum communication distances according to one of the distribution strategies described in Section~\ref{sec:model}: perturbed grid placement (pg) or random placement (rp).
After each node placement, communication links are added according to the UDG or QUDG model.
Nodes are added until an average node degree $d_{avg}$ is reached.
To generate holes, we apply hole patterns such as the ones in Figure~\ref{fig:nets}.
Our default layout uses perturbed grid placement, the UDG model, and average node degree $d_{avg}=12$.

\subsubsection{Considered Algorithms}
We compare the performance of our approaches EC-BR and MDS-BR to three well-known boundary recognition algorithms: the algorithm by Fekete \emph{et al.} \cite{fkpfb04} (labelled Fekete04) and the centralized and distributed algorithms by Funke~\cite{f05} and Funke \emph{et al.}~\cite{fk06} (labeled Funke05 and Funke06, respectively).
In addition, we show qualitative comparisons of these algorithms and the algorithm by Wang \emph{et al.} \cite{wgm06} in Section~\ref{ssec:visual_comparison}.
We apply our own implementation of these algorithms according to their description in the respective publications and use the recommended parameters.
For MDS-BR, $\alpha_{min} = 90^\circ$ and $\ r_{min} = 3$ are used throughout the simulations.
The refinement of EC-BR is used with threshold $\gamma=100\%$.
Further details on parameter selection are given in Section~\ref{ssec:parameter_tuning}.

Unfortunately, the number of existing approaches makes it impossible to include all of them in our comparison.
Thus, we tried to select algorithms that assume similar conditions and constraints as our approaches.
Many of the other existing algorithms would not allow for a fair comparison because of one of the following reasons:
they use additional information, like absolute or relative node positions or connectivity information of large neighborhoods or even of the whole network;
they rely on certain network properties, as high average node degrees or the UDG communication model;
or they require expensive operations, like flooding the whole network or centralized computation.
Some of these approaches might even achieve better classifications by utilizing more information or more expensive operations.
But our goal is to show that solely connectivity information of nearby nodes and a relatively low average node degree are sufficient to achieve very impressive classification results with simple yet highly efficient algorithms.

\subsubsection{Measurement Procedure}
Each setup is evaluated $100$ times for each hole pattern in Figure~\ref{fig:nets}.
Faces with circumference $h_{min} \geq 4$ are considered holes, e.g. a square of edge length $1$ with no communication link crossing it.
The analysis lists mean misclassification ratios (false negatives) in percent.
For optional boundary nodes, we give the percentage of nodes classified as interior nodes.
Best results for each setup are highlighted in bold.

As stated before, different approaches use slightly different hole and boundary definitions.
Thus, one might argue that these approaches will perform worse under our definition.
However, we believe that the distinction in mandatory and optional boundary nodes helps to allow for a fair comparison.
Only nodes that are immediately at the boundary or at least one hop away from the boundary are considered for determining the classification quality.
For the other nodes it depends on the scenario whether they should be classified as boundary nodes or not, thus they are not rated.

\subsection{Visual Comparison}\label{ssec:visual_comparison}

\begin{figure}[t]
\centering
\hfill
\subfigure[Truth]{\includegraphics[width=.21\columnwidth]{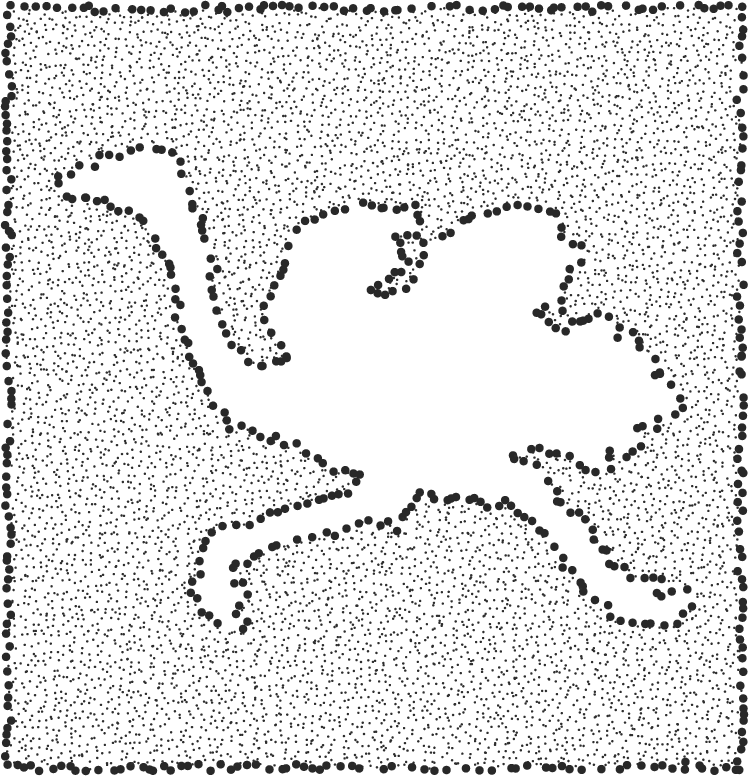}}
\hfill
\subfigure[MDS-BR]{\includegraphics[width=.21\columnwidth]{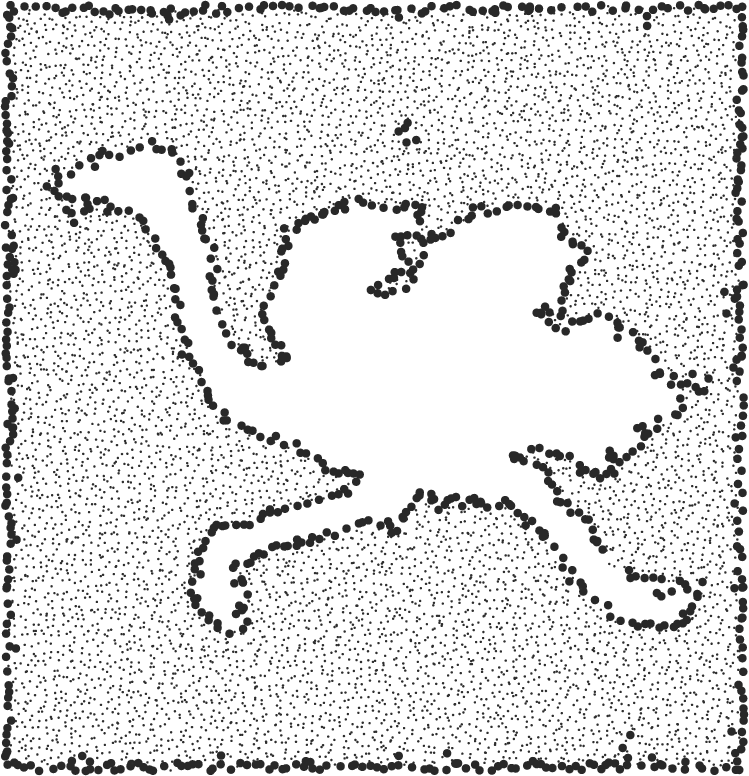}}
\hfill
\subfigure[EC-BR]{\includegraphics[width=.21\columnwidth]{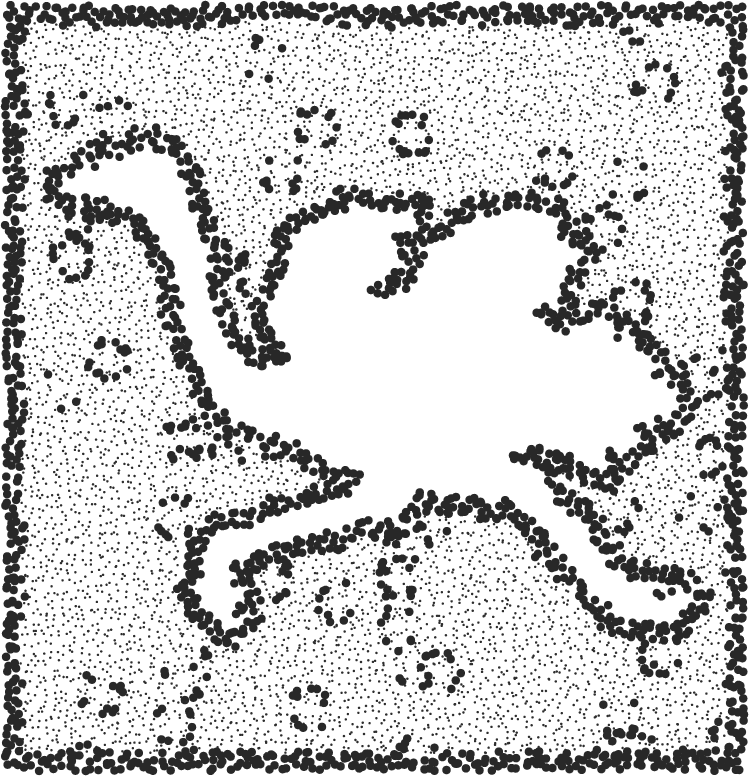}}
\hfill
\subfigure[EC-BR Ref.]{\includegraphics[width=.21\columnwidth]{figs_ds/ecbr2_bird.png}}
\hfill{}

\hfill
\subfigure[Fekete~04]{\includegraphics[width=.21\columnwidth]{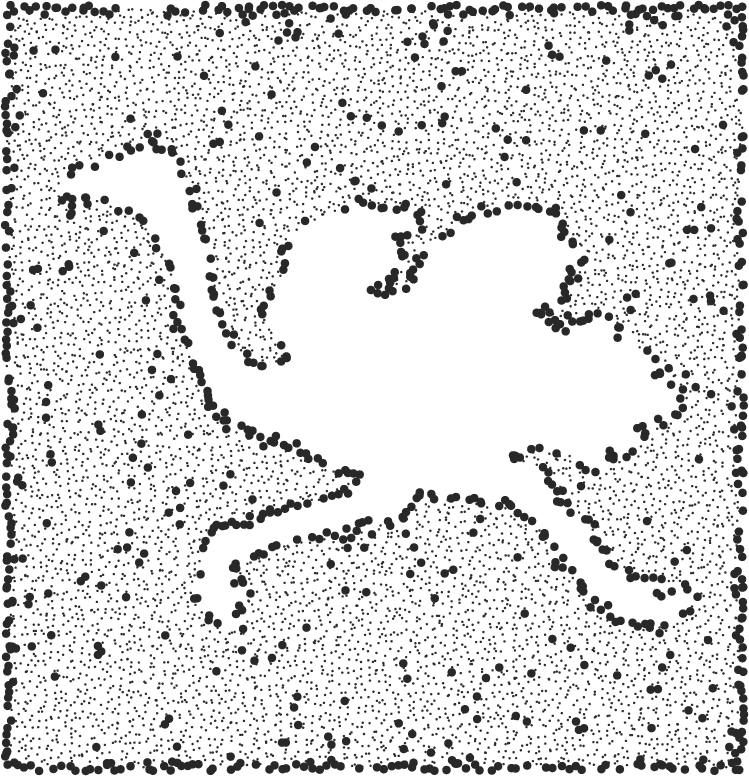}}
\hfill
\subfigure[Funke~05]{\includegraphics[width=.21\columnwidth]{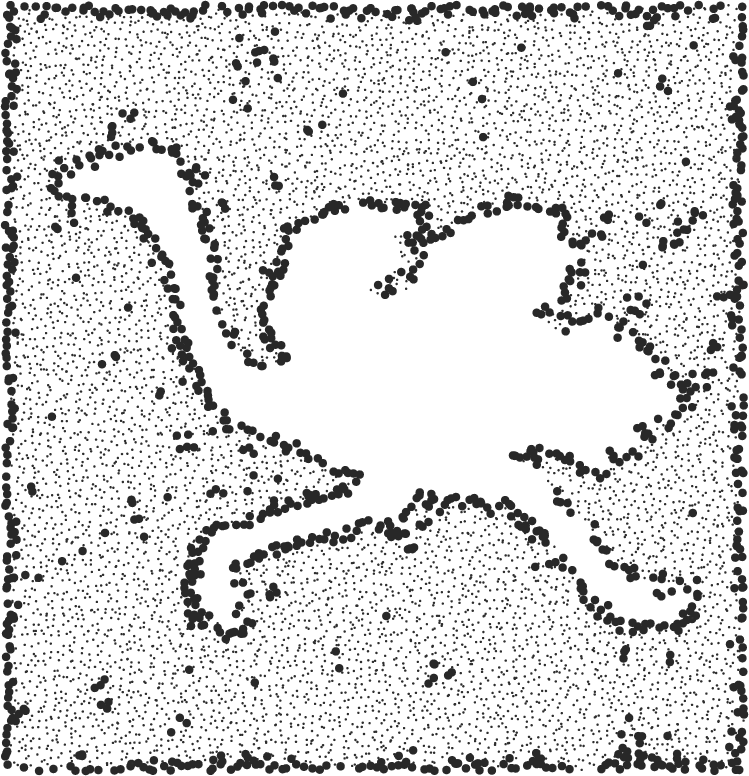}}
\hfill
\subfigure[Funke~06]{\includegraphics[width=.21\columnwidth]{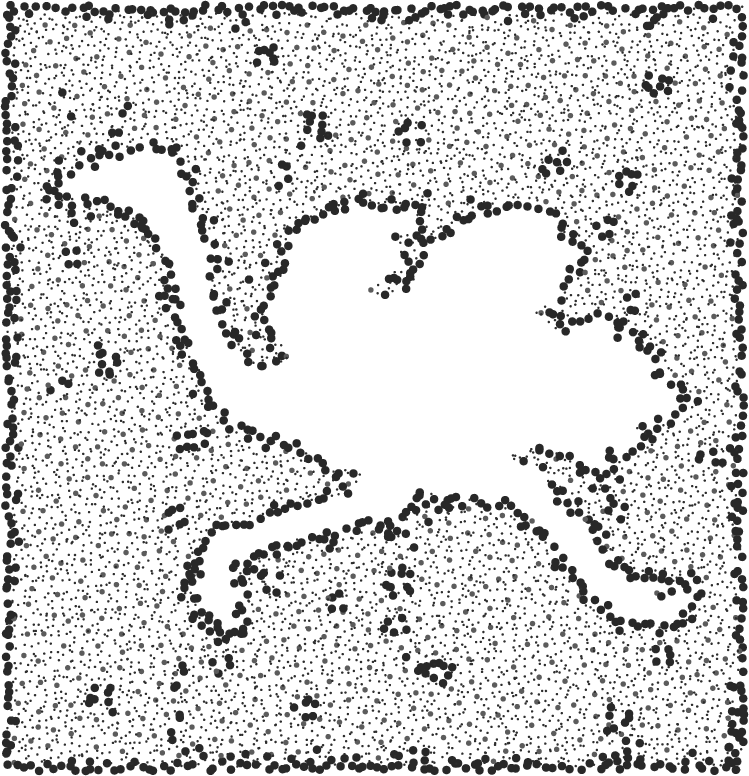}}
\hfill
\subfigure[Wang~06]{\includegraphics[width=.21\columnwidth]{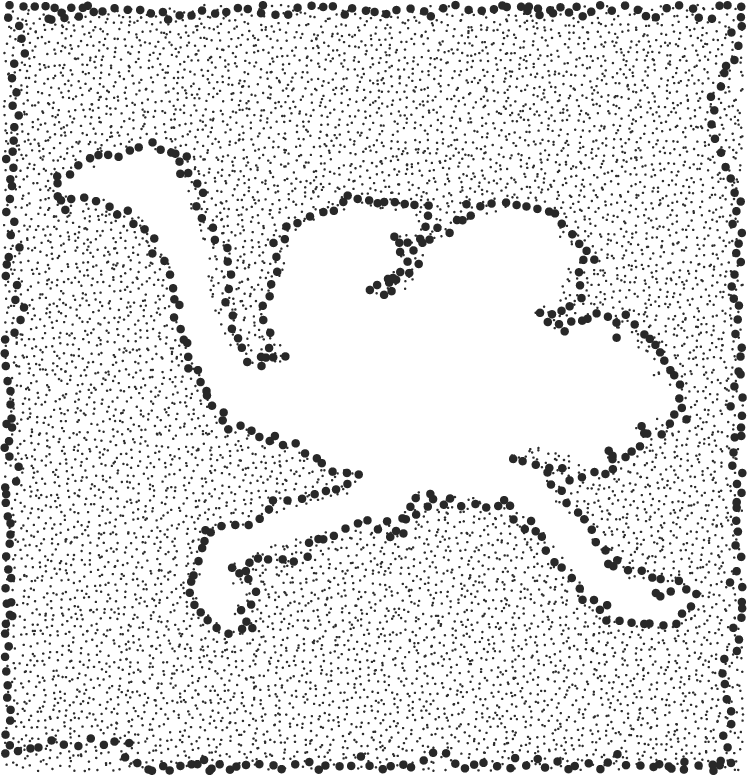}}
\hfill{}

\caption{Visual Comparison of several algorithms for boundary detection.}
\label{fig::visual_comp}

\end{figure}

\begin{figure}[p]
\centering
\hfill
\subfigure{\includegraphics[width=.19\columnwidth]{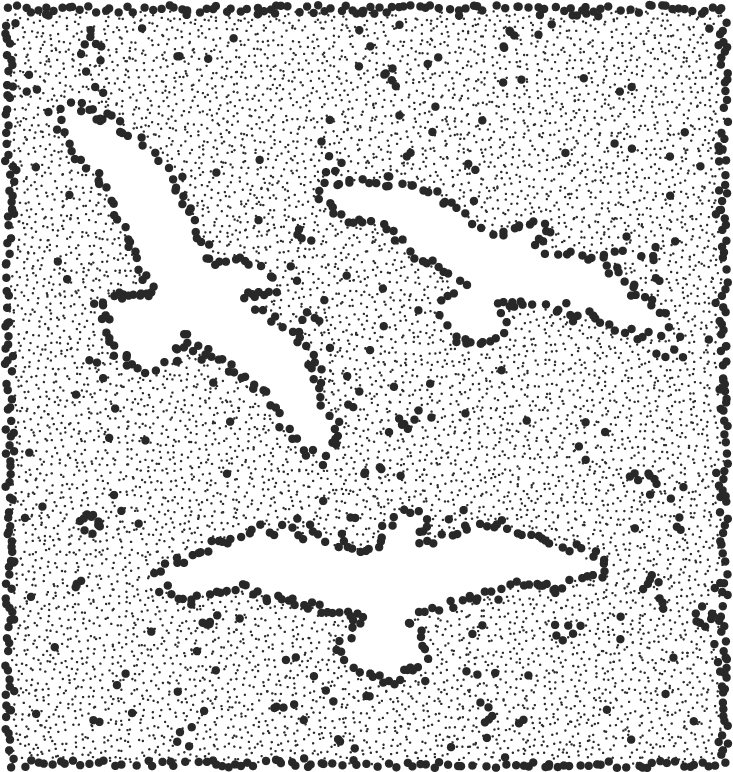}}
\hfill
\subfigure{\includegraphics[width=.19\columnwidth]{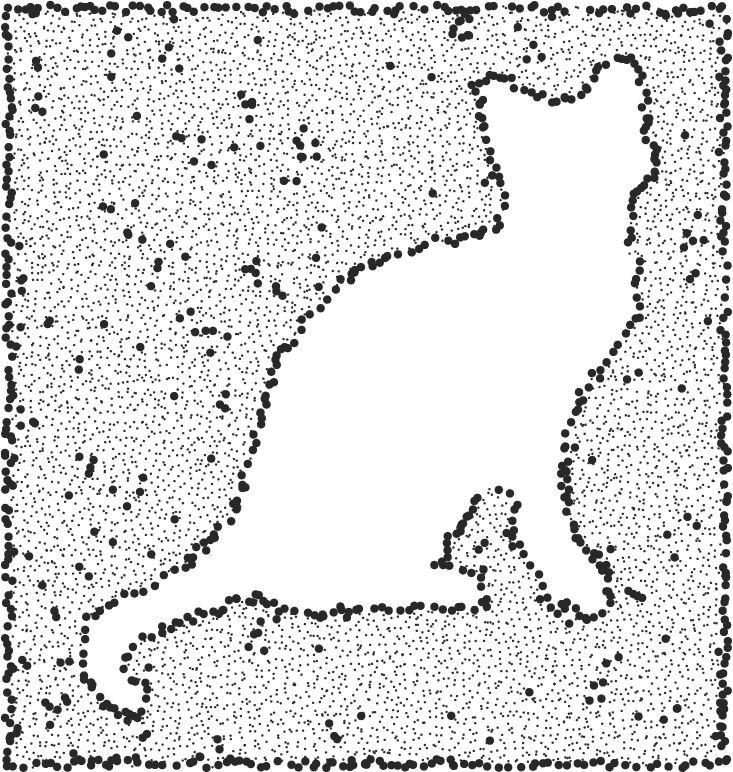}}
\hfill
\subfigure{\includegraphics[width=.19\columnwidth]{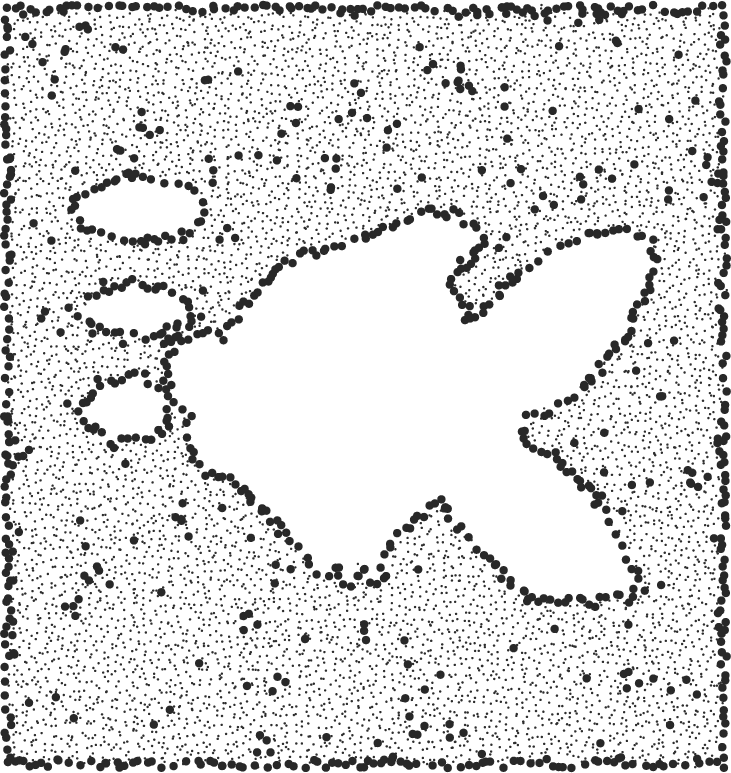}}
\hfill
\subfigure{\includegraphics[width=.19\columnwidth]{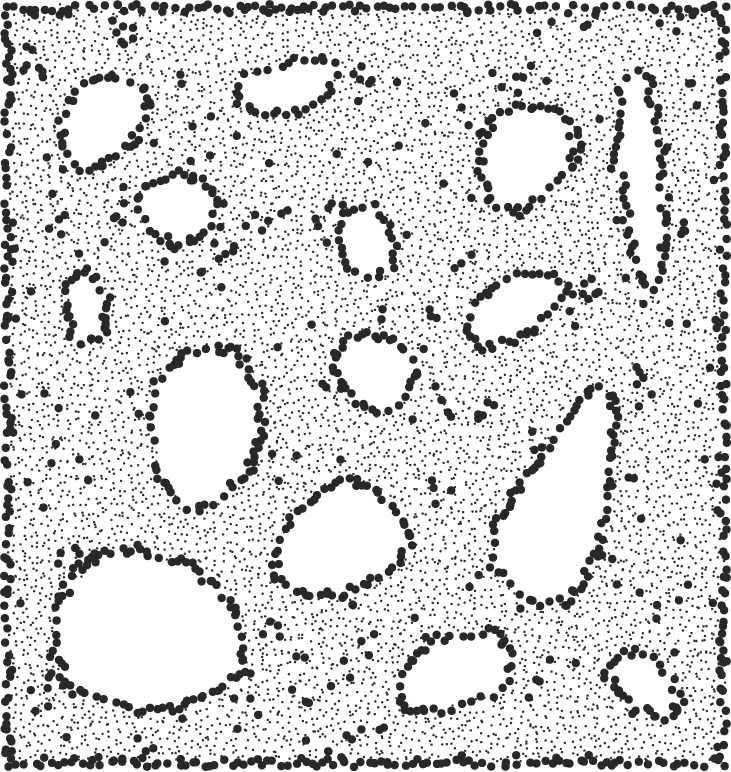}}
\hfill
\subfigure{\includegraphics[width=.19\columnwidth]{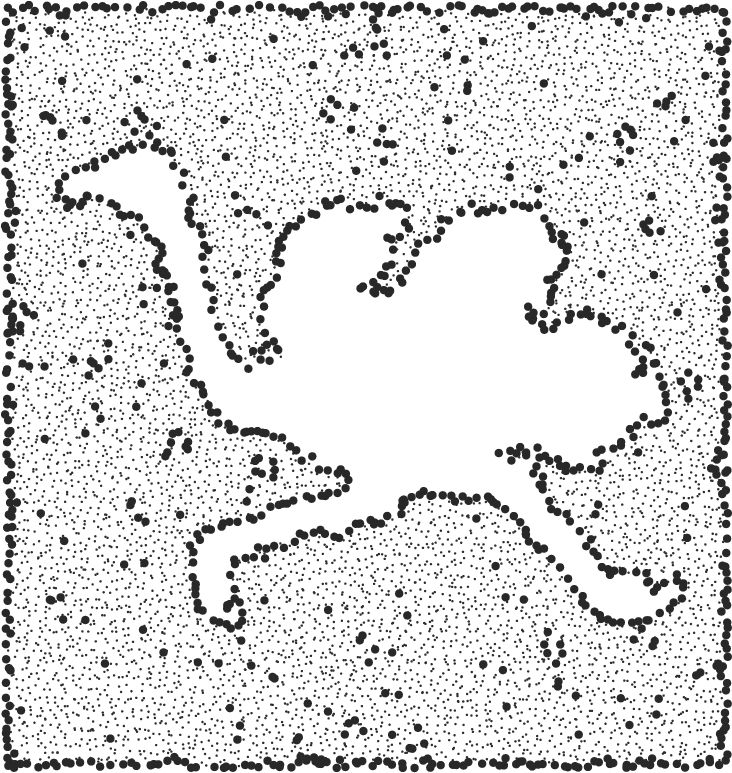}}
\hfill{}

\hfill
\subfigure{\includegraphics[width=.19\columnwidth]{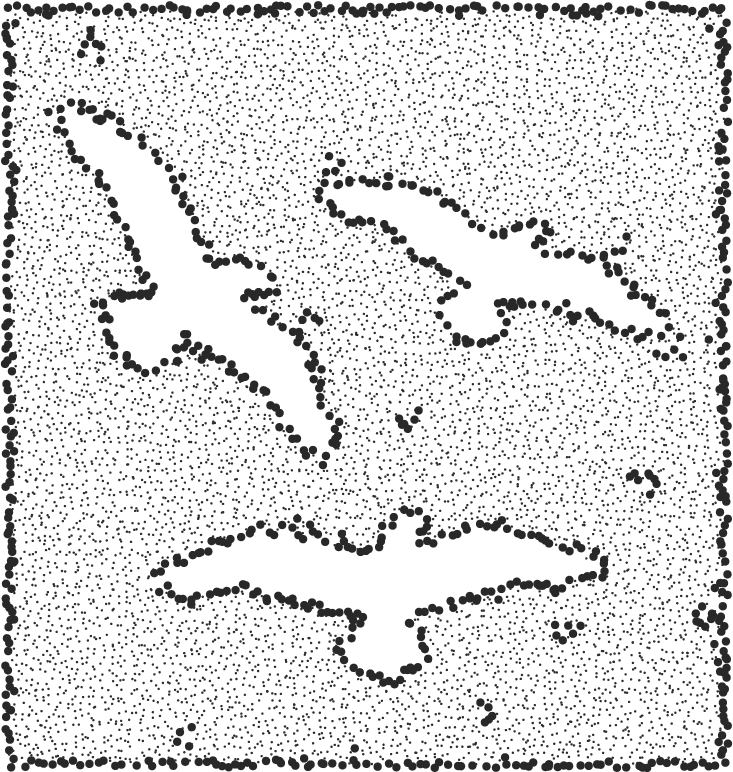}}
\hfill
\subfigure{\includegraphics[width=.19\columnwidth]{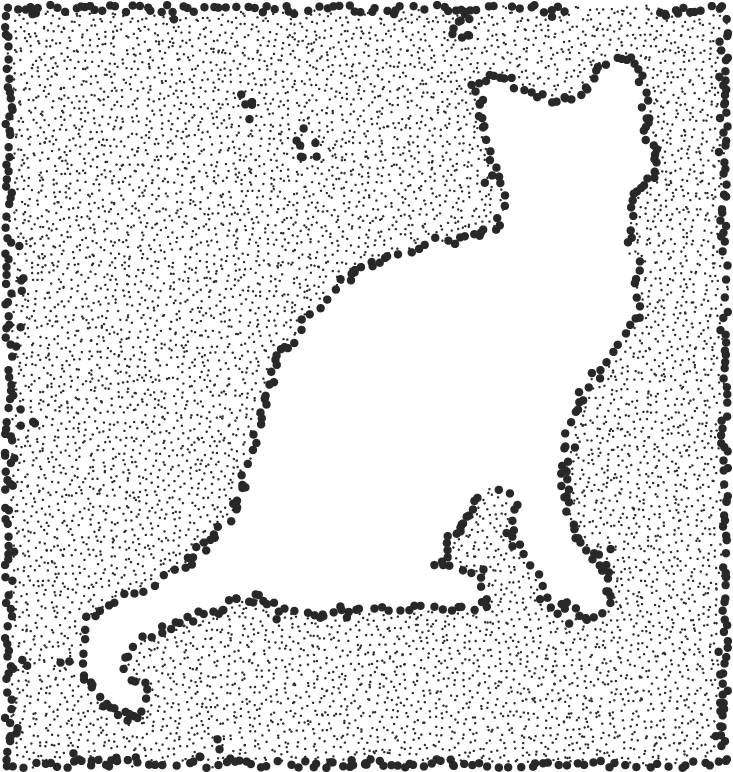}}
\hfill
\subfigure{\includegraphics[width=.19\columnwidth]{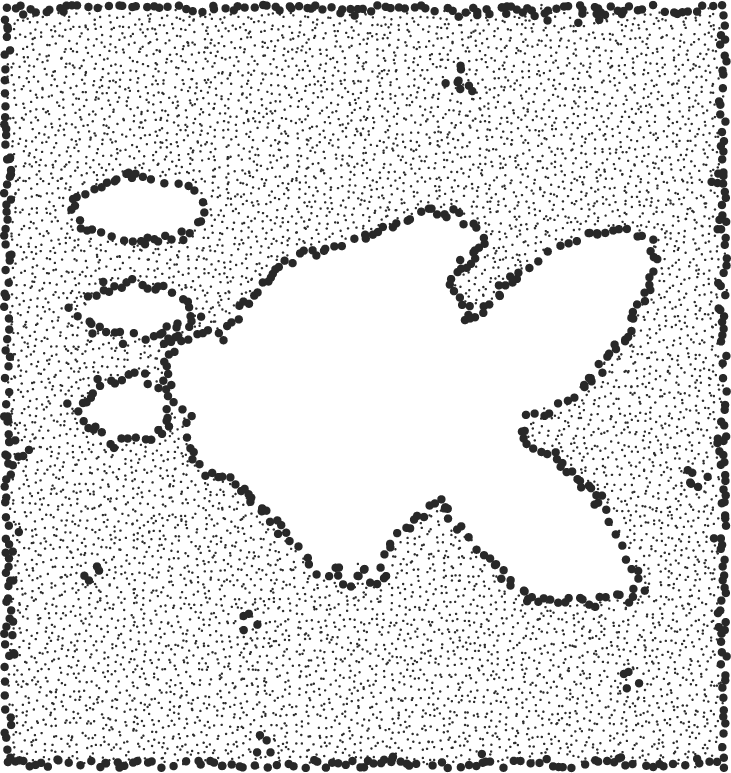}}
\hfill
\subfigure{\includegraphics[width=.19\columnwidth]{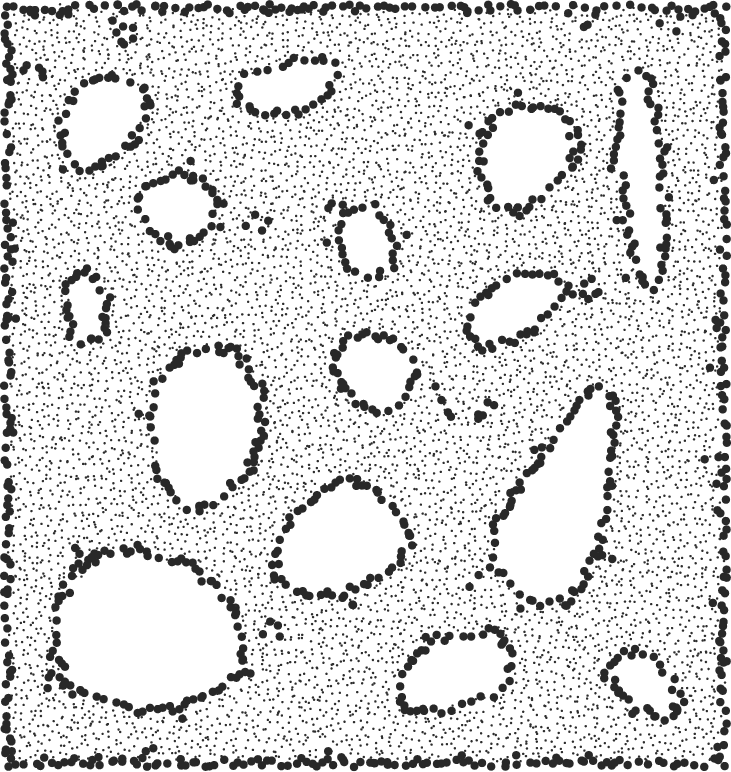}}
\hfill
\subfigure{\includegraphics[width=.19\columnwidth]{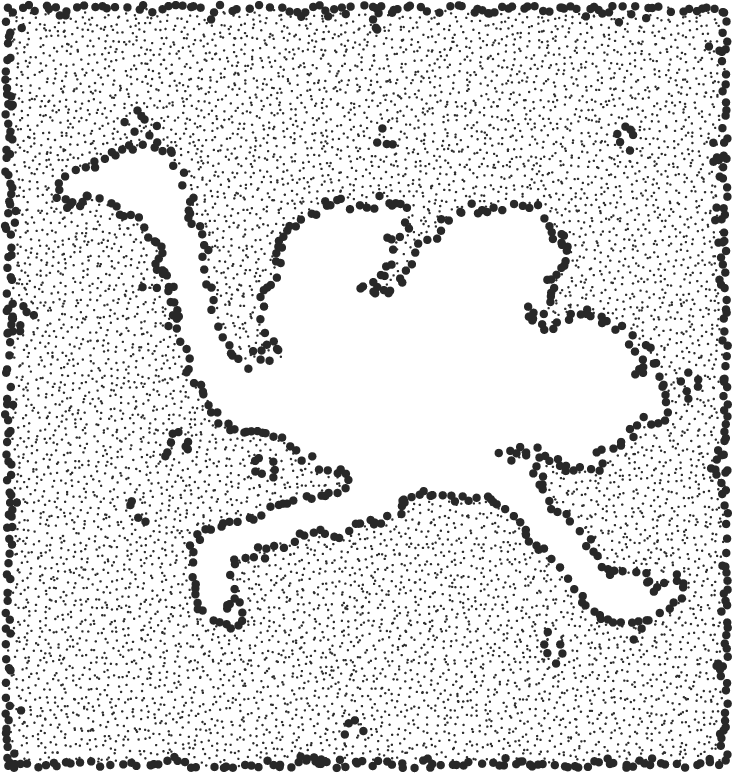}}
\hfill{}
\caption{Examples for MDS-BR on networks with average node degree 12 and different hole patterns. 
The top row shows the classification before refinement, the bottom row the classification with refinement.}
\label{fig::examples_mdsbr}
\end{figure}

\begin{figure}[p]
\centering
\hfill
\subfigure{\includegraphics[width=.19\columnwidth]{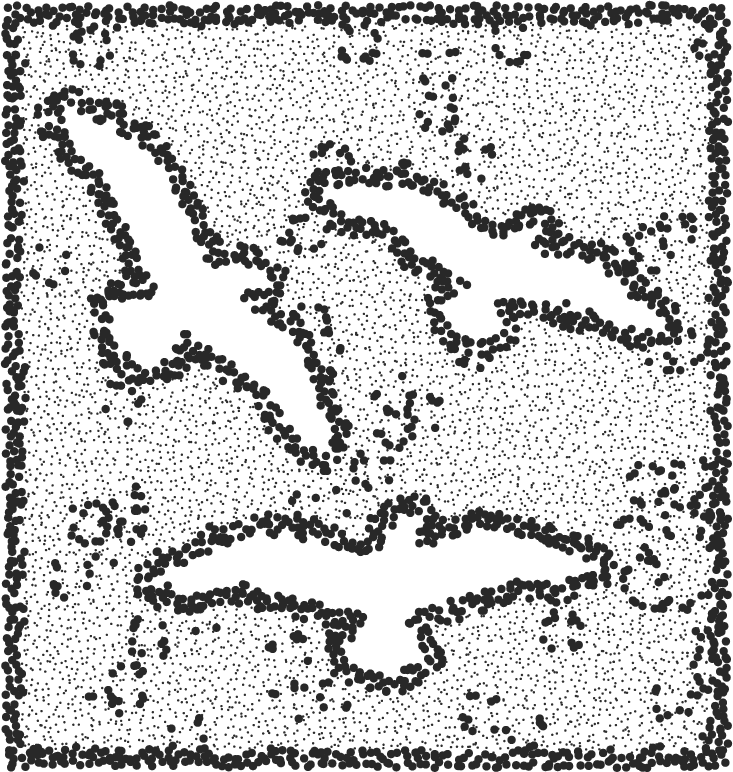}}
\hfill
\subfigure{\includegraphics[width=.19\columnwidth]{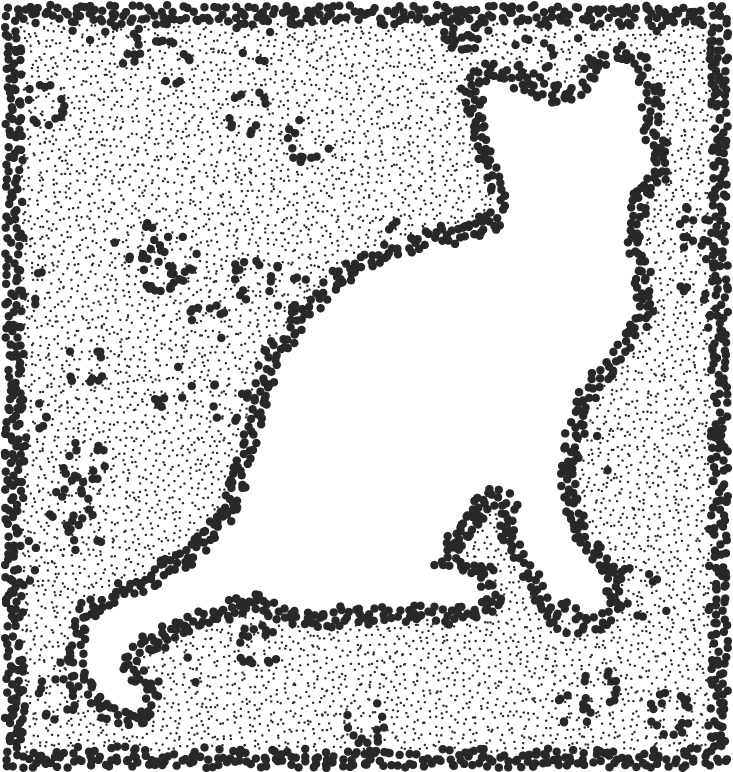}}
\hfill
\subfigure{\includegraphics[width=.19\columnwidth]{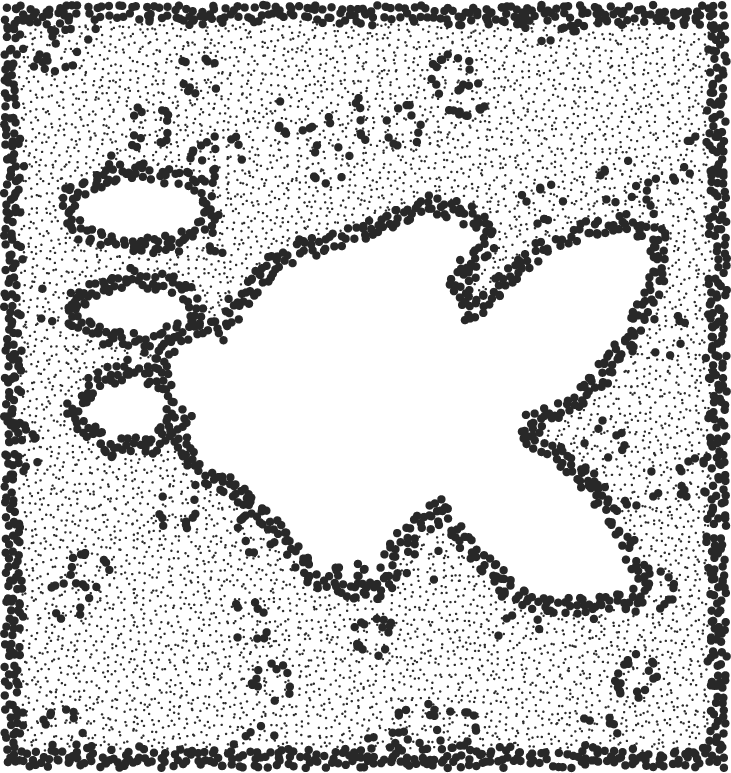}}
\hfill
\subfigure{\includegraphics[width=.19\columnwidth]{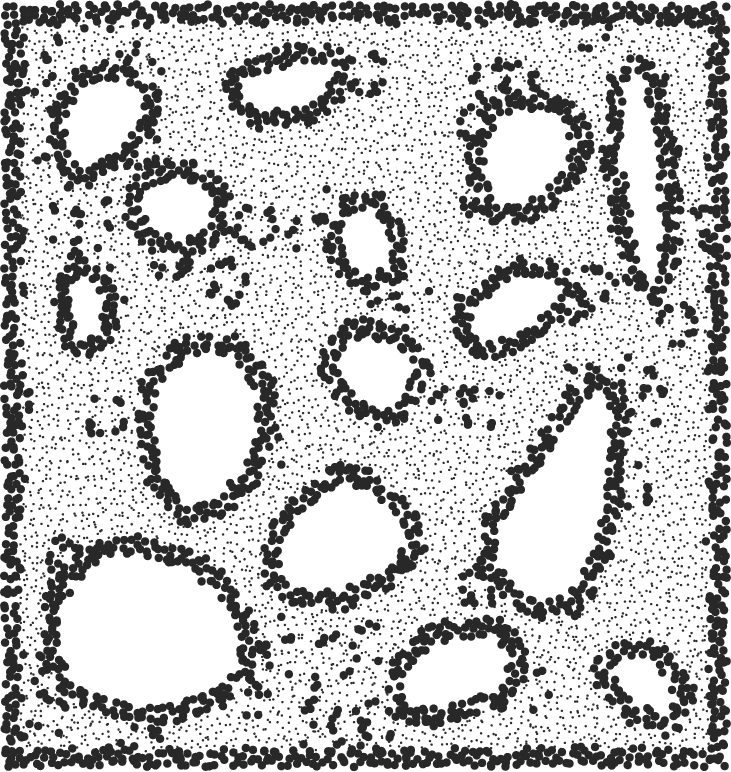}}
\hfill
\subfigure{\includegraphics[width=.19\columnwidth]{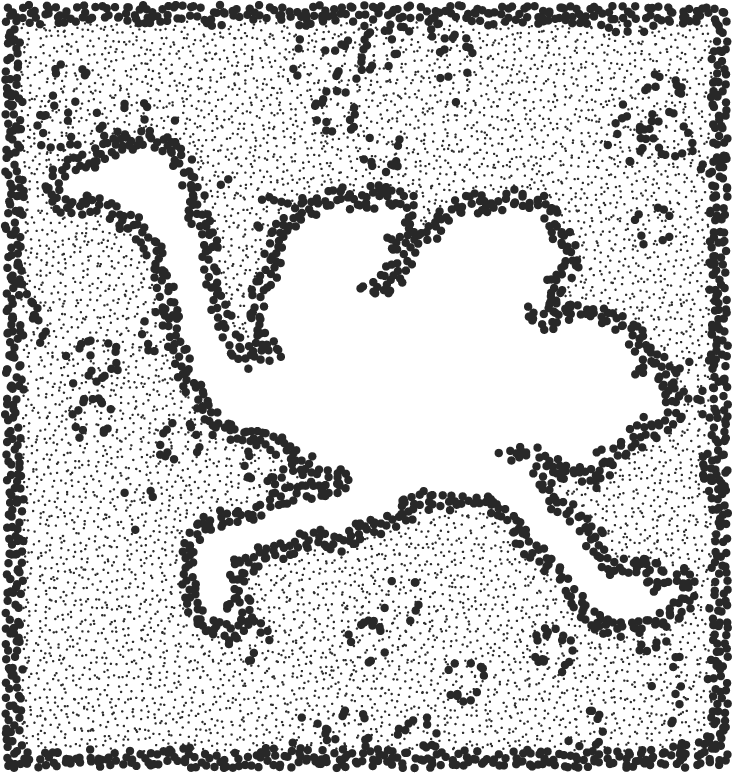}}
\hfill{}

\hfill
\subfigure{\includegraphics[width=.19\columnwidth]{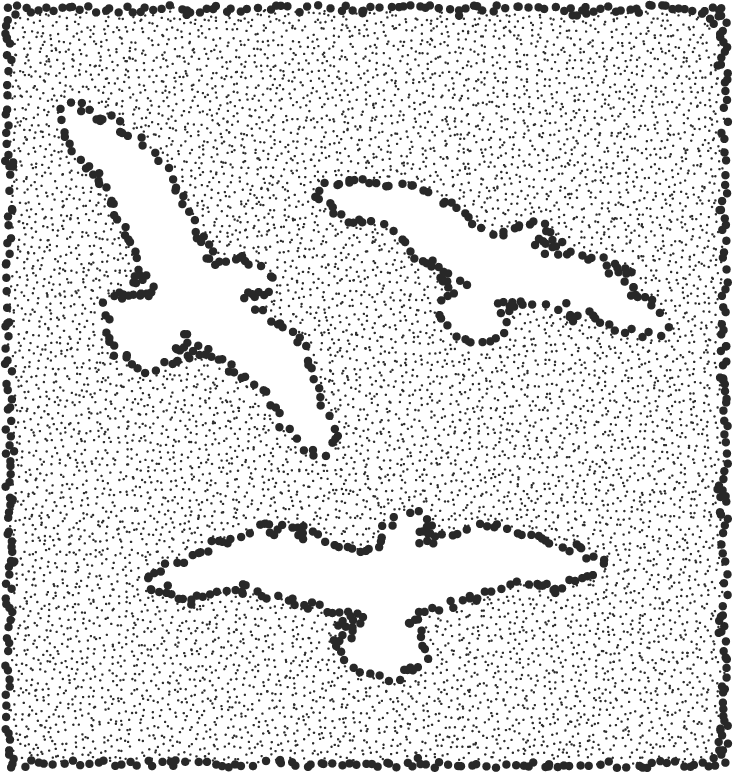}}
\hfill
\subfigure{\includegraphics[width=.19\columnwidth]{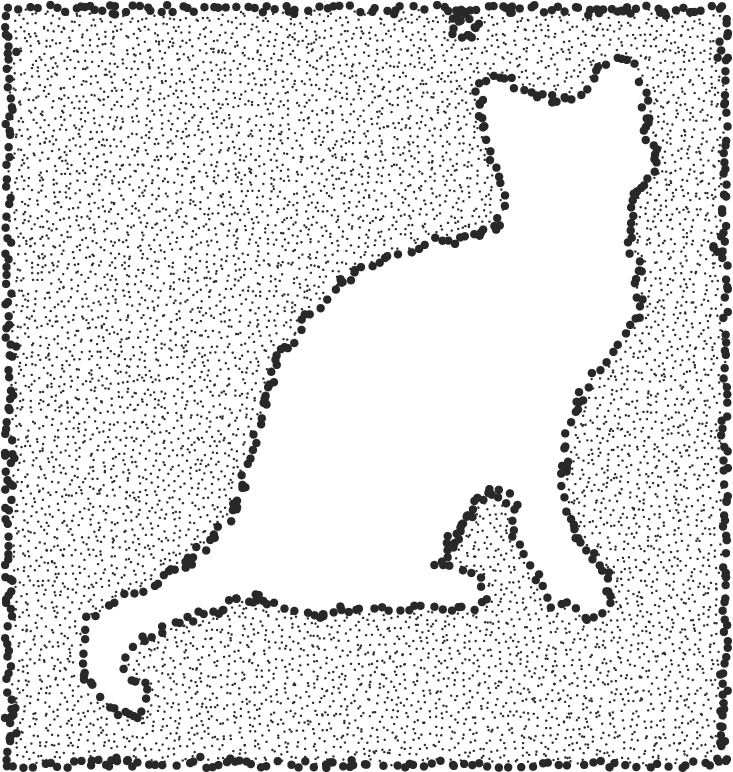}}
\hfill
\subfigure{\includegraphics[width=.19\columnwidth]{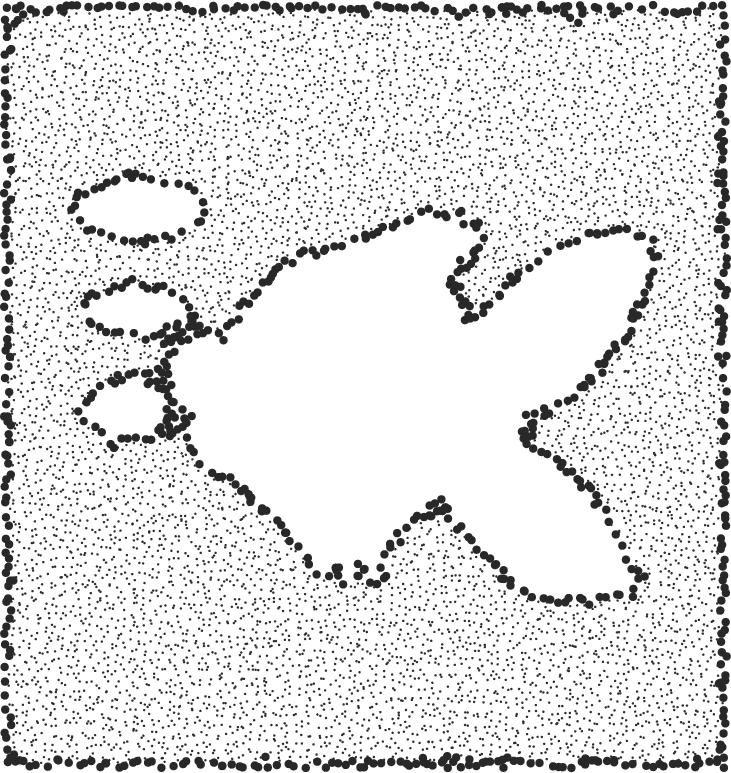}}
\hfill
\subfigure{\includegraphics[width=.19\columnwidth]{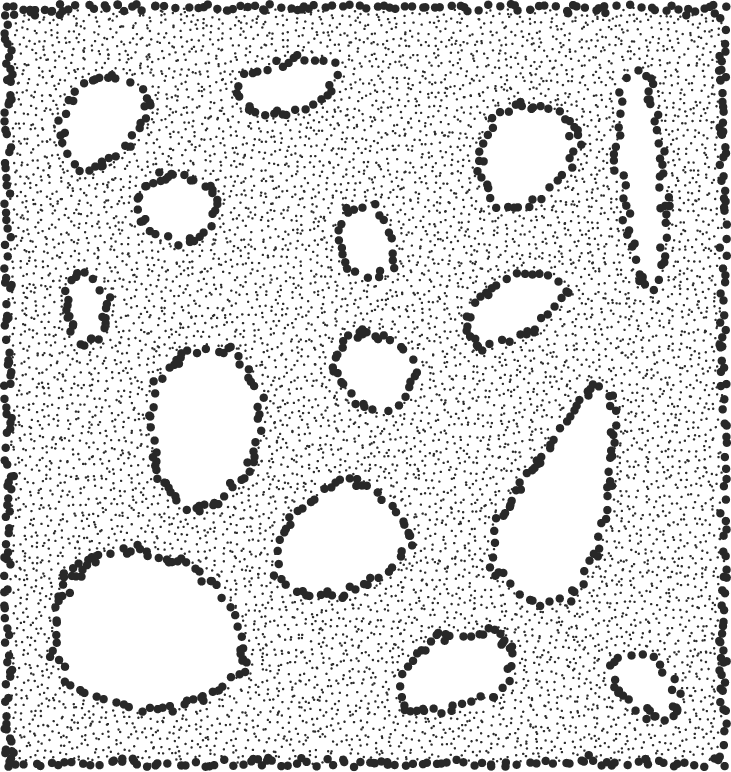}}
\hfill
\subfigure{\includegraphics[width=.19\columnwidth]{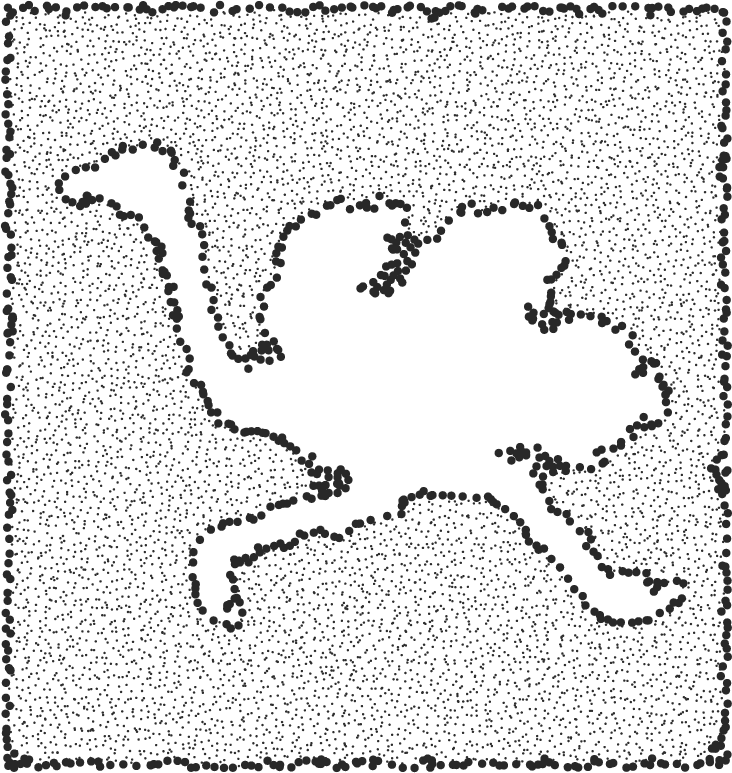}}
\hfill{}
\caption{Examples for EC-BR on networks with average node degree 12 and different hole patterns. 
The top row shows the classification before refinement, the bottom row the classification with refinement.}
\label{fig::examples_ecbr}
\end{figure}

We start with a visual comparison of the classification results of the considered algorithms in Figure~\ref{fig::visual_comp}.
The set of mandatory boundary nodes for the considered scenario is shown in~Figure~\ref{fig::visual_comp}(a).
Both, MDS-BR and EC-BR with refinement return precise outlines of the inner and outer border with almost no artifacts.
Figure~\ref{fig::visual_comp}(c) shows the result of EC-BR before the refinement.
One can clearly see the broader boundary and the circles around small-scale holes.
Funke06 also correctly identifies the boundaries with some artifacts.
Similar to EC-BR, the apparent noise is caused by small holes which are surrounded by marked nodes.
The results of Fekete04 and Funke05 show many artifacts.
Additionally, Fekete04 failed to detect some mandatory boundary nodes.
We also present a classification result of the global algorithm by Wang \emph{et al.}~\cite{wgm06}.
This algorithm produces closed boundary circles with no artifacts but, due to its nature, marked boundaries are not always at the true border but shifted inwards.
This characteristic was the reason why we did not include their algorithm into our quantitative analysis, as it would result in an unfairly poor rating compared to the other algorithms.

In order to convey a feeling for the influence of the refinement, we present additional classification results of our algorithms before and after refinement in Figures~\ref{fig::examples_mdsbr} and \ref{fig::examples_ecbr}. Later on, in Section~\ref{sec::sim::refinement}, we will deal with the questions whether our refinement procedures are interchangeable and if they can also be used to improve the classification results of the other algorithms.

\subsection{Network Density}
\begin{figure}[p]
\centering
\hfill
\subfigure{\includegraphics[width=.19\columnwidth]{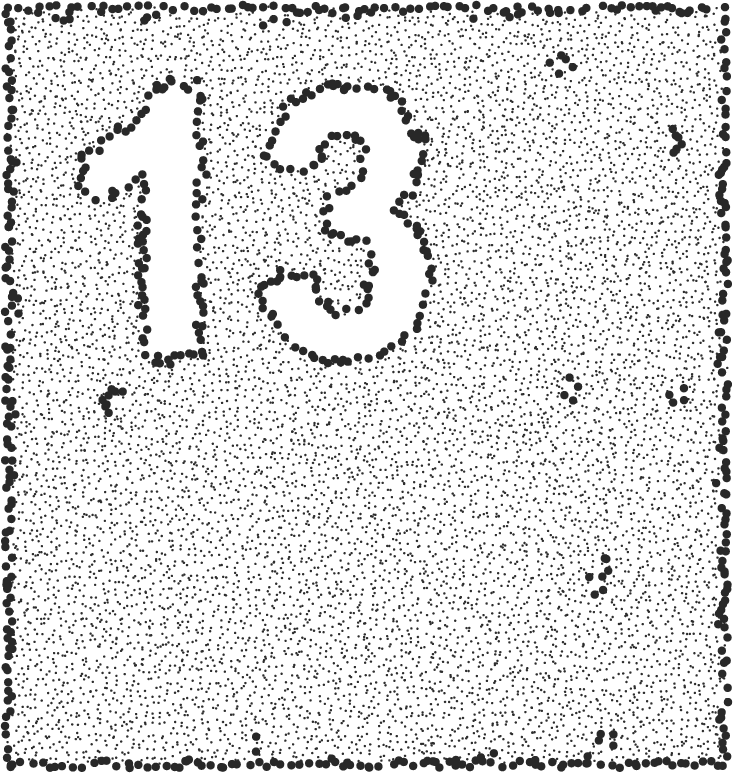}}
\hfill
\subfigure{\includegraphics[width=.19\columnwidth]{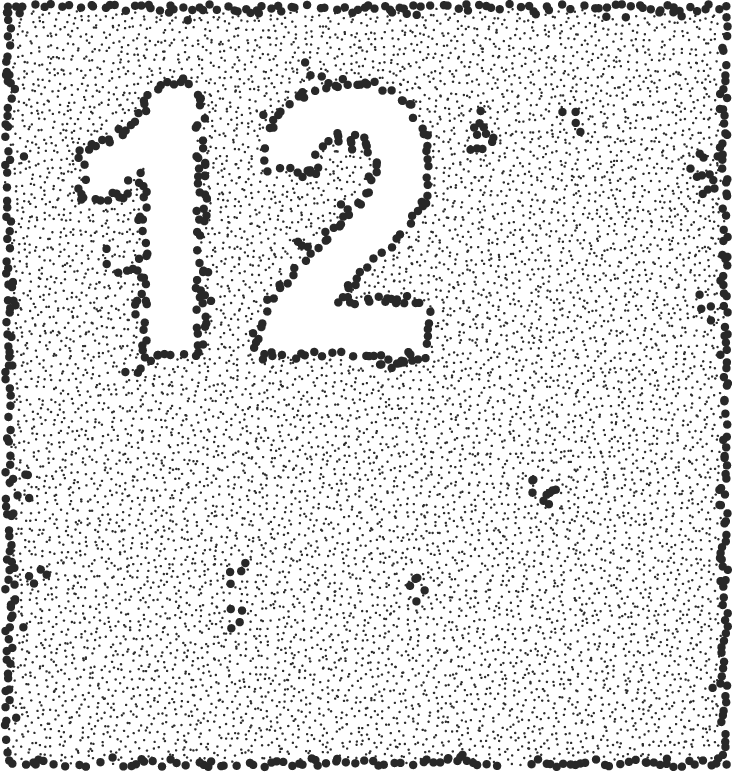}}
\hfill
\subfigure{\includegraphics[width=.19\columnwidth]{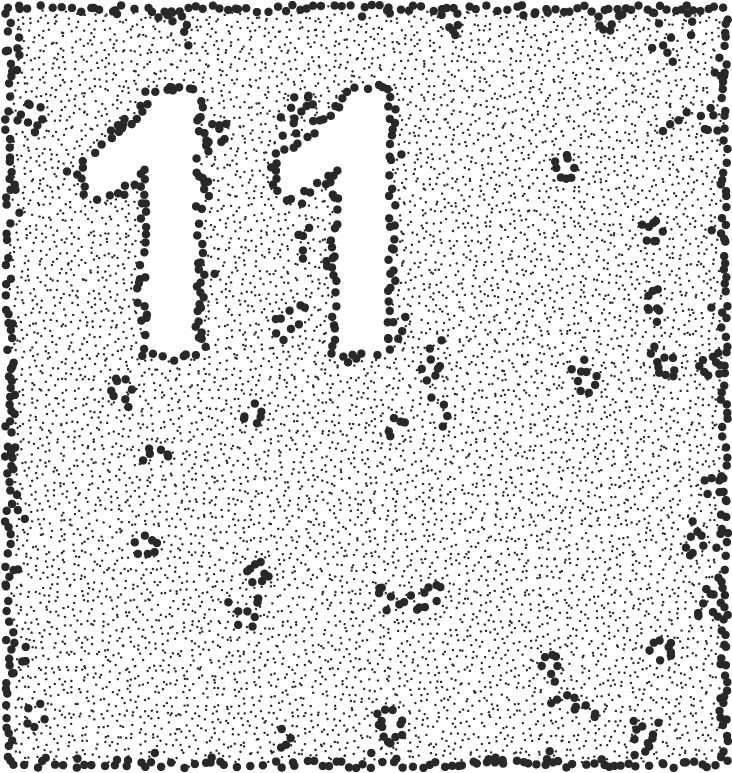}}
\hfill
\subfigure{\includegraphics[width=.19\columnwidth]{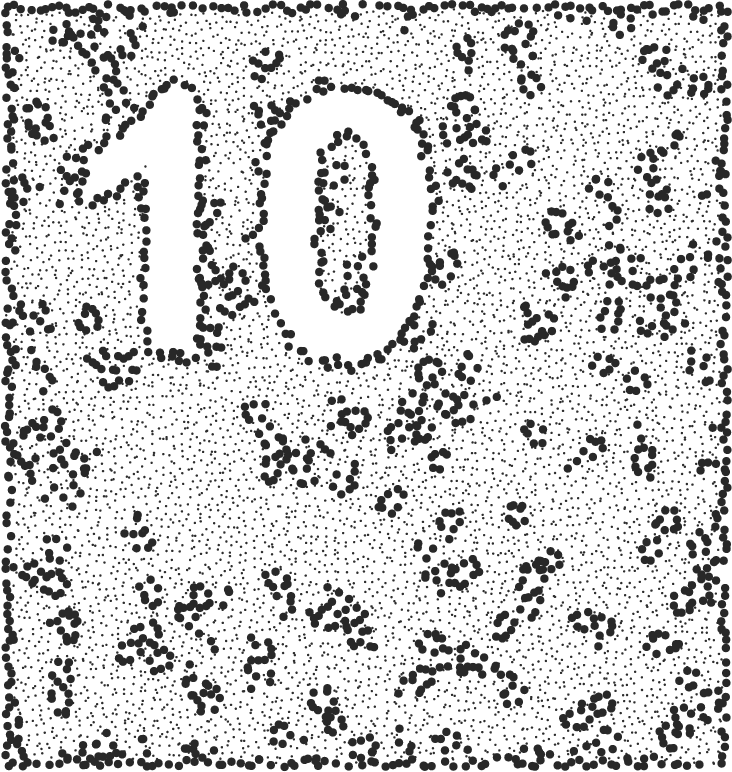}}
\hfill
\subfigure{\includegraphics[width=.19\columnwidth]{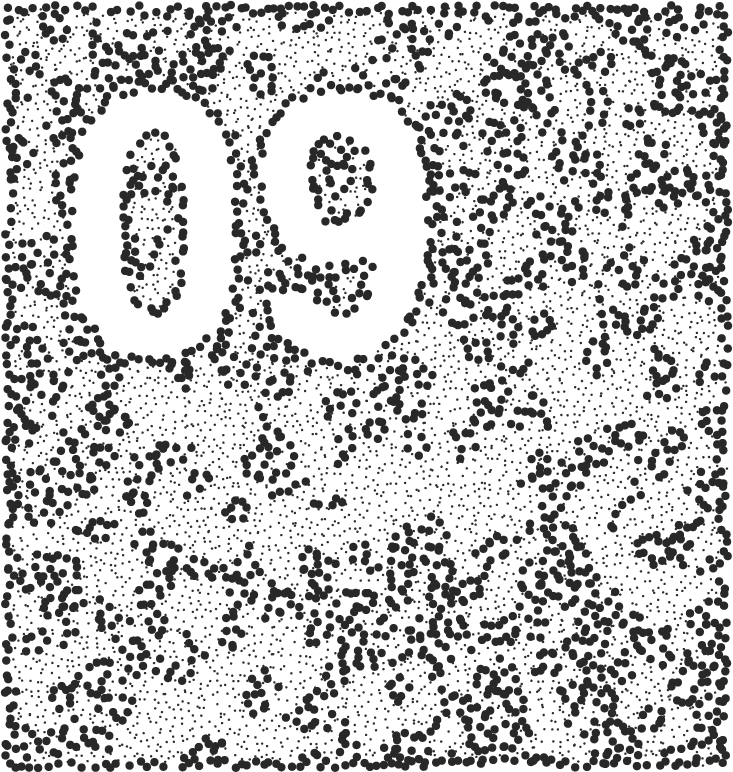}}
\hfill{}

\hfill
\subfigure{\includegraphics[width=.19\columnwidth]{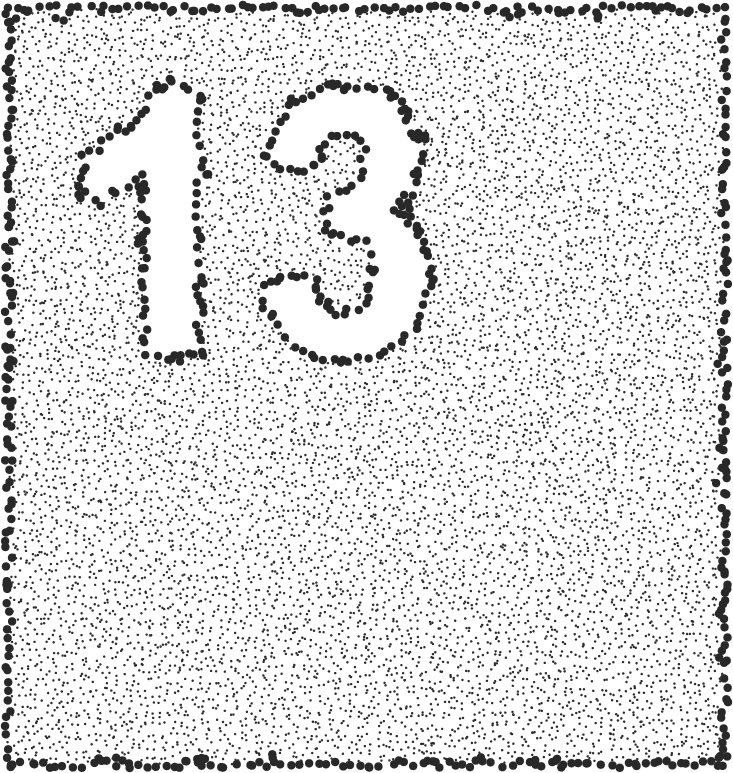}}
\hfill
\subfigure{\includegraphics[width=.19\columnwidth]{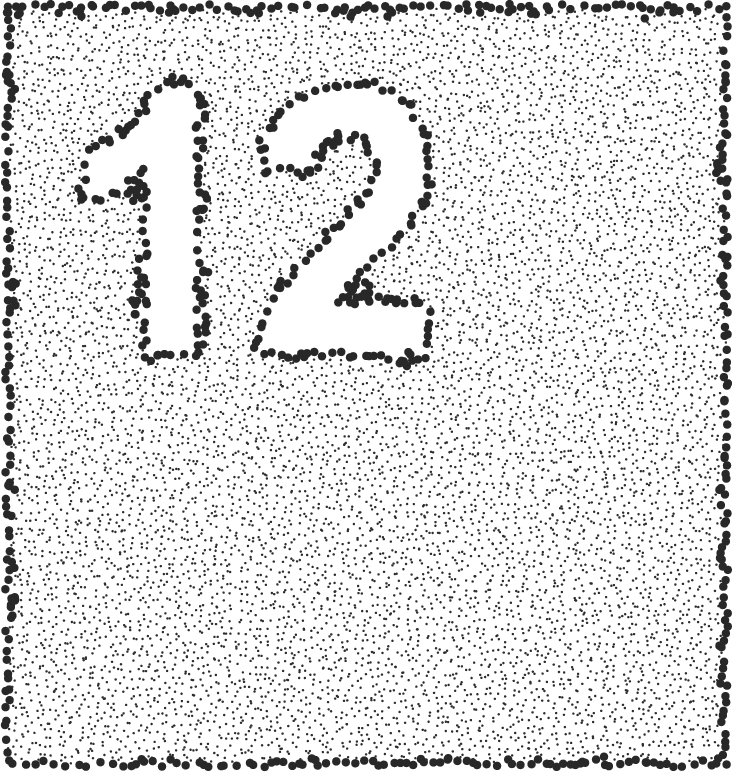}}
\hfill
\subfigure{\includegraphics[width=.19\columnwidth]{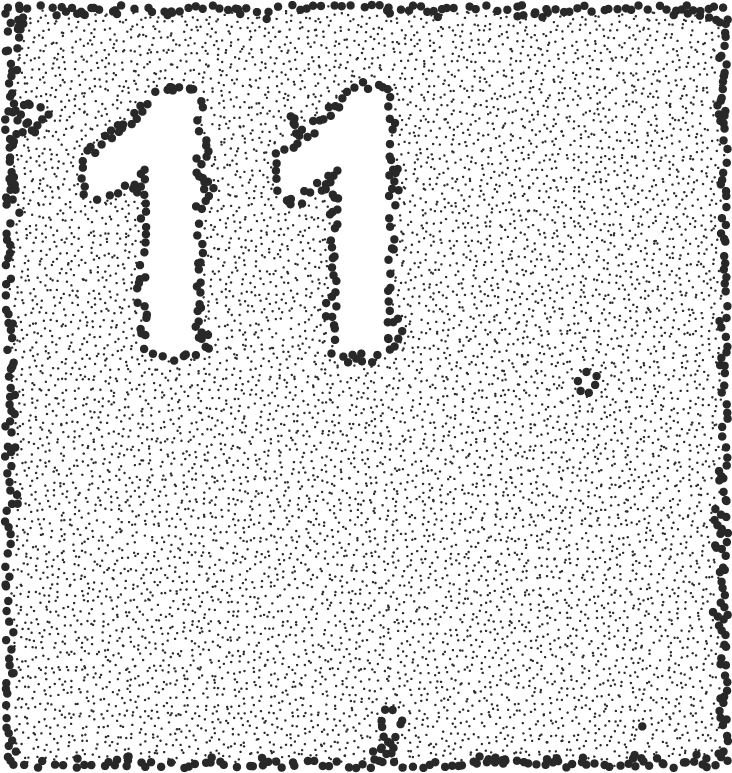}}
\hfill
\subfigure{\includegraphics[width=.19\columnwidth]{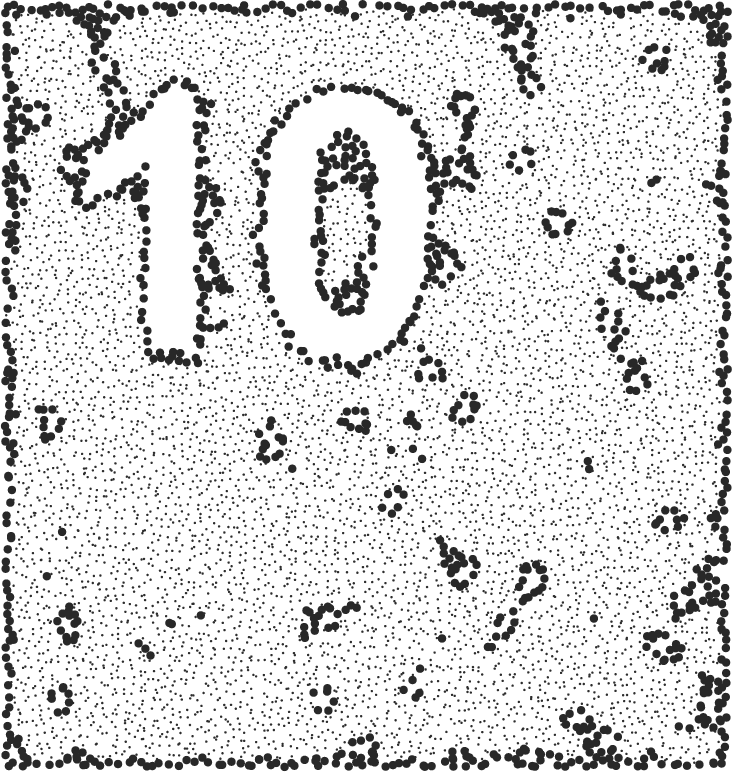}}
\hfill
\subfigure{\includegraphics[width=.19\columnwidth]{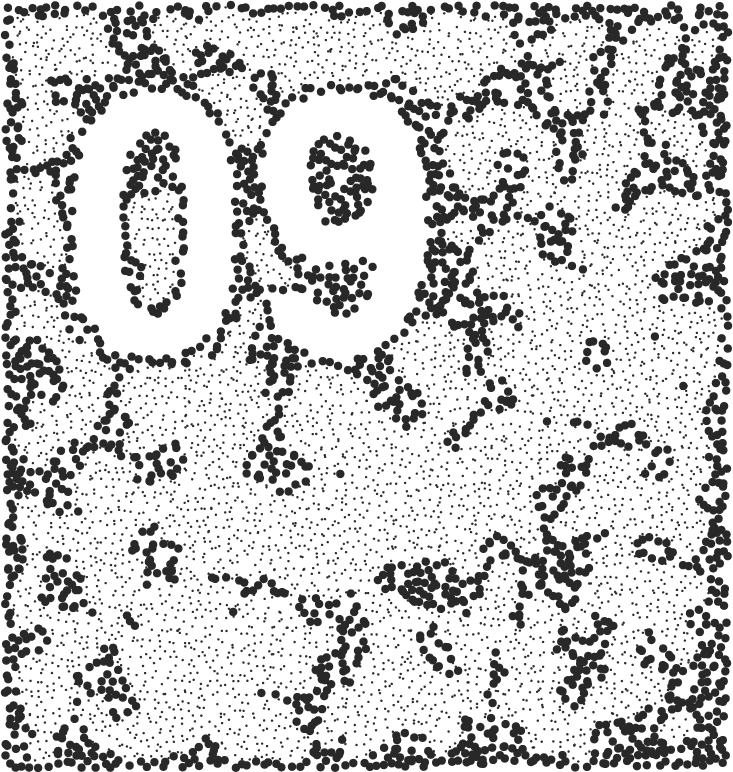}}
\hfill{}

\hfill
\subfigure{\includegraphics[width=.19\columnwidth]{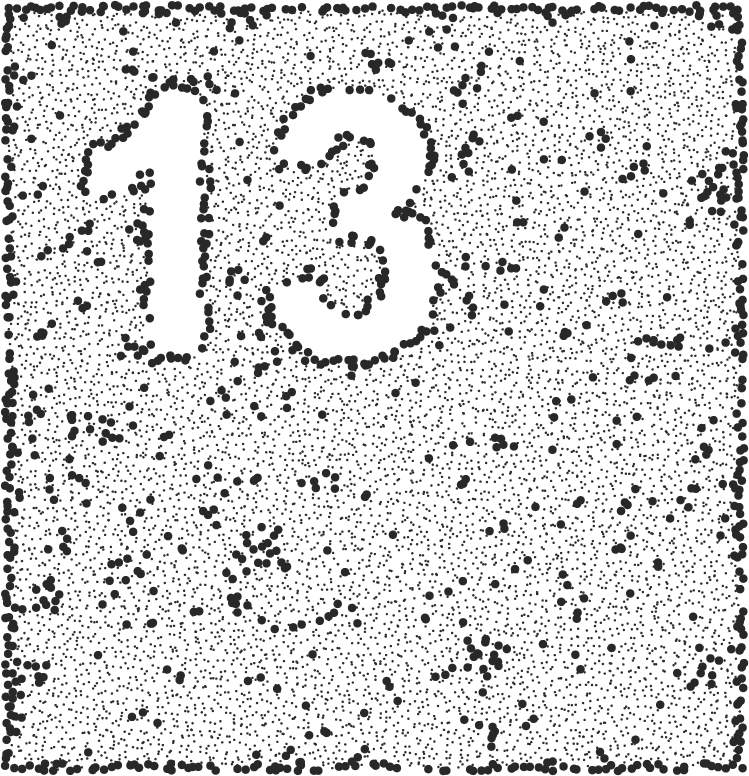}}
\hfill
\subfigure{\includegraphics[width=.19\columnwidth]{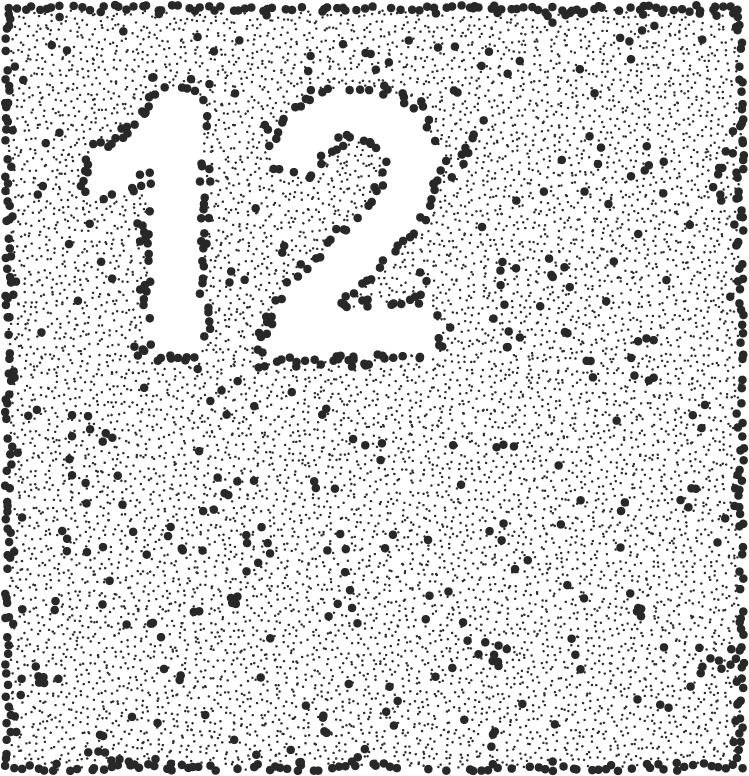}}
\hfill
\subfigure{\includegraphics[width=.19\columnwidth]{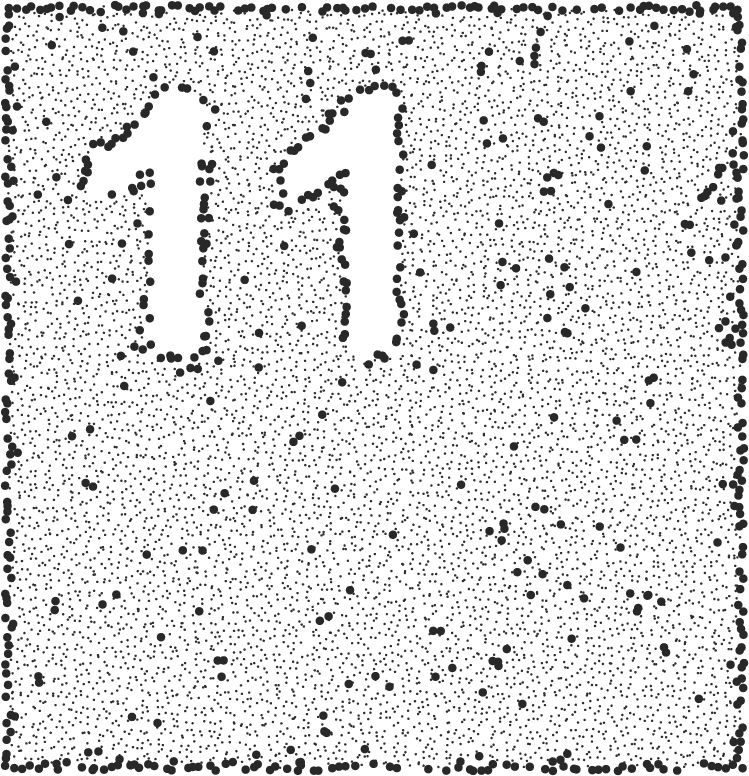}}
\hfill
\subfigure{\includegraphics[width=.19\columnwidth]{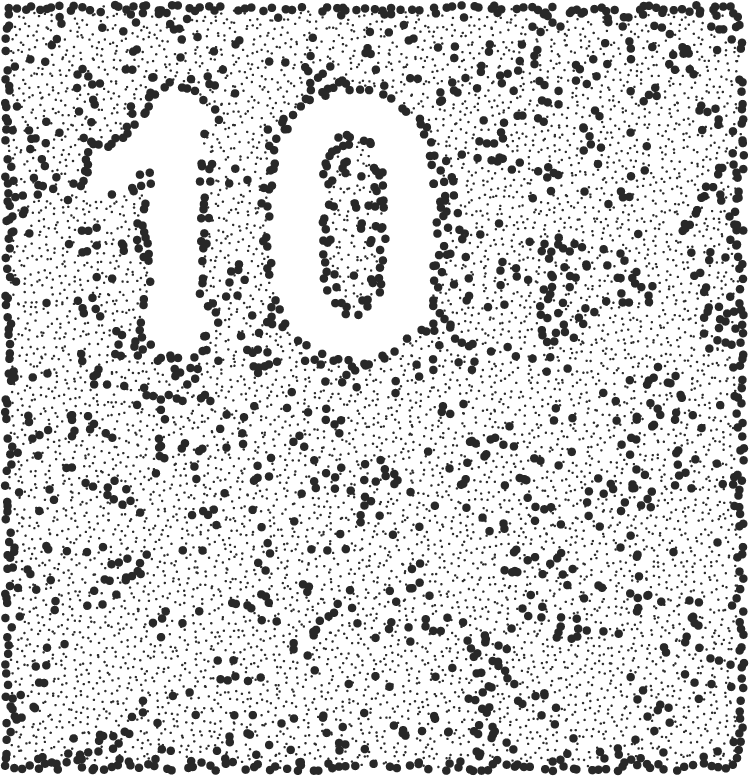}}
\hfill
\subfigure{\includegraphics[width=.19\columnwidth]{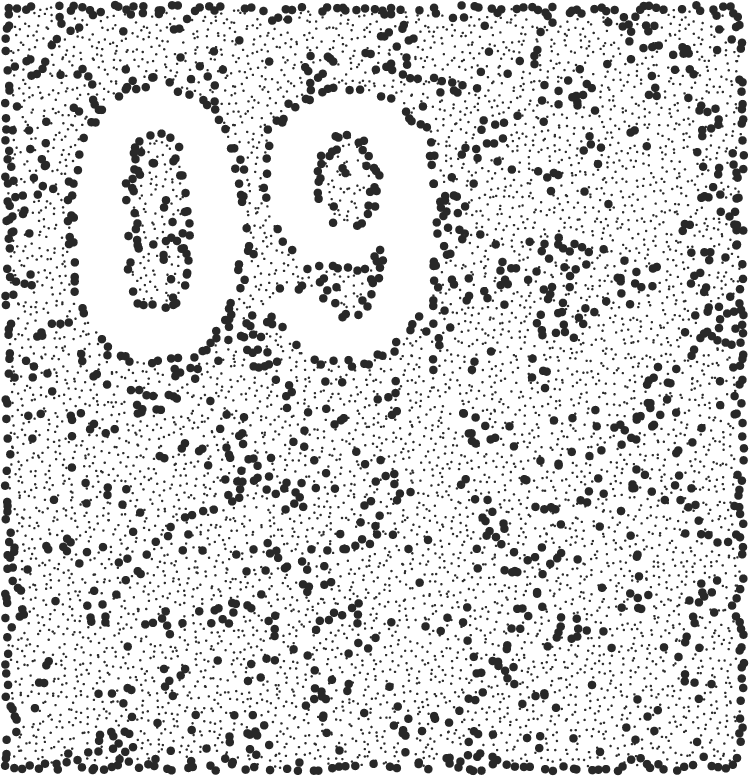}}
\hfill{}

\hfill
\subfigure{\includegraphics[width=.19\columnwidth]{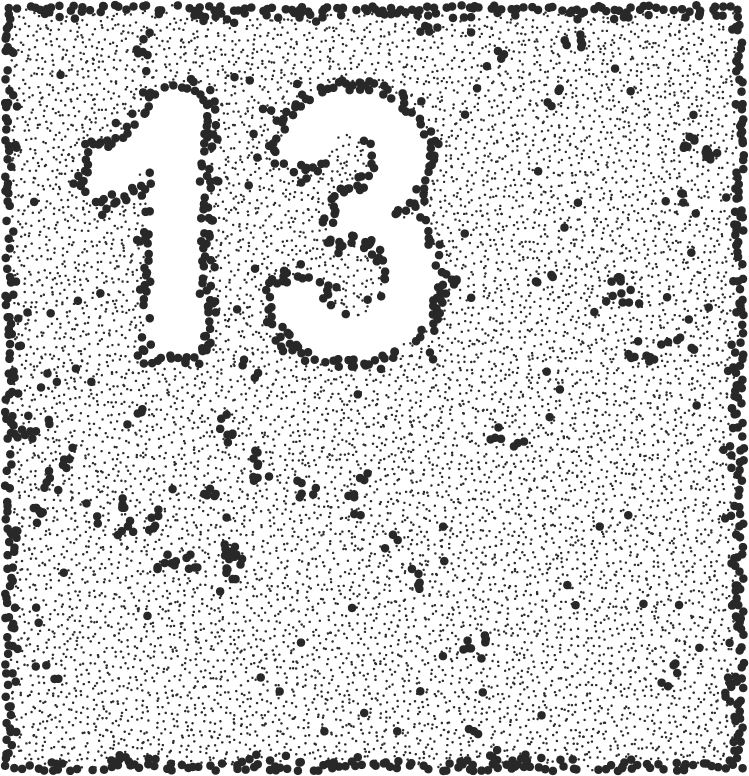}}
\hfill
\subfigure{\includegraphics[width=.19\columnwidth]{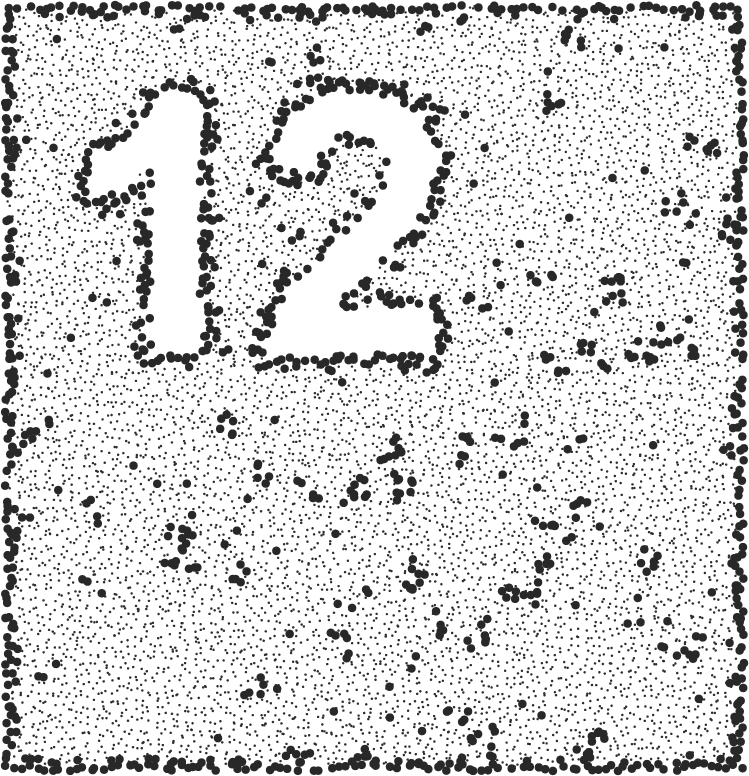}}
\hfill
\subfigure{\includegraphics[width=.19\columnwidth]{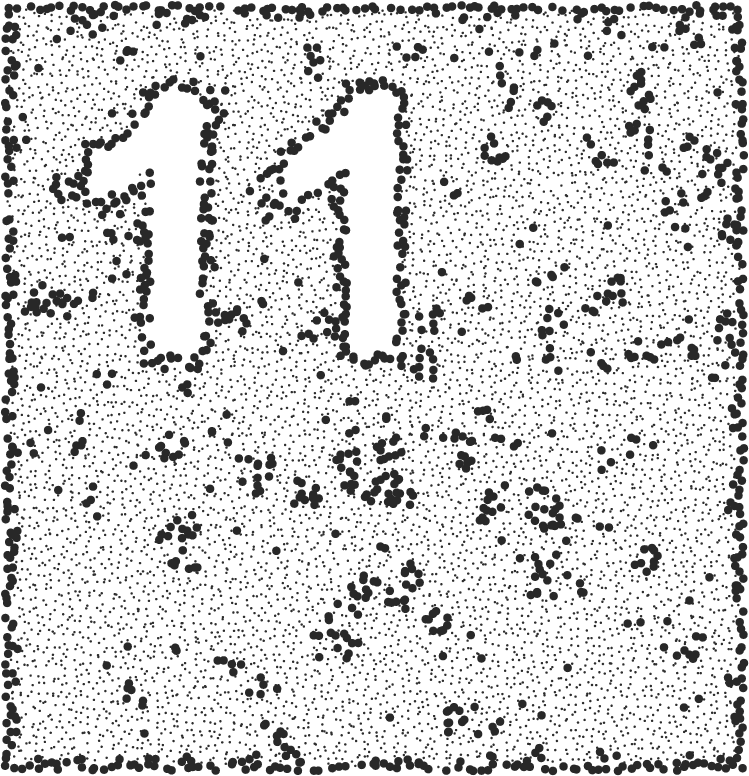}}
\hfill
\subfigure{\includegraphics[width=.19\columnwidth]{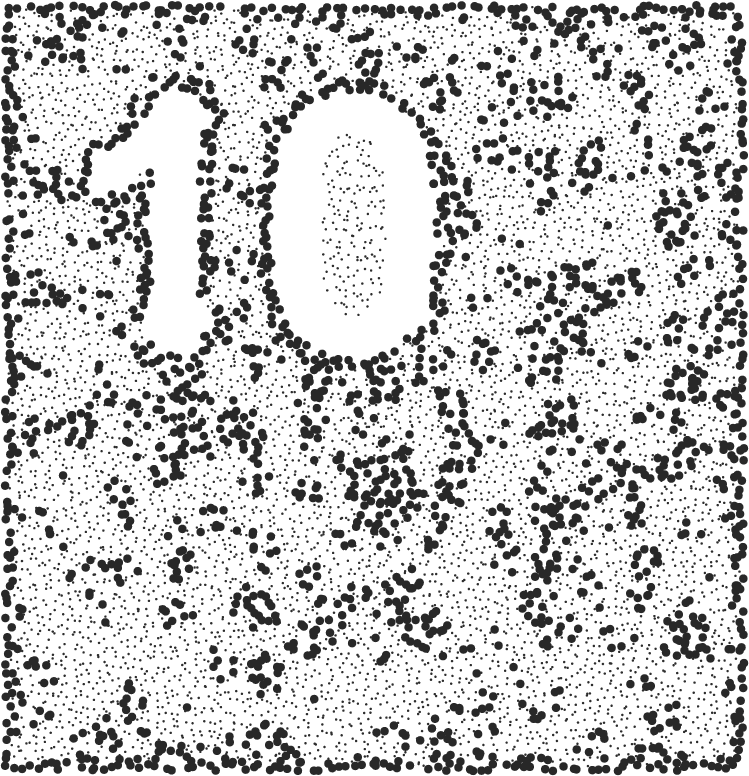}}
\hfill
\subfigure{\includegraphics[width=.19\columnwidth]{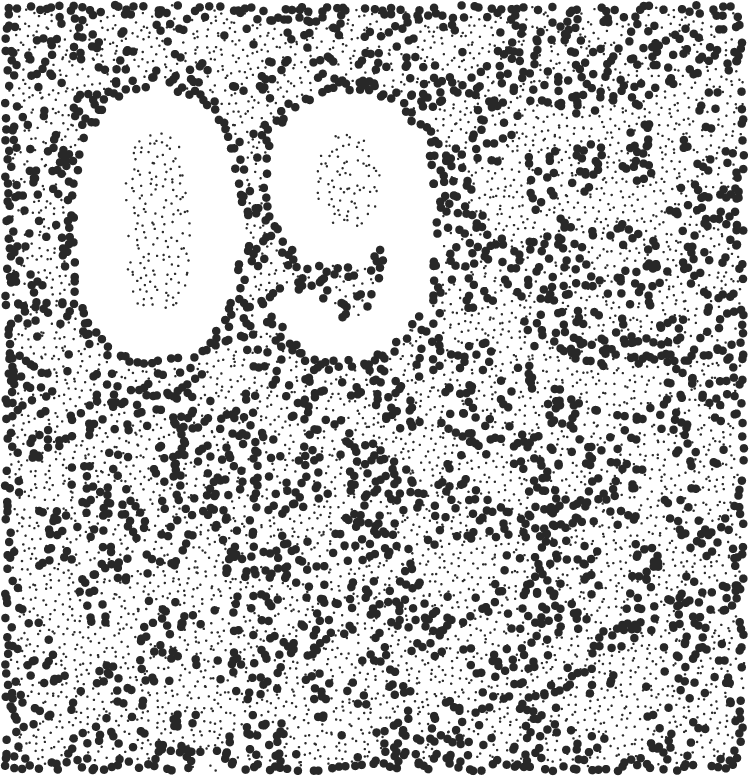}}
\hfill{}

\hfill
\subfigure{\includegraphics[width=.19\columnwidth]{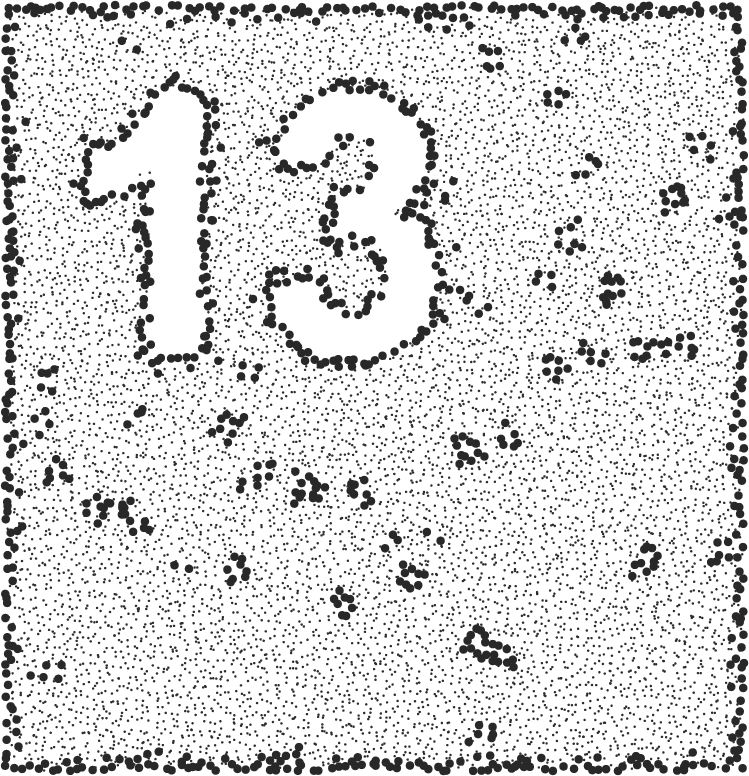}}
\hfill
\subfigure{\includegraphics[width=.19\columnwidth]{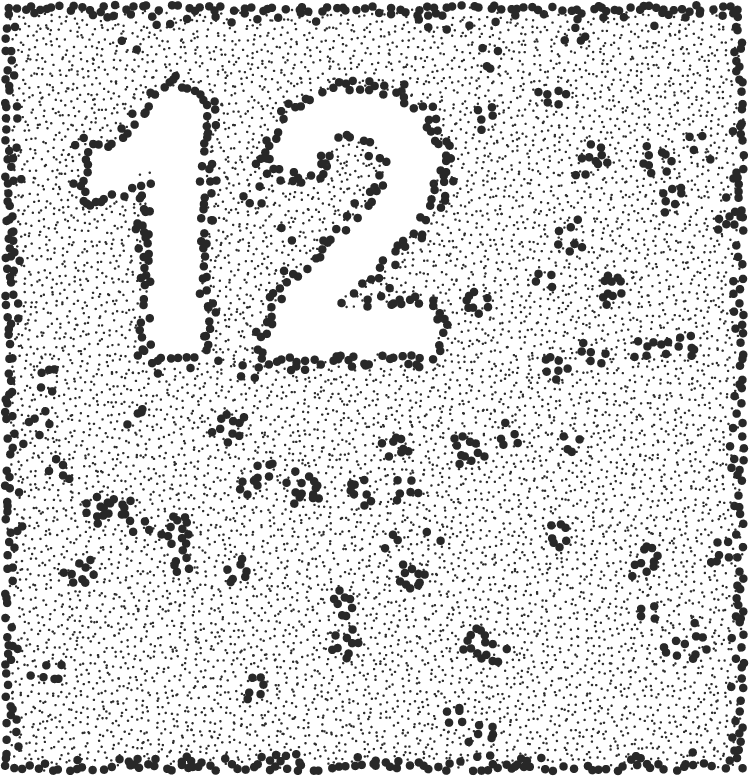}}
\hfill
\subfigure{\includegraphics[width=.19\columnwidth]{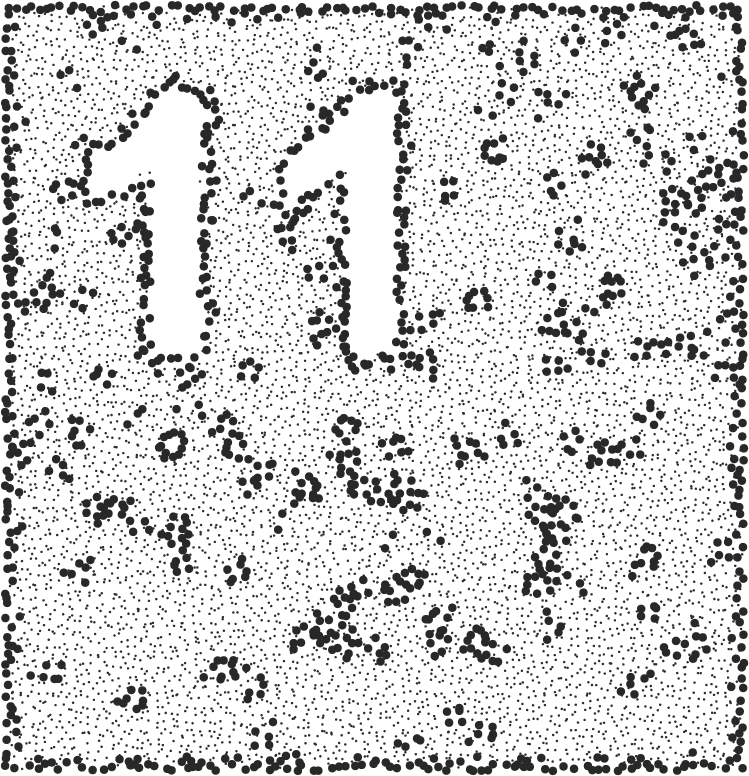}}
\hfill
\subfigure{\includegraphics[width=.19\columnwidth]{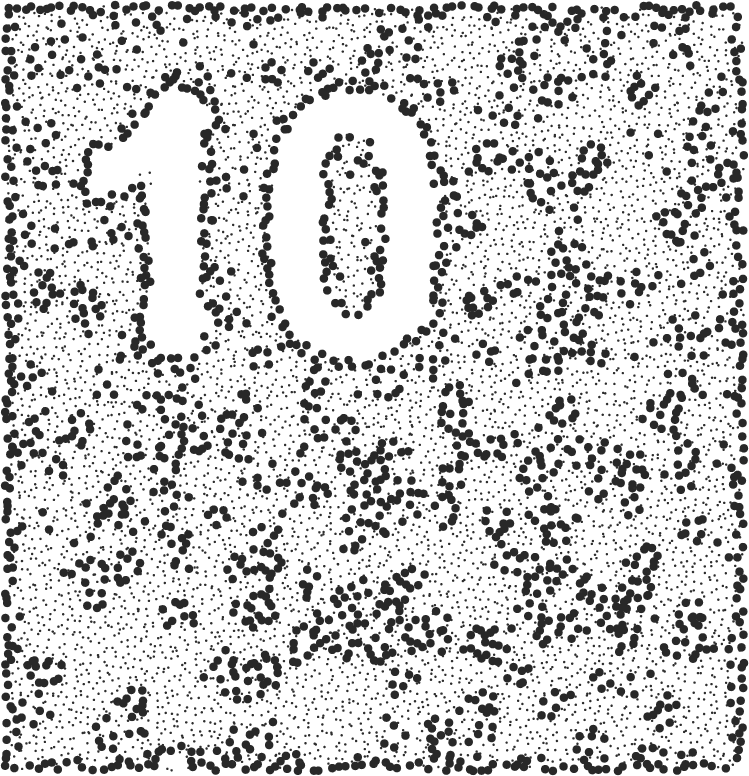}}
\hfill
\subfigure{\includegraphics[width=.19\columnwidth]{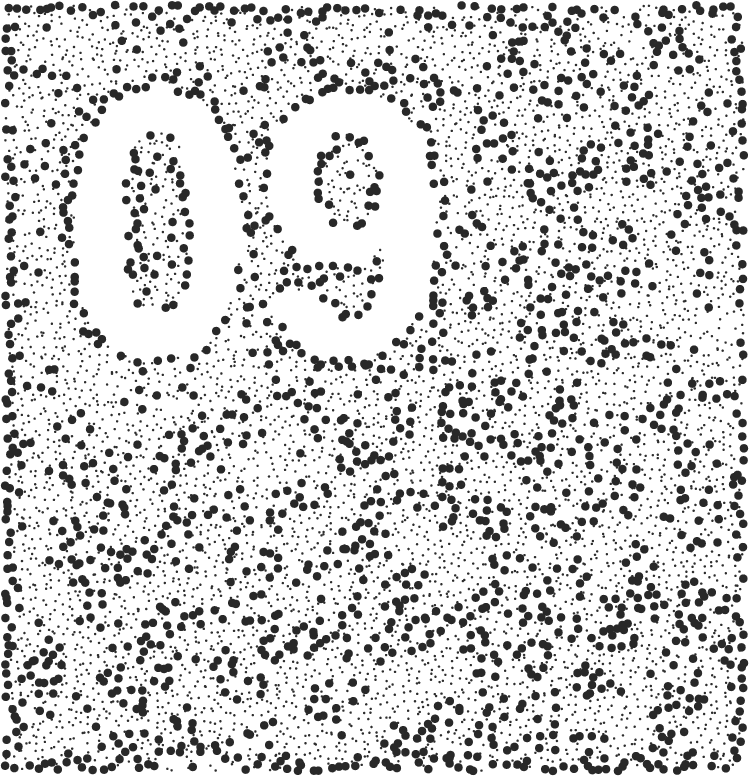}}
\hfill{}
\caption{Influence of average node degree on classification result. The number states the average node degree of the respective network. 
1st row: MDS-BR with refinement. 2nd row: EC-BR with refinement. 3rd row: Fekete04. 4th row: Funke05. 5th row: Funke06. }
\label{fig::node_degrees}
\end{figure}

In this section we consider how the classification performance depends on the average node degree $d_{avg}$ of the network.
A visual impression is given for the considered algorithms in Figure~\ref{fig::node_degrees}.
Obviously, for our algorithms the number of interior nodes classified as boundary nodes increases rapidly for average node degrees 10 and below.
The reason is that for low node densities many small-scale holes emerge.
Thus, most nodes are close to a hole and (correctly) classified as boundary nodes.
Accordingly, for such sparse networks different algorithms and boundary definitions might be more appropriate.
The other algorithms we are considering exhibit the same principle behaviour as our approaches, although they produce much more errors.
All classification results show a lot of ``noise'', interior nodes falsely marked as boundary nodes.
Furthermore, especially Fekete04 and Funke05 have problems detecting the boundary correctly.

A quantitative comparison is given in Table~\ref{tab:net_density}, where the percentage of false classifications for mandatory boundary nodes and interior nodes with increasing $d_{avg}$ is shown.
Results for optional boundary nodes state the percentage classified as interior nodes.
As can be seen, EC-BR with refinement and MDS-BR both classify almost all interior nodes correctly except for the smallest node degree.
But in this extreme scenario, all algorithms start having problems as additional small holes arise due to overall sparse connectivity.
The misclassifications of EC-BR before refinement are mostly due to the algorithm detecting very small holes that fall below our defined threshold.

The classification of mandatory boundary nodes is also excellent.
Here, EC-BR dominates all other algorithms.
The performance of MDS-BR fluctuates only slightly over the various node degrees.
With a misclassification of $3$-$4\%$, it is surpassed by Fekete04 on denser graphs.
The improvement of Fekete04 on these graphs is reasonable since, as a statistical approach, it is predestined for dense networks.

The numbers given for optional boundary nodes state how many nodes within maximum communication distance of a hole are not classified as boundary nodes, excluding mandatory boundary nodes.
Here, the results for EC-BR are particularly interesting.
Before refinement, the algorithm classifies almost all of the optional nodes as boundary nodes while still providing a strict separation to the interior nodes.
After refinement, EC-BR classifies less optional nodes as boundary nodes than all other approaches, while still recognizing almost all mandatory boundary nodes correctly.

\begin{table}[t]
\setlength\tabcolsep{5pt}
\centering
\caption{Misclassification ratios (false negatives) in percent for average node degrees between $9$ and $21$.}\label{tab:net_density}
\begin{tabular}{c||ccccc||ccccc||ccccc}
& \multicolumn{5}{c||}{\bfseries Mandatory}& \multicolumn{5}{c||}{\bfseries Optional}& \multicolumn{5}{c}{\bfseries Interior} \\
 & 9 & 12 & 15 & 18 & 21 & 9 & 12 & 15 & 18 & 21 & 9 & 12 & 15 & 18 & 21 \\
\hline
EC-BR & 2.1 & \bfseries 0.0 & \bfseries 0.0 & \bfseries 0.0 & \bfseries 0.0 & \bfseries 0.3 & \bfseries 0.1 & \bfseries 0.2 & \bfseries 0.2 & \bfseries 0.3 & 54.8 & 7.5 & 3.8 & 2.2 & 1.6  \\
EC-BR Ref & 4.4 & 0.4 & 0.6 & 1.0 & 1.3 & 51.2 & 80.4 & 81.3 & 82.8 & 84.3 & \bfseries 7.1 & \bfseries 0.0 & \bfseries 0.0 & \bfseries 0.0 & \bfseries 0.0  \\

MDS-BR & \bfseries 1.9 & 2.9 & 3.5 & 3.8 & 3.9 & 68.0 & 79.8 & 79.8 & 79.8 & 80.0 & 19.0 & 0.7 & 0.3 & 0.1 & \bfseries 0.0  \\
\hline
Fekete04 & 34.7 & 14.2 & 6.7 & 3.4 & 1.9 & 83.2 & 80.3 & 69.0 & 63.5 & 64.6 & 9.8 & 3.5 & 7.2 & 6.9 & 2.5 \\
Funke05 & 16.6 & 6.3 & 5.7 & 5.1 & 5.0 & 61.5 & 59.5 & 55.4 & 52.5 & 50.6 & 21.7 & 3.5 & 2.0 & 1.3 & 0.9  \\
Funke06 & 39.7 & 13.8 & 16.6 & 18.9 & 20.9 & 80.6 & 70.7 & 71.9 & 72.5 & 73.2 & 13.0 & 3.4 & 1.4 & 0.6 & 0.3
\end{tabular}
\end{table}

\subsection{Random Placement vs. Perturbed Grid}
Perturbed grid placement ensures low variance in node degree over the entire network and fewer very small holes.
These effects are especially well suited for statistical boundary recognition algorithms.
Thus, we particularly expect Fekete04 to yield many misclassifications in a random placement scenario whereas our algorithms should still produce reasonable results.

Table~\ref{tab:rp_pg} compares the performance of all algorithms for both placement strategies and $d_{avg} = 15$.
The performance of the existing algorithms decreases dramatically compared to perturbed grid placement.
On the other hand, the results of the proposed algorithms only decrease noteworthy for interior nodes.
But the classification of these nodes is still on par with or even better than for the other approaches.
For mandatory boundary nodes, our results remain much better than the next best competitor.
The increased misclassifications of interior nodes are partly caused due to our algorithms detecting very small holes with a circumference of less than $4$, which occur in random placement scenarios.

We also compare the influence of the network density when using random node placement.
Table~\ref{tab:rp_mand_int} shows the classification results for mandatory and interior nodes.
We see that all algorithms perform better with increasing network density.
Interestingly, Funke06 seems to perform a good deal worse than the other algorithms in this scenario.
Overall, our approaches dominate the other algorithms for both, mandatory boundary nodes and interior nodes.
For sparse networks, we see an increased misclassification of interior nodes.
As before, this is partly caused by recognizing very small holes.

\begin{table}[t]
\begin{minipage}[t]{0.48\columnwidth}
\setlength\tabcolsep{5pt}
\centering
\caption{Misclassifications for random placement (rp) and perturbed grid placement (pg).}\label{tab:rp_pg}
\begin{tabular}{c||cc||cc}
& \multicolumn{2}{c||}{\bfseries Mandatory}& \multicolumn{2}{c}{\bfseries Interior} \\
 & rp & pg & rp & pg \\
\hline
EC-BR & \bfseries 2.0 & \bfseries 0.0 & 48.7 & 3.8  \\
EC-BR Ref & 4.0 & 0.6 & \bfseries 4.1 & \bfseries 0.0  \\
MDS-BR & 5.2 & 3.5 & 12.2 & 0.3 \\
\hline
Fekete04 & 26.2 & 6.7 & 13.0 & 7.2 \\
Funke05 & 15.5 & 5.7 & 16.1 & 2.0  \\
Funke06 & 45.8 & 16.6 & 7.9 & 1.4
\end{tabular}
\end{minipage}
\hfill
\begin{minipage}[t]{0.48\columnwidth}
\setlength\tabcolsep{5pt}
\centering
\caption{Misclassifications dependent on average node degree for truly random placement.}\label{tab:rp_mand_int}
\begin{tabular}{c||ccc||ccc}
& \multicolumn{3}{c||}{\bfseries Mandatory}& \multicolumn{3}{c}{\bfseries Interior} \\
 & 15 & 20 & 25 & 15 & 20 & 25 \\
\hline
EC-BR & \bfseries 2.0 & \bfseries 1.6 & \bfseries 0.5 & 48.7 & 25.9 & 11.3  \\
EC-BR Ref & 4.0 & 3.1 & 1.9 & \bfseries 4.1 & \bfseries 0.5 & \bfseries 0.1  \\
MDS-BR & 3.8 & 5.2 & 5.8 & 13.4 & 5.2 & 1.7  \\
\hline
Fekete04 & 26.2 & 13.7 & 7.7 & 13.0 & 13.9 & 12.3 \\
Funke05 & 15.5 & 9.4 & 6.7 & 16.1 & 8.1 & 3.6  \\
Funke06 & 45.8 & 29.1 & 26.4 & 7.9 & 6.1 & 2.8
\end{tabular}
\end{minipage}
\end{table}

\subsection{Beyond Unit Disk Graphs}
Unit disk graphs are frequently used for theoretical analyses and in simulations.
They are motivated by the fact that under good-natured conditions every sender has a transmission range which is roughly fixed.
However, under realistic assumptions the transmission range also depends on environmental conditions and obstacles, as well as on unpredictable effects such as interference and signal reflections.
In this section, we evaluate the algorithms under more realistic conditions.
Uncertainties are taken into account by the use of the quasi unit disk graph model, which incorporates the observation that short-range transmissions are usually successful while long-range transmissions have some random behavior.

Table~\ref{tab::qudg1} shows the performance of the algorithms for average node degrees 12 and 15 in simulations on 0.75-QUDG networks.
The increased error rate of MDS-BR is due to the base algorithm producing a candidate set which is not necessarily connected.
In consequence, the refinement classifies many correct boundary candidates as interior nodes as the connected substructures are not sufficiently large.
Fekete04, Funke05 and Funke06 yield even higher misclassification rates and also perform significantly worse than in our UDG simulations.
For the refinement of EC-BR, the threshold $\gamma=70\%$ was used.
As mentioned before, a refinement with $\gamma=100\%$ would perform poorly because in QUDGs mandatory boundary nodes are not necessarily surrounded by boundary candidates.
With the adjusted refinement threshold, EC-BR  outperforms the other approaches significantly.
In Table~\ref{tab::qudg2}, we go a step further and compare 0.25-QUDGs with 0.75-QUDGs.
This means that there is a very high level of uncertainty.
As expected, all algorithms produce more misclassifications.
Again, EC-BR with refinement outperforms the other algorithms easily.

\begin{table}[t]
\begin{minipage}[t]{0.48\columnwidth}
\setlength\tabcolsep{5pt}
\centering
\caption{Misclassifications for the 0.75-QUDG model and different average node degrees.}\label{tab::qudg1}
\begin{tabular}{c||cc||cc}
& \multicolumn{2}{c||}{\bfseries Mandatory}& \multicolumn{2}{c}{\bfseries Interior} \\
& 12 & 15 & 12 & 15 \\
\hline
EC-BR & \bfseries 0.0 & \bfseries 0.0 & 28.5 & 7.7  \\
EC-BR Ref & \bfseries 0.0 & \bfseries 0.0 & \bfseries 4.9 & \bfseries 0.3  \\
MDS-BR & 8.3 & 11.2 & 8.3 & 1.6  \\
\hline
Fekete04 & 16.9 & 6.9 & 8.8 & 8.9  \\
Funke05 & 9.0 & 7.4 & 12.9 & 5.2  \\
Funke06 & 15.6 & 15.4 & 12.4 & 3.7
\end{tabular}
\end{minipage}
\hfill
\begin{minipage}[t]{0.48\columnwidth}
\setlength\tabcolsep{5pt}
\centering
\caption{Misclassifications for 0.25- and 0.75-QUDG models with average node degree 12.}\label{tab::qudg2}
\begin{tabular}{c||cc||cc}
& \multicolumn{2}{c||}{\bfseries Mandatory}& \multicolumn{2}{c}{\bfseries Interior} \\
& 0.25 & 0.75 & 0.25 & 0.75 \\
\hline
EC-BR & \bfseries 3.0 & \bfseries 0.0 & 41.5 & 28.5  \\
EC-BR Ref. & 12.7 & \bfseries 0.0 & \bfseries 1.7 & \bfseries 4.9  \\
MDS-BR & 27.7 & 8.3 & 11.8 & 8.3  \\
\hline
Fekete04 & 14.6 & 16.9 & 13.6 & 8.8  \\
Funke05 & 11.5 & 9.0 & 17.8 & 12.9  \\
Funke06 & 24.2 & 15.6 & 2.8 & 12.4
\end{tabular}
\end{minipage}
\end{table}

\subsection{MDS Variants}\label{ssec:variants}
The MDS-BR algorithm requires a good approximation of relative angles between nodes to produce premium results.
Actual coordinates are not necessary.
We now analyze the three MDS embedding variants presented in Section~\ref{sec:mdsbr} and study their impact on the classification quality.
The following naming scheme is used: MDS is our default variant, OPT denotes the usage of real coordinates, MDS3 applies MDS on a $3$-hop neighborhood, and SSMDS incorporates signal strength information into the MDS computation.\footnote{We assume strong signals between nodes that are less than half the maximum communication distance apart, weak signals if they are at least within maximum communication distance, and no signals otherwise.}

Classification results of MDS-BR and the three variants are listed in Table~\ref{tab:mdsbr_ss} for three scenarios.
All networks have an average node degree $d_{avg}=15$, \emph{rnd} additionally applies random node placement and \emph{qudg} uses the 0.75-QUDG model.
We find that all variants classify mandatory boundary nodes much better than the default MDS-BR algorithm with a similar misclassification rate of interior nodes.
Surprisingly, OPT yields the worst classification of mandatory nodes over all variants in the \emph{rnd} scenario.
This is due to the other algorithms generally marking more nodes as boundary nodes which leads to better results for boundary nodes but worse results for interior nodes.

Incorporating even just very coarse information on signal strength into the model improves the classification quality significantly.
Considering these impressive results, we infer that utilizing signal strength information is a desirable extension of the algorithm, even more so than using the additional data of a larger neighborhood.

Comparing the results to EC-BR, we see that MDS-BR can close the gap to EC-BR.
Thus, both algorithms can be considered as good choices for distributed high-quality boundary detection.

\begin{table}[b]
\setlength\tabcolsep{5pt}
\centering
\caption{Misclassifications of different embedding variants for MDS-BR.}\label{tab:mdsbr_ss}
\begin{tabular}{c||ccc||ccc}
& \multicolumn{3}{c||}{\bfseries Mandatory}& \multicolumn{3}{c}{\bfseries Interior} \\
 & norm & rnd & qudg & norm & rnd & qudg \\
\hline
MDS & 3.6 & 3.8 & 11.3 & 0.3 & 13.5 & 1.6  \\
MDS3 & 1.8 & 2.6 & 6.1 & 0.3 & 13.2 & 1.4  \\
SSMDS & 0.6 & \bfseries 1.0 & 2.6 & 0.4 & 15.9 & 2.2  \\
OPT & \bfseries 0.4 & 5.3 & \bfseries 0.8 & \bfseries 0.1 & \bfseries 11.7 & \bfseries 0.9
\end{tabular}
\end{table}

In this context, we also analyzed the quality of our embeddings by comparing the computed maximum angles to the true values.
For MDS, the average deviation was about $11^\circ$.
By differentiating between interior nodes and mandatory boundary nodes we gain further information.
While the deviation for interior nodes is only about $8^\circ$, it is up to $39^\circ$ for mandatory boundary nodes.
This error decreases to about $30^\circ$ using MDS3 and down to $23^\circ$ using SSMDS, respectively.
Using random node placement or quasi unit disk graphs only slightly impairs these numbers.
Even though these deviations seem to be quite large, we have already seen in our analysis that the maximum opening angle is still an excellent classification criterion.

\subsection{Parameter Selection}\label{ssec:parameter_tuning}
So far, we applied the same set of parameter values, $\alpha_{min} = 90^\circ$ and $r_{min} = 3$, for all simulation runs of MDS-BR.
These values yield good results for all tested instances, but one might ask if this was merely by chance and how sensitive our algorithm reacts to varying the values of $\alpha_{min}$ and $r_{min}$.

First, we discuss changing the value of the minimum opening angle $\alpha_{min}$ at which a node is classified as boundary node.
Smaller values yield more nodes classified as boundary nodes and vice-versa.
For sparse graphs, larger values of $\alpha_{min}$ are required if we do not want to misclassify too many interior nodes as boundary nodes.
For dense graphs, starting at about $d_{avg}=15$, even small values of $\alpha_{min}$ yield a good classification of boundary nodes with little impact on the results of the interior nodes.
Table~\ref{tab:mdsbr_angle} provides results for networks with $d_{avg} \in \{12,15\}$.
As expected, very large values of $\alpha_{min}$ lead to many nodes falsely classified as interior nodes while almost no node is classified as boundary nodes that is not.
The opposite behavior is true for very small values of $\alpha_{min}$.
Good results are achieved for a relatively wide range of $\alpha_{min}$ between $70^\circ$ and $100^\circ$.
Since the results represent averages over many setups, we assert that the value of $\alpha_{min}$ is robust to the respective network layout.

\begin{table}[b]
\setlength\tabcolsep{5pt}
\centering
\caption{Misclassifications of MDS-BR on sensor networks with $d_{avg} \in \{12,15\}$ for different values of $\alpha_{min}$.}\label{tab:mdsbr_angle}
\begin{tabular}{cl||ccccccc}
& $\alpha_{min}$ & 54$^\circ$ & 72$^\circ$ & 81$^\circ$ & 90$^\circ$ & 99$^\circ$ & 108$^\circ$ & 126$^\circ$ \\
\hline
12 & Mandatory &  0.1 & 0.6 & 1.3 & 2.8 & 5.8 & 12.0 & 41.1 \\
   & Interior  & 41.1 & 6.7 & 2.1 & 0.7 & 0.2 &  0.1 &  0.0 \\
15 & Mandatory &  0.2 & 0.9 & 1.7 & 3.5 & 6.8 & 12.9 & 42.5 \\
   & Interior  & 26.9 & 2.6 & 0.8 & 0.3 & 0.1 &  0.0 &  0.0   
\end{tabular}
\end{table}

Secondly, we analyzed the impact of modifying the value of the $r_{min}$ parameter of the refinement step. 
Results for networks with $d_{avg} \in \{12, 15\}$ are given in Table~\ref{tab:mdsbr_rmin}.
Here, $r_{min}=0$ corresponds to not performing the refinement step at all.
We see that the actual value of $r_{min}$ has little impact on the classification quality.
As soon as a refinement is performed, misclassification of interior nodes decreases considerably with only marginal impact on the classification quality of the mandatory boundary nodes.
We assert that the refinement successfully filters the ``noise'' while retaining the correctly classified boundary nodes.
During this analysis, we also kept track of the average size of the marked neighborhood that is analyzed by each node in the refinement step.
As expected, it increases only slightly with growing node density from $7.2$ nodes at $d_{avg}=12$ to $10.1$ nodes at $d_{avg}=21$, both times using $r_{min} = 3$.
Changing the value of $r_{min}$ by $1$ alters the respective neighborhood size by about $2.5$ nodes for all considered values of $d_{avg}$.

\begin{table}[tb]
\setlength\tabcolsep{5pt}
\centering
\caption{Misclassifications of MDS-BR on sensor networks with $d_{avg} \in \{12,15\}$ varying the $r_{min}$ value of the refinement step. The value $r_{min}=0$ corresponds to not performing a refinement.}\label{tab:mdsbr_rmin}
\begin{tabular}{cl||ccccc}
   & $r_{min}$ & 0 & 2 & 3 & 4 & 5 \\
\hline
12 & Mandatory & 2.8 & 2.8 & 2.9 & 2.9 & 3.0 \\
   & Interior  & 3.2 & 1.0 & 0.7 & 0.4 & 0.3 \\
15 & Mandatory & 3.5 & 3.5 & 3.5 & 3.6 & 3.6 \\
   & Interior  & 1.4 & 0.4 & 0.3 & 0.2 & 0.1
\end{tabular}
\end{table}

To complete the analysis, we examined varying $\alpha_{min}$ and $r_{min}$ in conjunction.
We found that the effects of varying the value of each parameter add up nicely.
Given an arbitrary but fixed value of $\alpha_{min}$, increasing the value of $r_{min}$ always resulted in slightly more nodes falsely classified as interior nodes and slightly less nodes falsely classified as boundary nodes.
The respective behavior can be observed for an arbitrary but fixed value of $r_{min}$ and modifying the value of $\alpha_{min}$.

On the whole, we saw that our original values, $\alpha_{min}=90^\circ$ and $r_{min} = 3$, already yield overall best results, with a possible alternative $\alpha_{min}=81^\circ$ and $r_{min} = 4$ emphasizing slightly more on correctly classifying mandatory boundary nodes than interior nodes.
MDS-BR is robust against variations in $r_{min}$ as long as the refinement step is performed.
On the other hand, $\alpha_{min}$ must be chosen carefully within the range of $70^\circ$ to $100^\circ$ to yield good results.
Fortunately, this value does not seem to depend on the analyzed network but on basic geometric considerations.

Regarding EC-BR in this context, a very nice feature of this algorithm is that a single fixed parameter setting already works well for all scenarios. The circle length of $6$ as a threshold to distinguish inner nodes from boundary nodes is mostly independent of node degree, kind of placement, and communication model. And for the refinement, checking whether $70\%$ of the neighbors are marked as being on the boundary works well in all situations. For the special case of unit disk graphs, a threshold of $\gamma=100\%$ can be used to obtain a thinner boundary. Hence, we do not go further into detail regarding the parameters of EC-BR.

\subsection{Refinement}
\label{sec::sim::refinement}
In this report, we have presented two procedures to generate initial boundary classifications and corresponding refinement heuristic to eliminate false positives.
One might ask why we need two different refinement routines or how well each of them works on the solution provided by the other base algorithm.

\begin{figure}[p]
\centering
\hfill
\subfigure[]{\includegraphics[width=.21\columnwidth]{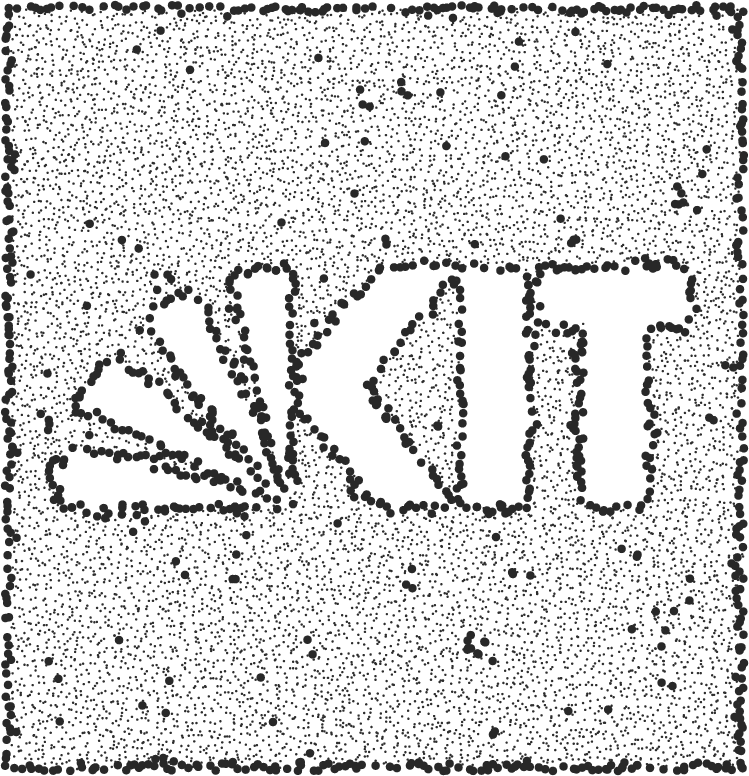}}
\hfill
\subfigure[]{\includegraphics[width=.21\columnwidth]{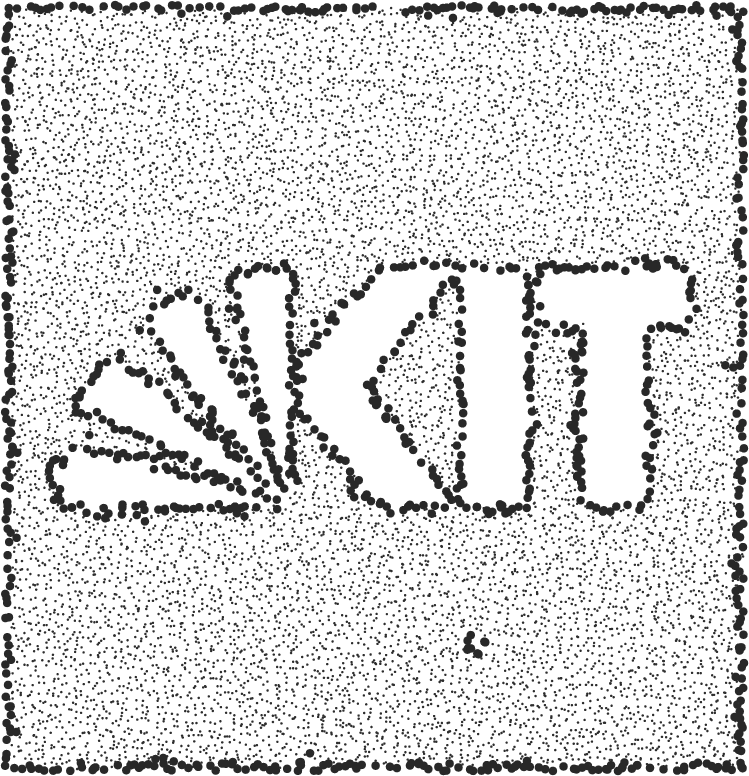}}
\hfill
\subfigure[]{\includegraphics[width=.21\columnwidth]{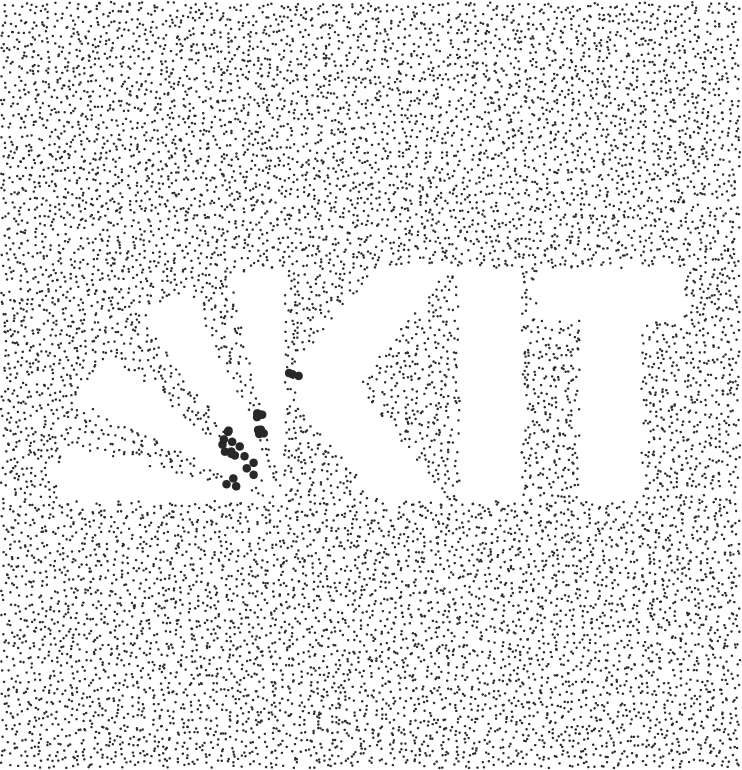}}
\hfill
\subfigure[]{\includegraphics[width=.21\columnwidth]{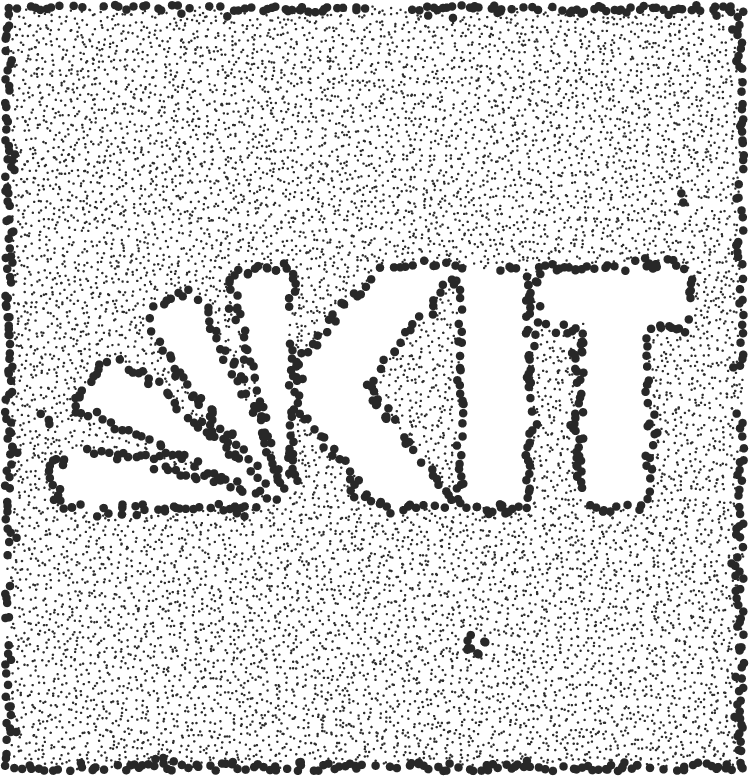}}
\hfill{}
\caption{Impact of using different refinement heuristics on the classification results of MDS-BR. (a) No Refinement. (b) MDS-BR Refinement. (c) EC-BR Refinement ($\gamma=100\%$). (d) EC-BR Refinement ($\gamma=20\%$).}
\label{fig:refinement_mdsbr}
\end{figure}

\begin{figure}[p]
\centering
\hfill
\subfigure[]{\includegraphics[width=.21\columnwidth]{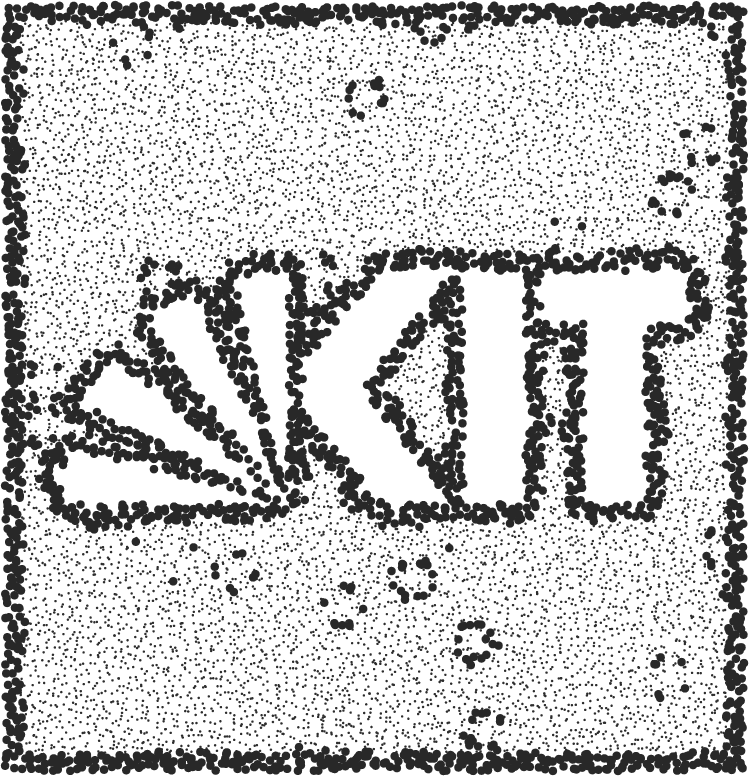}}
\hfill
\subfigure[]{\includegraphics[width=.21\columnwidth]{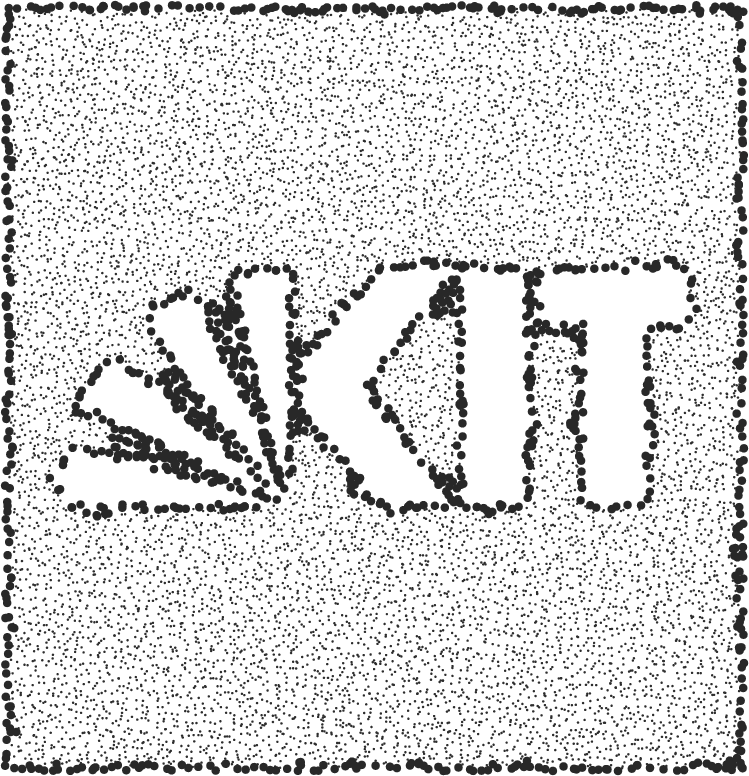}}
\hfill
\subfigure[]{\includegraphics[width=.21\columnwidth]{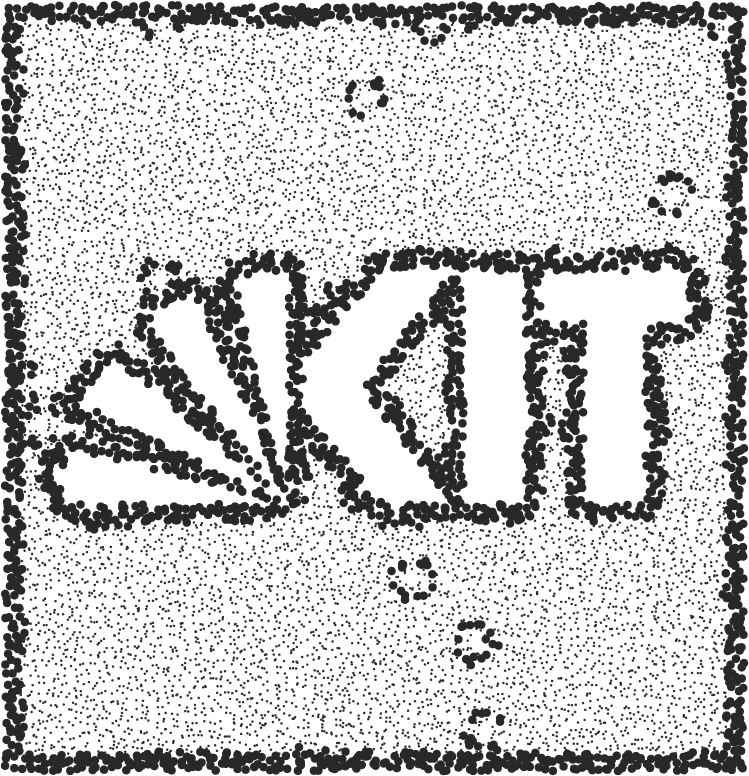}}
\hfill
\subfigure[]{\includegraphics[width=.21\columnwidth]{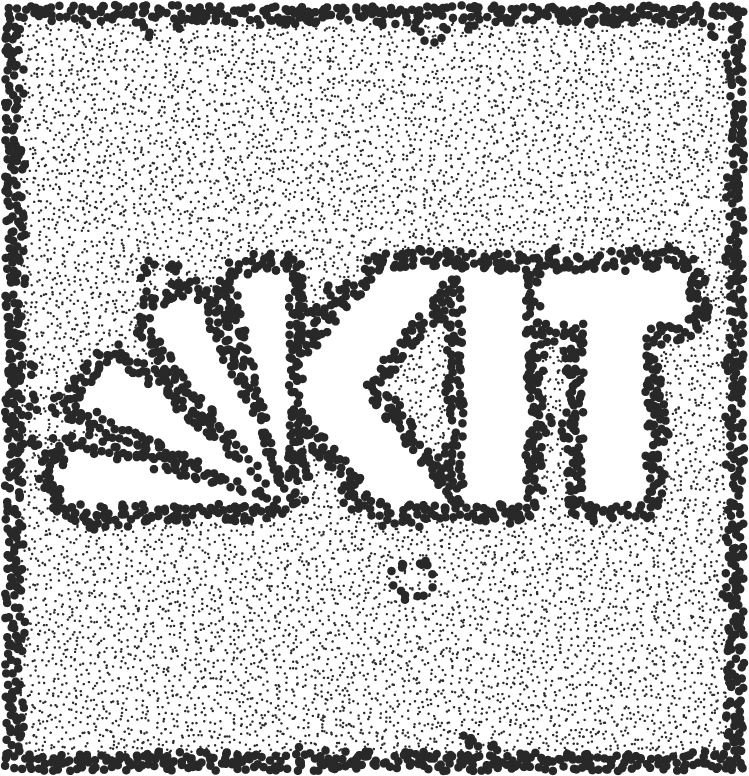}}
\hfill{}
\caption{Impact of using different refinement heuristics on the classification results of EC-BR. (a) No Refinement. (b) EC-BR Refinement. (c) MDS-BR Refinement ($r_{min} = 3$). (d) MDS-BR Refinement ($r_{min} = 10$).}
\label{fig:refinement_ecbr}
\end{figure}

In short, each refinement step is tailored to the particular results of each base algorithm and does perform poorly with the other one.
To illustrate this behavior, Figure~\ref{fig:refinement_mdsbr} and Figure~\ref{fig:refinement_ecbr} show the results of our two algorithms with each refinement heuristics.
First, using the refinement step of EC-BR on the results of the base algorithm of MDS-BR, we retain almost no node classified as boundary node.
This occurs as MDS-BR already yields a thin outline of the boundary but the refinement of EC-BR requires boundary nodes to be completely surrounded by other boundary nodes to remain a boundary node.
We can tune the refinement step similarly to what was done for quasi unit disk graphs to obtain better results.
To obtain best results in this scenario, we require a node to be surrounded by only $\gamma=20\%$ of boundary nodes.
This considerably improves the classification, but there are still sections of the boundary missing while ``noise'' already starts to emerge.
Secondly, applying the refinement step of MDS-BR to the classification results of EC-BR, we see that only few boundary nodes are switched to interior nodes.
As EC-BR yields a broad halo around each boundary, the refinement of MDS-BR almost always finds a path within this halo of the required expansion.
Using even much more aggressive parameter values for the refinement step does not improve the results by much.
Some additional small structures are removed but the boundary remains a broad halo.

\begin{figure}[p]
\centering
\hfill
\subfigure[]{\includegraphics[width=.21\columnwidth]{figs_ds/mdsbr-mdsbr-ref.png}}
\hfill
\subfigure[]{\includegraphics[width=.21\columnwidth]{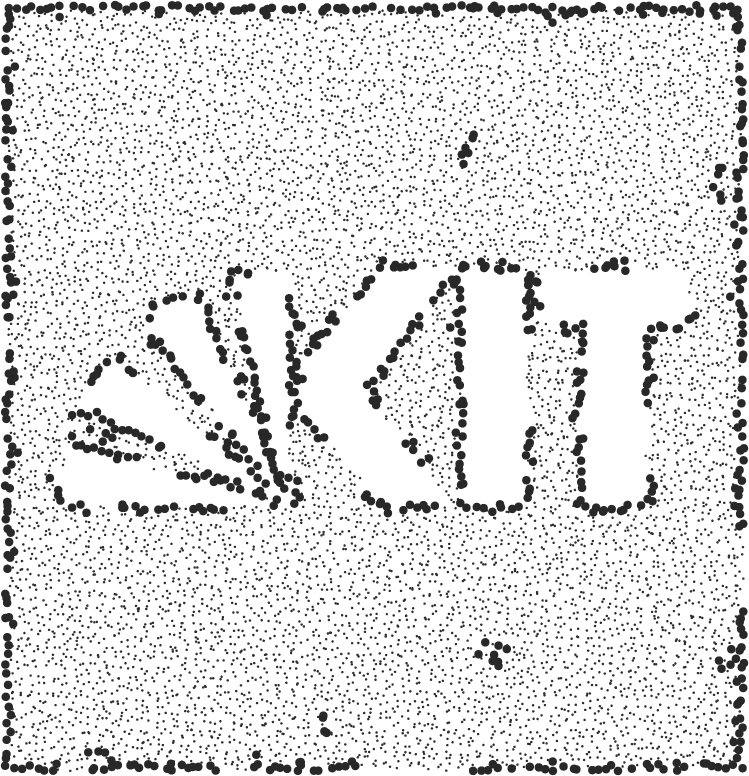}}
\hfill
\subfigure[]{\includegraphics[width=.21\columnwidth]{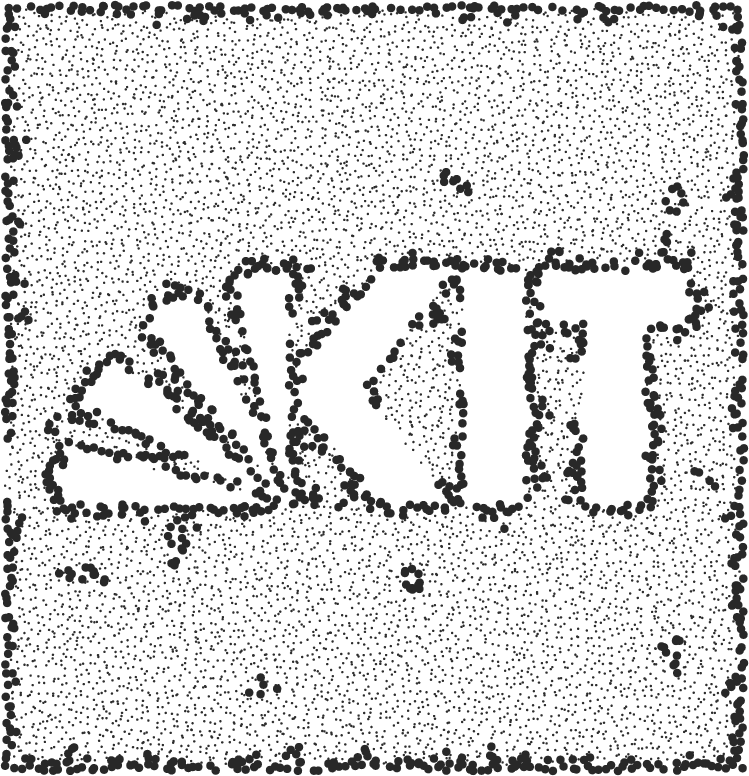}}
\hfill
\subfigure[]{\includegraphics[width=.21\columnwidth]{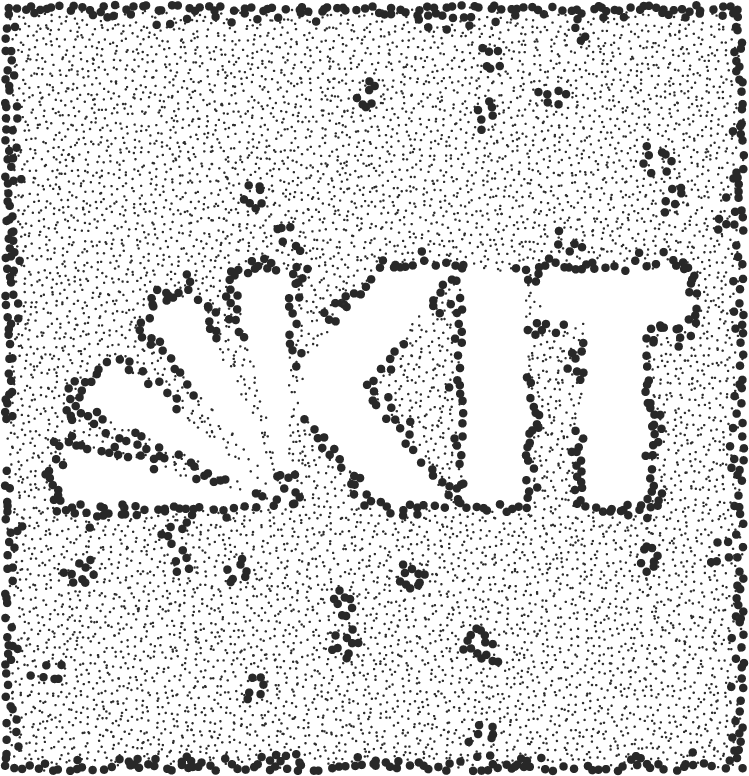}}
\hfill{}
\caption{Impact of MDS-BR refinement on different algorithms. (a) MDS-BR. (b) Fekete04. (c)Funke05. (d)~Funke06.}
\label{fig:refinements}
\end{figure}

One might wonder whether our refinement procedures could be used in conjunction with the other algorithms to improve their results.
We now analyze the performance when MDS-BR refinement is applied to their results.
As seen in Figure~\ref{fig:refinements}, the refinement removes some of the initial noise, but the results remain worse than MDS-BR.
In case of Kroeller04, the noise is almost completely gone but so are large parts of the boundary.
The results of Funke05 retain some of the original noise but still show a broad boundary structure at the border.
Similar to EC-BR with MDS-BR refinement, Funke06 retains boundaries around micro-holes, but boundaries around small structures are completely lost.
We do not present classifications for these algorithms using EC-BR refinement as the results would be similar to MDS-BR with EC-BR refinement for the same reasons.

In light of the already poor visual results, we refrained from performing extensive quantitative simulations.
Overall, it became evident that our refinement procedures are tailored for their respective base algorithms.
In conjunction with them they yield formidable results, but if applied to the results of a different algorithm, the final classification turned out to be quite poor.
This also refutes the assumption that our algorithms only perform so well due to the refinement step and others could close the gap using a similar refinement.
Essentially both, the base algorithms and the refinement procedures, contribute to the impressive results.

\section{Conclusion}
We proposed two novel \emph{distributed} algorithms for \emph{location-free} boundary recognition in wireless networks.
Both of them only depend on \emph{connectivity information} of small \emph{local neighborhoods} around each node, at most 3 hops for MDS-BR and only 2 hops for EC-BR.
This is a huge improvement over most existing approaches.
The resulting low communication overhead makes both algorithms excellent choices for boundary recognition in large-scale sensor networks.
This also makes them well suited for scenarios which include mobility or dynamic changes of the network topology.

We showed in extensive simulations that both algorithms are very robust to different network densities, communication models, and node distributions.
Despite their simplicity and low communication overhead, they outperformed the other considered approaches significantly.
Additionally, they have much lower computational complexity than most existing approaches.
Especially EC-BR should be applicable even on the weakest sensor nodes.

One might ask why we propose two algorithms for the same problem.
First of all, both ideas seemed to be very promising, so we decided to examine both and compare their results.
It turned out that both algorithms performed surprisingly well, especially considering their simplicity and low communication requirement.
Summarizing, one can say that EC-BR provides slightly better classifications with less communication and computation complexity, whereas MDS-BR offers more flexibility to adjust the classification results.

Possible future work includes comparison to a broader selection of existing boundary detection approaches.
Regarding MDS-BR, its classification conditions could be further improved as already suggested.
Exploiting angular information of the $2$-hop neighborhood of each node seems to be a promising approach.
So far, both refinement steps only turn boundary nodes into interior nodes.
An improved refinement that can also correct falsely classified interior nodes would be another worthwhile future investigation.


\section{Acknowledgments}
This work was supported by the German Research Foundation (DFG) within the Research Training Group GRK 1194 ''Self-organizing Sensor-Actuator-Networks``.
We would like to thank Bastian Katz for useful discussions, as well as Yue Wang and Olga Saukh for providing us with the topologies of their respective sensor networks.


\bibliographystyle{splncs03}
\bibliography{references2}



\end{document}